\documentclass[11pt,english]{article}
\usepackage{jheppub}
\usepackage{amsmath,amsfonts,epsfig}
\usepackage{amssymb}
\usepackage{mathrsfs}
\usepackage{hyperref}

\title{Exact finite volume expectation values of $\bar{\Psi} \Psi$ in the Massive Thirring model 
from light-cone lattice correlators}

\author{\'Arp\'ad Heged\H us}

\affiliation{MTA Lend\"ulet Holographic QFT Group,
Wigner Research Centre,\\
H-1525 Budapest 114, P.O.B. 49, Hungary}

\emailAdd{hegedus.arpad@wigner.mta.hu}

\abstract{In this paper, using the light-cone lattice regularization, we compute the finite volume expectation 
values of the composite operator $\bar{\Psi}\Psi$ between pure fermion  states in the Massive Thirring Model. 
In the light-cone regularized picture, this expectation value is related to 2-point functions of 
lattice spin operators being located at neighboring sites of the lattice. The operator $\bar{\Psi}\Psi$ is proportional 
to the trace of the stress-energy tensor. This is why the continuum finite volume expectation values can be computed also
from the set of non-linear integral equations (NLIE) governing the finite volume spectrum of the theory. 
Our results for the expectation values coming from the computation of lattice correlators agree with 
those of the NLIE computations. 
Previous conjectures for the LeClair-Mussardo-type series representation of the expectation values  
are also checked.}

\begin{document}
\numberwithin{equation}{section}
\setlength{\unitlength}{.8mm}

\maketitle



\section{Introduction}

Finite volume matrix elements of local operators play an important role in several applications 
of integrable quantum field theories. Namely, they are fundamental building blocks of the form 
factor perturbation theory \cite{DMS96} and their determination is indispensable for the computation of 
the string field theory vertex \cite{BJsftv} and of the
heavy-heavy-light 3-point functions \cite{Bjhhl} in the planar $AdS_5/CFT_4$ correspondence.

In the past decade a remarkable progress has been made in the computation of finite volume form 
factors in integrable quantum field theories \cite{PT08a,PT08b,Pmu,FT,FPT11,Palmai13,Pozsg13,PST14}. 
Most of the methods use the infinite volume form factors \cite{Smirnov} as a starting point 
and the finite volume form factors are to be determined 
in the form of a systematic large volume series. As a first step the large volume corrections, that decay 
with a power of the volume were determined \cite{PT08a,PT08b} and not much later a method 
was proposed \cite{Pmu} for computing some special type exponential in volume corrections.
These investigations shed light on the fact that the computation of diagonal matrix elements 
is a much simpler task than that of the non-diagonal ones. Recently, in \cite{BajnokWu} it 
has been shown, that the leading term in the large volume series representation of 
the diagonal form-factors in \cite{PT08b,saleur} can be derived from the 
formulas for the non-diagonal form-factors of \cite{PT08a}, by taking
the diagonal limit appropriately.

Though the structure of exponentially
 small in volume corrections for the non-diagonal matrix elements is still unknown, inspired
 by \cite{LM99} for the diagonal matrix elements, a nice series representation was proposed 
in \cite{Pozsg13,PST14}.
However the proposal is valid only to purely elastic scattering theories and its extension to 
non-diagonally scattering theories is still unknown in general.

Recently, in the Massive Thirring (sine-Gordon) model a similar series representation was proposed to describe 
the finite volume diagonal form factors of the theory \cite{En}. The conjecture was based on the computation 
of the diagonal solitonic (fermionic) matrix elements of the $U(1)$ current
 from the light-cone lattice regularization \cite{ddvlc} of the theory.

The purpose of this paper is two-fold. On the one hand we would like to demonstrate that the light-cone 
lattice approach admits an appropriate framework for computing the finite volume form factors 
of the Massive Thirring (sine-Gordon) model and on the other hand we would like to give further 
justification for the validity of the LeClair-Mussardo 
type series representation conjectured in \cite{En}.

To do so we compute the diagonal form factors of the composite operator $\bar{\Psi}\Psi$ 
from the light-cone lattice approach. There are several advantages of the choice of this operator.
First of all, this operator is proportional to the trace of the stress-energy tensor.
Thus the results of reference \cite{zamiz} imply, that up to a constant factor 
these expectation values can be computed simply from the non-linear integral equations 
(NLIE) governing the finite volume spectrum of the model \cite{KP1}-\cite{Fevphd}. 
This makes possible to check the results 
coming from the lattice computations against a result coming from a completely 
different method. 
Second of all, the operator $\bar{\Psi}\Psi$ is still simple enough not to mix with other 
operators under renormalization. Nevertheless, contrary to the case of the $U(1)$-current \cite{En}, 
in this case an infinite renormalization constant arises in accordance with field theoretical computations
\cite{ZinnBook}.

In the present paper, using the framework of Quantum Inverse Scattering Method \cite{FST79}-\cite{LukTer}, 
we compute such special spin-spin 2-point functions on the lattice, in which the spin operators
are located at neighboring sites of the lattice. A straightforward computation shows, that 
the discretized version of the continuum operator $\bar{\Psi}\Psi$ corresponds to 
the lattice operator: $\sigma^+_n \sigma^-_{n+1}+\sigma^-_n \sigma^+_{n+1}.$ 
We compute the expectation values of these operators between those Bethe eigenstates which 
correspond to the pure fermion (soliton) states in the continuum theory.
 Then we show, that in the continuum limit these fermionic expectation values (as expected) 
are proportional to the fermionic diagonal matrix elements of the trace of the stress-energy tensor.
 Latter can be computed purely \cite{zamiz} from the NLIE description of the sandwiching states.
Our method, by nature accounts for the lattice artifacts, as well.

Our results also show, 
that in the continuum limit, when the lattice constant tends to zero,
 the leading order divergence arising in the fermionic expectation values of $\bar{\Psi}\Psi$ is of the 
same form as that
expected 
from the renormalization group analysis of the Massive Thirring (sine-Gordon) model.  
Finally, we also checked that the all order conjecture \cite{En} for the systematic large volume series 
representation of the diagonal fermionic (solitonic) form-factors of the Massive Thirring (sine-Gordon) 
model is also valid for this operator. 

The outline of the paper is as follows:
In section 2. we recall the most important properties of the
 Massive Thirring and sine-Gordon models and their light-cone lattice regularizations. 
 This section contains the pure NLIE computation of 
 the fermionic (solitonic) expectation values of the trace of the stress-energy tensor.
In section 3. we summarize the Quantum Inverse Scattering Method framework and
the lattice part of the computation of the special spin-spin 2-point functions of interest.
The continuum limit procedure is described in section 4.
In section 5. we rephrase our results in the form  of a systematic large volume series and check the 
validity of the conjecture of \cite{En}.
Our summary and outlook closes the body of the paper in section 6. 
The paper contains three appendices, as well. In appendix \ref{appSUM} we rewrite the sums entering the 
lattice formulas for the two-point functions into integral expressions. In appendix \ref{appArg} 
we describe how to compute the lattice cutoff tend to zero limit within these integral expressions. 
Finally, appendix \ref{largeARG} contains the large argument series representations 
of the convolution integrals being necessary for the computations.

\section{Light-cone lattice approach to the Massive-Thirring and sine-Gordon models}

The Massive Thirring (MT) model  is defined by the Lagrangian:
\begin{equation}
\label{mTh_Lagrangian}
{\cal L}_{MT}= \bar{\Psi }(i\gamma _{\nu }\partial ^{\nu }-m_{0})\Psi -\displaystyle\frac{g}{2}\bar{\Psi }\gamma^{\nu }\Psi \bar{\Psi }\gamma _{\nu }\Psi \,,
\end{equation}
where $m_0$ and $g$ denotes the bare mass and the coupling constant of the theory, respectively.
As usual, $\gamma_\mu$s stand for the $\gamma$-matrices. They satisfy 
the algebraic relations: $\{\gamma^\mu,\gamma^\nu\}=2 \eta^{\mu \nu}$ with $\eta^{\mu \nu}=\text{diag}(1,-1)$. 
Throughout the paper we use the chiral representation for the fermions as follows:
\begin{equation}\label{gammamus}
 \Psi=\begin{pmatrix} \psi_L \\ \psi_R \end{pmatrix}, \quad \gamma^0=\begin{pmatrix} 0 & 1 \\ 1 & 0 \end{pmatrix}, \quad 
\gamma^1=\begin{pmatrix} 0 & 1 \\ -1 & 0 \end{pmatrix}, \qquad
\gamma^5=\gamma^0 \gamma^1=-\eta=\begin{pmatrix} -1 & 0 \\ 0 & 1 \end{pmatrix}.
\end{equation} 
It is well known \cite{s-coleman}, that this fermion model can be mapped to the sine-Gordon (SG) model:
\begin{equation}
\label{sG_Lagrangian}
{\cal L}_{SG}= \displaystyle\frac{1}{2}\partial _{\nu }\Phi \partial ^{\nu }\Phi +\displaystyle \alpha_0  \left( \cos \left( \beta \Phi \right)-1 \right), \,  \qquad 0<\beta^2<8 \pi,
\end{equation}
provided the coupling constants of the two theories are related by the formula:
\begin{equation} \label{gbeta}
1+\frac{g}{4 \pi}=\frac{4 \pi}{\beta^2}.
\end{equation}
A more detailed investigation of this equivalence \cite{klassme} pointed out, that the two models are identical only in the even topological charge sector of 
their Hilbert-spaces and they differ in the odd topological charge sector.

The operator we study in this paper is the fermion bilinear $\bar{\Psi}\Psi$ in the MT model. 
To be more precise, here $\bar{\Psi}\Psi$ means the bare (unrenormalized) fermion bilinear of the model.
According to the equivalence \cite{s-coleman} it is proportional to the potential of the sine-Gordon model
\cite{ZinnBook}:
\begin{equation}
\label{opekv}
\bar{\Psi}\Psi \leftrightarrow \frac{1}{\pi a} \, \cos(\beta \Phi),
\end{equation}
with $a$ being a cutoff in coordinate space.
The perturbing operator $\cos(\beta \Phi)$ of the SG model is related to the trace of the stress-energy 
tensor $\Theta_T$ as follows\footnote{In this sine-Gordon - Massive Thirring correspondence, the components of the stress energy tensors of the two models are mapped onto each other.}:
\begin{equation}\label{tracerel}
\Theta_T=4 \pi \alpha_0 \, (1-\tfrac{\beta^2}{8 \pi})\, \cos(\beta \, \Phi).
\end{equation}
From (\ref{opekv}) and (\ref{tracerel}) the fermion bilinear can be expressed in terms of the trace of the stress-energy tensor as follows:
\begin{equation}\label{ppT0}
\bar{\Psi}\Psi\sim \, \frac{\beta^2}{4 \pi^2 \,(1-\beta^2/8 \pi)}\,    \frac{\Theta_T}{a \, \alpha_0}. 
\end{equation}
Due to renormalization effects $\alpha_0$ scales with the coordinate space cutoff $a$ as $ \alpha_0 \sim a^{-{\beta^2}/{4 \pi}}$ \cite{ZinnBook}, thus
\begin{equation}\label{ppT}
\bar{\Psi}\Psi\sim  a^{{\beta^2}/{4 \pi}-1} \, \Theta_T.
\end{equation}
The minimal length $a$ can be thought of as a lattice constant, as well. From (\ref{ppT}) it can be seen that the matrix elements of 
$\bar{\Psi}\Psi$ are divergent in the attractive regime ($\beta^2<4  \pi$) and the operator valued 
coefficient of the leading order divergence in $a$ is 
proportional to the trace of the stress-energy tensor{\footnote{The trace of the stress-energy tensor 
is a conserved quantity in the continuum quantum field theory, this is why it is not subjected to 
multiplicative renormalization. Consequently, its matrix elements are finite in the continuum limit.}}.
 In this paper we show that our light-cone lattice computations  account for the scaling behavior (\ref{ppT}) and 
up to a constant factor, allow one to compute the diagonal matrix elements of $\Theta_T$.

\subsection{The light-cone lattice regularization}

\begin{figure}[htb]
\begin{flushleft}
\hskip 15mm
\leavevmode
\epsfxsize=120mm
\epsfbox{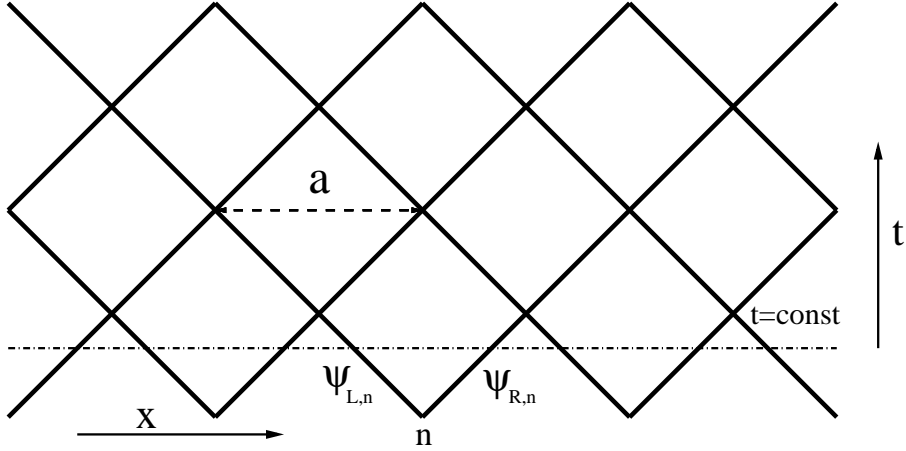}
\end{flushleft}
\caption{{\footnotesize
The pictorial representation of the light-cone lattice. 
}}
\label{figure}
\end{figure}

The light-cone lattice regularization scheme \cite{ddvlc} admits an appropriate lattice approach to the even topological charge sector of the MT model. 
In this description the space-time is discretized along the light-cone directions: $x_{\pm}=x\pm t$ with an even number of lattice sites in the spatial direction.
The sites of the light-cone lattice correspond to the discretized points of space-time. The left- and right-mover fermion fields live on the left- and right-oriented
edges of the lattice. In this manner a left- and a right-mover  fermion field can be associated to each site of the lattice (See figure \ref{figure}.).
Lattice Fermi operators satisfy the discretized version of the usual anti-commutation 
relations:
 \begin{equation} \label{ferm1}
 \{\psi_{A,n},\psi_{B,m}\}=0, \quad \{\psi_{A,n},\psi^{+}_{B,m}\}=\delta_{AB}\, \delta_{nm}, 
\quad A,B=R,L, \quad 1\leq m,n\leq N.
 \end{equation}
 As figure \ref{figure}. indicates, the chirality of the Fermi operators is related to 
the parity of the lattice-site index.
 Namely, left-mover fields live on the odd- and right-mover fields live on the even-edges of the lattice, respectively:
 \begin{equation} \label{ferm2}
 \psi_{R,n}=\psi_{2 n}, \quad \psi_{L,n}=\psi_{2n-1}, \quad 1\leq n \leq \tfrac{N}{2}.
 \end{equation}
 For later purposes it is worth to rewrite the lattice Fermi operators in terms of the spin-operators of the lattice.
 This can be achieved  by a Jordan-Wigner transformation:
 \begin{equation} \label{JWtf}
 \psi^+_n=\sigma^+_n \prod\limits_{l=1}^{n-1} \sigma^z_l, \qquad \psi_n=\sigma^-_n \prod\limits_{l=1}^{n-1} \sigma^z_l,
 \end{equation}
where $\sigma^\pm=\tfrac{1}{2}(\sigma^x \pm i\, \sigma^y)$ with $\sigma^{x,y,z}$ being 
the Pauli-matrices.

 The dynamics of the regularized model is given by light-cone evaluation operators: $U_L$ and $U_R.$
 They are given by transfer matrices of an inhomogeneous 6-vertex model \cite{ddvlc}:
 \begin{equation} \label{URL}
 U_L=e^{i\tfrac{a}{2}(H-P)}={\cal T}(\xi_2|\vec{\xi}), \qquad U_R^+=e^{-i\tfrac{a}{2}(H+P)}={\cal T}(\xi_1|\vec{\xi}),
\end{equation}
 where ${\cal T}$ is the trace of the monodromy matrix over the auxiliary space $V_0 \simeq \mathbb{C}^2,$
 \begin{equation} \label{trans}
{\cal T}(\lambda|\vec{\xi})=\text{Tr}_0 \, T(\lambda|\vec{\xi}), \qquad \left[ {\cal T}(\lambda|\vec{\xi}),{\cal T}(\lambda'|\vec{\xi}) \right] =0.
\end{equation}
The monodromy matrix is given by the $R$-matrix of the 6-vertex model in the usual way \cite{KMT00},
 \begin{equation} \label{monodromy}
 T(\lambda|\vec{\xi})=R_{01}(\lambda-\xi_1)\, R_{02}(\lambda-\xi_2)\, ...R_{0N}(\lambda-\xi_N)=\begin{pmatrix} A(\lambda) & B(\lambda) \\
 C(\lambda) & D(\lambda) \end{pmatrix}_{[0]},
 \end{equation}
 \begin{equation} \label{Rmatrix}
 R(\lambda)=\begin{pmatrix} 1 & 0 & 0 & 0 \\
 0 & \tfrac{\sinh(\lambda)}{\sinh(\lambda-i \gamma)} & \tfrac{\sinh(-i \gamma)}{\sinh(\lambda-i \gamma)} & 0 \\
0 & \tfrac{\sinh(-i \gamma)}{\sinh(\lambda-i \gamma)}  & \tfrac{\sinh(\lambda)}{\sinh(\lambda-i \gamma)} & 0 \\
0 & 0 & 0 & 1 
\end{pmatrix},
\end{equation}
such that $\xi_n$s denote the inhomogeneities of the model.
 The entries of the $2 \times 2$ monodromy matrix act on the quantum space of the model ${\cal{H}}={{\otimes}_{i=1}^N} \, V_i$ with $V_i \simeq \mathbb{C}^2$ and they 
 play crucial role in the algebraic Bethe-Ansatz solution of the model. In (\ref{URL}) $H$, $P$ and $a$ denote the Hamiltonian, the momentum
 and the lattice constant of the model, respectively. In order to get a massive interacting quantum field theory as the continuum limit of this 
 lattice model, the inhomogeneities of the vertex-model must be chosen as follows \cite{ddvlc}:
  \begin{equation}\label{xis}
 \xi_n=\rho_n -i \tfrac{\gamma}{2}, \qquad \rho_n=(-1)^n \rho_0, \qquad n=1,..,N,
 \end{equation}
such that the parameter $\rho_0$ must be tuned with the lattice constant $a$, or equivalently 
with the number of lattice sites $N$ according to the formula as follows:
\begin{equation} \label{rho}
\rho_0=\tfrac{\gamma}{\pi} \, \ln \tfrac{4}{{\cal M} \, a}=\tfrac{\gamma}{\pi} \, \ln \tfrac{2 \, N}{{\cal M} \, L},
\end{equation}
where ${\cal M}$ denotes the physical mass of fermions (solitons) of the MT (SG) model, $L$ stands for the 
finite volume and $N$ is the 
number{\footnote{In this convention, in the light-cone lattice the number of lattice sites in spatial direction is $\tfrac{N}{2}.$ See figure \ref{figure}.}} 
of lattice sites of the 6-vertex model.  

The parameters of the regularized lattice model are the inhomogeneities, the number of lattice sites and the anisotropy parameter $\gamma.$ 
In (\ref{xis}) and (\ref{rho}) we described how to choose the inhomogeneities to obtain a massive interacting integrable quantum field theory 
in the continuum limit. 
The infinite volume solution of the model{\footnote{The infinite volume solution consists of two steps.
 First, the $N\to \infty$ limit is taken with $a$ kept finite. Equation (\ref{rho}) implies that 
this means that the inhomogeneity $\rho_0$ is also kept finite. Then the $a \to 0$ limit is taken by 
tuning $\rho_0$ in the large $N$ result according to (\ref{rho}).}} shows \cite{deVegaIJMP}
that this massive continuum quantum field theory is nothing but the 
MT or SG model, provided the following relation holds  between the anisotropy parameter of the vertex model and the coupling constants of the 
Lagrangians (\ref{mTh_Lagrangian}) and (\ref{sG_Lagrangian}):
\begin{equation}\label{csatrel}
 \frac{\beta^2}{4 \pi}=\frac{1}{1+\tfrac{g}{4 \pi}}=2 (1-\tfrac{\gamma}{\pi}).
 \end{equation}
 For later purpose it is worth to introduce a new parameterization for the anisotropy parameter:
\begin{equation} \label{csatrel1}
\gamma=\tfrac{\pi}{p+1}, \quad \text{with} \quad 0<p< \infty, \quad  \text{then:} \quad \frac{\beta^2}{4 \pi}=\frac{2 p}{p+1}.
\end{equation}
 We note, that the regimes $0<p<1$ and $1<p$ correspond to the attractive and repulsive regimes of the quantum field 
theory, respectively.

The definition (\ref{URL}) embeds the light-cone evolution operators of our model into the hierarchy of mutually commuting set of 
transfer matrices of the 6-vertex model. This implies that the Hamiltonian 
and the momentum of the model can be diagonalized 
via the Algebraic Bethe Ansatz method \cite{FST79}.

\subsection{Algebraic Bethe Ansatz}

In the framework of algebraic Bethe Ansatz, 
the eigenvectors of the transfer matrix (\ref{trans}) 
 are constructed by successive application of creation operators on the 
bare vacuum of the model. The bare vacuum or reference state $|0 \rangle$ is 
 the completely ferromagnetic state with all spins up. The role of creation 
operators are played by the $12$-matrix element of the monodromy matrix (\ref{monodromy}):
$T_{12}(\lambda)=B(\lambda)$ which decreases the $S_z$ quantum number of a state by $1.$
A state constructed in this manner:
\begin{equation} \label{sajatvec}
|\vec{\lambda}\rangle=|\lambda_1,\lambda_2,..,\lambda_m \rangle=B(\lambda_1)\, B(\lambda_2)\, ...B(\lambda_m)\,|0 \rangle, \qquad  S_z|\vec{\lambda}\rangle=(\tfrac{N}{2}-m)|\vec{\lambda}\rangle,
\end{equation}
is an is an eigenvector of the transfer matrix, provided the spectral parameters in the arguments of the creation 
operators satisfy the Bethe equations:
\begin{equation} \label{BAE}
\prod\limits_{i=1}^N \, \frac{\sinh(\lambda_a-\xi_i-i \gamma)}{\sinh(\lambda_a-\xi_i)} \, 
\prod\limits_{b=1}^m \, \frac{\sinh(\lambda_a-\lambda_b+i \gamma)}{\sinh(\lambda_a-\lambda_b-i \gamma)}=-1, \qquad a=1,...,m.
\end{equation}
In the Algebraic Bethe Ansatz approach the solutions of the Bethe equations play central role, since all physical quantities can be expressed in terms of these roots.
The eigenvalue of the transfer matrix (\ref{trans}) on a Bethe-eigenvector (\ref{sajatvec}) is given by the 
formula: 
\begin{equation} \label{transsajat}
{\cal T}_{\vec{\lambda}}(\mu|\vec{\xi})=\prod\limits_{k=1}^m \, \frac{\sinh(\mu-\lambda_k+i \gamma)}{\sinh(\mu-\lambda_k)}+
\prod\limits_{i=1}^N \, \frac{\sinh(\mu-\xi_i)}{\sinh(\mu-\xi_i-i \gamma)} \,
\prod\limits_{k=1}^m \, \frac{\sinh(\mu-\lambda_k-i \gamma)}{\sinh(\mu-\lambda_k)}.
\end{equation}
For the cases, when the number of Bethe-roots is large, it is more convenient to reformulate the Bethe-equations
(\ref{BAE}) in their logarithmic form. The central object of this formulation is the so-called counting-function. 
For the choice of inhomogeneities (\ref{xis}) it is  
defined by the formula \cite{ddv97}:
\begin{equation} \label{countfv}
Z_\lambda(\lambda)=\frac{N}{2} \, \left( \phi_1(\lambda-\rho_0)+\phi_1(\lambda+\rho_0)\right)-\sum\limits_{k=1}^{m} \, \phi_2(\lambda-\lambda_k),
\end{equation}
where $\phi_\nu(\lambda)$ is an odd function on the whole complex plane with all discontinuities 
running parallel to the real axis. In its fundamental domain $|\text{Im} \, \lambda| < \nu$, it is given by the analytic formula:
\begin{equation} \label{phinu}
\phi_\nu(\lambda)=-i \, \log \frac{\sinh(i \tfrac{\gamma}{2} \nu -\lambda)}{\sinh(i \tfrac{\gamma}{2} \nu +\lambda)}, \qquad 0<\nu, \qquad \phi_{\nu}(0)=0, \qquad |\text{Im} \, \lambda| < \nu.
\end{equation}
The counting-function allows one to reformulate the Bethe-equations (\ref{BAE}) 
in the form as follows:
\begin{equation} \label{BAEZ}
Z_{\lambda}(\lambda_a)=2 \pi \, I_a, \quad  \quad I_a\in \mathbb{Z}+\tfrac{1+\delta}{2}, \qquad \delta=m \,\, (\text{mod} \, 2),
\qquad a=1,..,m.
\end{equation}
In this formulation, depending on the value of $\delta,$ an integer or half-integer quantum number 
$I_a$ can be assigned to each Bethe-root. When one considers states formed by only real Bethe-roots, then 
all these quantum numbers are different{\footnote{Due to the appropriate choice of branch cuts for $\phi_\nu(\lambda).$}} 
and they characterize the state uniquely.

The true vacuum corresponding to the ground state of the quantum field theory, is the $S_z=0,$ anti-ferromagnetic vacuum with $\delta=0$.
This state is formed by real Bethe-roots such that the 
quantum numbers of the Bethe-roots fill completely the whole allowed range 
$\left[Z_{\lambda}(-\infty)/ 2 \pi,Z_{\lambda}(\infty)/ 2 \pi \right].$ 
The excitations above this sea of real roots are characterized by complex Bethe-roots and holes.
In this paper we will consider only hole excitations, since they correspond to fermion or soliton excitations 
of the continuum quantum field theory \cite{fioravanti,ddv97,FRT1,FRT2,FRT3,Fevphd}. 
The holes are such special real solutions of
 (\ref{BAEZ}), which are not Bethe-roots{\footnote{Namely, they do not enter in the definition of
 $Z_{\lambda}(\lambda)$ in (\ref{countfv})}}. 
Holes can be interpreted as missing Bethe-roots and the quantum numbers of the missing Bethe-roots
 can be assigned to them:
\begin{equation} \label{BAEH}
Z_{\lambda}({h}_k)=2 \pi \, I_k, \quad  \quad I_k\in \mathbb{Z}+\tfrac{1+\delta}{2}, \qquad k=1,..,m_H,
\end{equation} 
where $h_k$ denotes the positions of the holes and their number is denoted by $m_H.$


\subsection{NLIE for the finite volume spectrum}

When one has to deal with a large number of Bethe-roots, it is worth to rephrase 
the Bethe-equations (\ref{BAE}) or equivalently (\ref{BAEZ}) in a form of 
a set of nonlinear-integral equations (NLIE) \cite{KP1}-\cite{FRT3}.

Here we present the equations only for the pure hole sector of the theory \cite{fioravanti}
 and here we will use the rapidity convention for the equations. This means a simple
 rescaling of the spectral parameter: $\theta=\tfrac{\pi}{\gamma}\lambda.$

In the pure hole sector, the counting-function in rapidity variable
 $Z_N(\theta)=Z_{\lambda}(\tfrac{\gamma}{\pi}\theta)$ satisfy 
the nonlinear-integral equations as follows:
\begin{equation} \label{DDVlat}
\begin{split}
Z_N(\theta)=\frac{N}{2}\left\{\arctan\left[\sinh(\theta-\Theta)\right]+\arctan\left[ \sinh(\theta+\Theta)\right]  \right\}+\sum\limits_{k=1}^{m_H} \, \chi(\theta-H_k) \\
+ \, \int\limits_{-\infty}^{\infty} \frac{d\theta'}{2 \pi i} \, G(\theta-\theta'-i\eta) \, L_N^{(+)}(\theta'+i \eta)
-\, \int\limits_{-\infty}^{\infty} \frac{d\theta'}{2 \pi i} \, G(\theta-\theta'+i\eta) \, L_N^{(-)}(\theta'-i \eta),
\end{split}
\end{equation}
where $\chi(\theta)$ is the soliton-soliton scattering 
phase and $G(\theta)$ denotes its derivative;
\begin{equation} \label{CHI}
\chi(\theta)=\! 2 \int\limits_{0}^{\infty} \! d\omega \, 
\frac{\sin(\omega \, \theta)}{\omega} \, \frac{\sinh(\tfrac{(p-1) \,\pi \omega}{2})}{2 \cosh( \tfrac{\pi \omega}{2}) 
\,\sinh(\tfrac{p \,  \pi \, \omega}{2})},
\end{equation}
\begin{equation} \label{G}
G(\theta)= \frac{d}{d\theta} \chi(\theta)=\! \int\limits_{-\infty}^{\infty} \! d\omega \, 
e^{-i \, \omega \theta} \, \frac{\sinh(\tfrac{(p-1) \,\pi \omega}{2})}{2 \cosh( \tfrac{\pi \omega}{2}) 
\,\sinh(\tfrac{p \,  \pi \, \omega}{2})},
\end{equation}
$0<\eta<\text{min}(p \pi,\pi)$ is an arbitrary positive contour-integral parameter,
$\Theta=\ln \tfrac{2 \, N}{{\cal M} \, L}$ is the inhomogeneity parameter of the vertex-model 
and $H_k=\tfrac{\pi}{\gamma} h_k$ denote the 
positions of the holes in the rapidity convention. 
They are subjected to the quantization equations:
\begin{equation} \label{QHk}
Z_N(H_k)=2 \pi \, I_k,\, \qquad I_k\in \mathbb{Z}+\tfrac{1+\delta}{2}, \qquad k=1,..,m_H.
\end{equation}
The nonlinearity of the equations is encoded into the form of $L_N^{(\pm)}(\theta),$
 which takes the form:
\begin{equation} \label{LNpm}
L_N^{(\pm)}(\theta)=\ln\left(1+(-1)^\delta \, e^{\pm i \, Z_N(\theta)} \right).
\end{equation}
The number of holes is not independent of the $S_z$ quantum number of the state. 
The connection between these two quantum numbers is given by the 
counting-equation{\footnote{Here we present the equations without the presence of special objects.
For a more detailed description see for example \cite{ddv97,Fevphd}.} \cite{ddv97}:
\begin{equation} \label{countlat}
m_H=2 \, S_z-2 \left[ \tfrac{1}{2}+\tfrac{S_z}{p+1} \right],
\end{equation}
where here $[...]$ stands for integer part. This equation immediately implies that 
on a lattice with even number of sites, only states with even number of holes can exist.

The main advantage of formulating the spectral problem in terms of the counting function is that it has a well-defined continuum limit.
If one keeps the hole quantum numbers fixed, it is just the $N \to \infty$ limit of the 
lattice counting-function \cite{ddv92,ddv95}:
\begin{equation} \label{Zcontlim}
Z(\theta)=\lim_{N \to \infty} Z_N(\theta), \qquad
L_{\pm}(\theta)=\lim_{N \to \infty}\, L_N^{(\pm)}(\theta)=\ln\left(1+(-1)^\delta \, e^{\pm i \, Z(\theta)} \right).
\end{equation}
The continuum counting-function satisfy the nonlinear-integral equations as follows \cite{fioravanti}-\cite{FRT3}: 
\begin{equation} \label{DDVcont}
\begin{split}
Z(\theta)=\ell \sinh \theta +\sum\limits_{k=1}^{m_H} \, \chi(\theta-H_k) 
+ \, \int\limits_{-\infty}^{\infty} \frac{d\theta'}{2 \pi i} \, G(\theta-\theta'-i\eta) \, L_{+}(\theta'+i \eta) \\
-\, \int\limits_{-\infty}^{\infty} \frac{d\theta'}{2 \pi i} \, G(\theta-\theta'+i\eta) \, L_-(\theta'-i \eta),
\end{split}
\end{equation}
where $\ell={\cal M} \, L$ with $L$ being the volume and ${\cal M}$ is the fermion (soliton) mass.
The holes formally satisfy exactly the same quantization equations as in the lattice model:
\begin{equation} \label{contQHk}
Z(H_k)=2 \pi \, I_k,\, \qquad I_k\in \mathbb{Z}+\tfrac{1+\delta}{2}, \qquad 
\delta \in \{0,1\},\qquad k=1,..,m_H.
\end{equation}
 The energy and momentum of the pure hole states in the continuum theory read as:
\begin{equation} \label{Econt}
\begin{split}
E={\cal M} \, \sum_{k=1}^{m_H} \, \cosh H_k -\frac{{\cal M}}{2 \pi i} \sum\limits_{\alpha=\pm}  \alpha \! \cdot \!\!\!\!
\int\limits_{-\infty}^{\infty} \! \! d\theta \, \sinh(\theta+i \, \alpha\, \eta) \, L_{\alpha}(\theta+i \, \alpha \, \eta),
\end{split}
\end{equation}
\begin{equation} \label{Pcont}
\begin{split}
P={\cal M} \, \sum_{k=1}^{m_H} \, \sinh H_k -\frac{{\cal M}}{2 \pi i} \sum\limits_{\alpha=\pm}  \alpha \! \cdot \!\!\!\!
\int\limits_{-\infty}^{\infty} \! \! d\theta \, \cosh(\theta+i \, \alpha\, \eta) \, L_{\alpha}(\theta+i \, \alpha \, \eta).
\end{split}
\end{equation}
The counting-equation (\ref{countlat}) also changes non-trivially in the continuum limit \cite{ddv97,En}: 
\begin{equation} \label{contcont}
m_H=Q,
\end{equation}
where $Q$ is the $U(1)$ (topological) charge of the continuum model.

The NLIE (\ref{DDVcont}) can be solved iteratively in the large volume limit. From this solution it follows, 
that the nonlinear terms $L_{\pm}(\theta \pm i \,\eta)$ are exponentially small in the volume. As a consequence 
in (\ref{Econt}) and (\ref{Pcont}) the integral terms can be dropped in the infinite volume limit and
one ends up with the energy  and momentum formulas of $m_H$ pieces of fermions (solitons) with rapidities $\{H_j\}_{j=1...m_H}.$ This implies that the holes in the sea of real roots describe the fermions (solitons) 
of the MT (SG) model. This is why in the sequel we will refer to holes as fermions or solitons.

Finally, we note that the actual value{\footnote{On the lattice the actual value of $\delta$ can be influenced by the parity of $\tfrac{N}{2}.$}} of the quantum number $\delta$ is important from the 
point of view of the continuum theory. Its value can make difference between fermions ($\delta=1$)
 of the MT model and the solitons ($\delta=0$) of the SG model in the odd $U(1)$ charge sector
 of the theory \cite{FRT1}-\cite{FRT3}.  In the even charge sector only the $\delta=0$ value is physical and there 
is no difference between MT fermions and SG solitons \cite{FRT1}-\cite{FRT3}.

From the discussion above it follows that only the even charge sector of the MT and SG 
models can be regularized by the twistless 6-vertex model. The description of the 
odd charge sector requires a twisted vertex-model with a twist angle $\omega=\tfrac{\pi}{2}$
 \cite{BHsg}. However, in this paper we restrict ourselves to the twistless case.

\subsection{Expectation values of the trace of the stress-energy tensor}

In this subsection using the NLIE description of the finite volume spectrum
given by (\ref{DDVcont}) and (\ref{Econt}), 
we compute the fermionic (solitonic) expectation values of 
 the trace of the stress-energy tensor $\Theta_T.$ 
The finite temperature 1-point functions, which correspond to the finite volume 
vacuum expectation value, has been previously computed and discussed
 in \cite{BS6} and \cite{BucT}.

It has been shown in \cite{zamiz} that the diagonal matrix 
elements of $\Theta_T$ can be computed from the volume dependence of the energy 
of the sandwiching state by the following formula:
\begin{equation}\label{tetaT}
\langle \Theta_T \rangle=\langle \Theta_T^\infty \rangle+2 \pi \, {\cal M} 
\left( \frac{E(\ell)}{\ell}+\frac{d E(\ell)}{d \ell}  \right).
\end{equation}
In the sequel we compute $\langle \Theta_T \rangle$ when the sandwiching state is 
an $m_H$-fermion state described by the equations (\ref{DDVcont}) and (\ref{Econt}).


As a starting point, it is worth to compute the infinite volume 
or in other words the bulk expectation value: 
 $\langle \Theta_T^\infty \rangle.$ Using Zamolodchikov's argument \cite{zamiz},
 it can be expressed 
in terms of the eigenstate independent bulk energy of the model by the formula:
\begin{equation}\label{tetaTinf}
\langle \Theta_T^\infty \rangle=2 \pi \, {\cal M} 
\left( \frac{E_{bulk}(\ell)}{\ell}+\frac{d E_{bulk}(\ell)}{d \ell}  \right).
\end{equation}
In the MT (SG) model the bulk energy term is of the form \cite{ddv95}:
\begin{equation} \label{Ebulk}
\begin{split}
E_{bulk}(\ell)=-\tfrac{{\cal M} \ell}{4} \, \tan(\tfrac{p \pi}{2}).
\end{split}
\end{equation}
Inserting (\ref{Ebulk}) into (\ref{tetaTinf}) one obtains:
\begin{equation} \label{tetaTinf1}
\begin{split}
\langle \Theta_T^\infty \rangle=-\pi {\cal M}^2 \, \tan(\tfrac{p \pi}{2}).
\end{split}
\end{equation}
As a next step we express the non-bulk part of $\langle \Theta_T \rangle$ in (\ref{tetaT}) 
in terms of the solution of the NLIE (\ref{DDVcont}). 
To do so, it is worth to introduce some useful notations.
Let ${\cal F}_{\pm}(\theta)$ denote the nonlinear combinations as follows:
\begin{equation} \label{calF}
{\cal F}_{\pm}(\theta)=\frac{(-1)^\delta \, e^{\pm i \, Z(\theta)}}
{1+(-1)^\delta \, e^{\pm i \, Z(\theta)}}. 
\end{equation}
Then the derivative of $L_{\pm}(\theta)$ with respect to any parameter ${\cal P}$ 
is given by the formula:
\begin{equation} \label{dLpm}
\begin{split}
\frac{d L_{\pm}(\theta)}{d {\cal P}}=\pm i \, \frac{d Z(\theta)}{d {\cal P}} \, 
{\cal F}_{\pm}(\theta).
\end{split}
\end{equation}
In practice ${\cal P}$ can denote one of the parameters of the NLIE equations (\ref{DDVcont}). 
Namely, it can be the dimensionless volume $\ell$, the spectral parameter $\theta$ or one 
of the positions of the holes $H_j$.
The second term in the right hand side of (\ref{tetaT}) consists of two terms. The first term 
$\sim  \tfrac{E(\ell)}{\ell}$
can be expressed in terms
of  $\frac{d Z(\theta)}{d {\cal \theta}}$ and of ${\cal F}_{\pm}(\theta)$, 
while the second term 
$\sim \frac{d E(\ell)}{d \ell}$
turns out to be the functional of $\frac{d Z(\theta)}{d {\cal \ell}}$
 and of ${\cal F}_{\pm}(\theta).$

Integrating the right hand side of (\ref{Econt}) by parts,  $\tfrac{E(\ell)}{\ell}$ can be rephrased as follows:
\begin{equation} \label{EL/L}
\begin{split}
\frac{E(\ell)}{\ell}=\frac{{\cal M}}{\ell} \, \sum\limits_{k=1}^{m_H} \, \cosh(H_k) 
X_k^{(d)}+
\frac{\cal M}{ \ell} \, \sum\limits_{\alpha =\pm} \int\limits_{-\infty}^{\infty}
 \! \frac{d\theta}{2 \pi} \,\cosh(\theta+i \, \alpha \, \eta)
\, {\cal G}_d(\theta+i \, \alpha \, \eta) \, {\cal F}_{\alpha} (\theta+i \, \alpha \, \eta),
\end{split}
\end{equation}
where ${\cal G}_d(\theta)=Z'(\theta),$ $X_k^{(d)}=\frac{{\cal G}_d(H_k)}{Z'(H_k)}=1.$  
Differentiating (\ref{DDVcont}) with respect to $\theta,$ one can show, that 
they satisfy the set of linear integral equations as follows:
\begin{equation} \label{llie_d}
\begin{split}
{\cal G}_{d}(\theta)&-\sum\limits_{\alpha =\pm} \int\limits_{-\infty}^{\infty} \! \frac{d\theta'}{2 \pi} G(\theta-\theta'-i \, \alpha \, \eta)
{\cal G}_{d}(\theta'+i \, \alpha \, \eta) \, {\cal F}_{\alpha} (\theta'+i \, \alpha \, \eta)
=\\
&=\ell \, \cosh(\theta)+\sum\limits_{j=1}^{m_H} \, G(\theta-H_j) \, X_j^{(d)}, \\
X_j^{(d)}&=\frac{{\cal G}_{d}(H_j)}{Z'(H_j)}, \qquad j=1,...,m_H. \\
\end{split}
\end{equation}
Taking the derivative of (\ref{Econt}) and (\ref{DDVcont}) with respect to $\ell$, leads the following
 expression for $\frac{d E(\ell)}{d \ell}:$
\begin{equation} \label{dEL/dL}
\begin{split}
\frac{d E(\ell)}{d \ell}=-{\cal M} \, \sum\limits_{k=1}^{m_H} \, \sinh(H_k) \,
X_k^{(\ell)}-{\cal M} \, \sum\limits_{\alpha =\pm} \int\limits_{-\infty}^{\infty}
 \! \frac{d\theta}{2 \pi} \,\sinh(\theta+i \, \alpha \, \eta)
\, {\cal G}_{\ell}(\theta+i \, \alpha \, \eta) \, {\cal F}_{\alpha} (\theta+i \, \alpha \, \eta),
\end{split}
\end{equation}
where ${\cal G}_{\ell}(\theta)=\frac{d Z(\theta)}{d \ell},$ $X_k^{(\ell)}=\frac{{\cal G}_{\ell}(H_k)}{Z'(H_k)}=-H^{\prime}_k(\ell).$  
They are solutions of the set of linear integral equations as follows{\footnote{The $X_k^{(\ell)}=-H^{\prime}_k(\ell)$ 
 equation can be derived by taking the derivative of the hole quantization equation (\ref{contQHk}) with respect to $\ell.$ }}:
\begin{equation} \label{llie_l}
\begin{split}
{\cal G}_{\ell}(\theta)&-\sum\limits_{\alpha =\pm} \int\limits_{-\infty}^{\infty} \! \frac{d\theta'}{2 \pi} G(\theta-\theta'-i \, \alpha \, \eta)
{\cal G}_{\ell}(\theta'+i \, \alpha \, \eta) \, {\cal F}_{\alpha} (\theta'+i \, \alpha \, \eta)
=\\
&= \sinh(\theta)+\sum\limits_{j=1}^{m_H} \, G(\theta-H_j) \, X_j^{(\ell)}, \\
X_j^{(\ell)}&=\frac{{\cal G}_{\ell}(H_j)}{Z'(H_j)}, \qquad j=1,...,m_H. \\
\end{split}
\end{equation}
Plugging (\ref{EL/L}) and (\ref{dEL/dL}) into (\ref{tetaT}) we obtain our final 
formula for the fermionic (solitonic) expectation values of the trace of the stress-energy tensor: 
\begin{equation} \label{tetaTfinal}
\begin{split}
&\langle \Theta_T \rangle=\langle \Theta_T^\infty \rangle+
2 \pi {\cal M}^2 \, \sum\limits_{k=1}^{m_H} \left\{ \, \cosh(H_k) \,
\frac{X_k^{(d)}}{\ell}-\sinh(H_k) \, X_k^{(\ell)} \right\}+ \\
 \,&+{\cal M}^2 \, \sum\limits_{\alpha =\pm} \int\limits_{-\infty}^{\infty}
 \! d\theta \,\left[ \cosh(\theta+i \, \alpha \, \eta)
\, \frac{{\cal G}_d(\theta+i \, \alpha \, \eta)}{\ell}
-\sinh(\theta+i \, \alpha \, \eta)
\, {\cal G}_{\ell}(\theta+i \, \alpha \, \eta) \right]\,
 {\cal F}_{\alpha} (\theta+i \, \alpha \, \eta).
\end{split}
\end{equation}
Representation (\ref{tetaTfinal}) for $\langle \Theta_T \rangle$ should be used as follows.
First one has to solve the NLIE equations (\ref{DDVcont}) for the sandwiching fermion (soliton)
 state. Then the linear integral equations (\ref{llie_d}) and (\ref{llie_l}) should be solved. 
Finally, inserting these solutions into (\ref{tetaTfinal}) gives the required expectation value.
 Though, this representation for $\langle \Theta_T \rangle$ might seem strange for the first sight, but 
in the later sections it will turn out, that it fits very well for the structure of the lattice results. 

In the rest of the paper our main goal is to reproduce the formula (\ref{tetaTfinal})
 from the  light-cone lattice computation of the multi-fermion (soliton) expectation 
values of $\bar{\Psi}\Psi.$

\subsection{The lattice counterpart of $\bar{\Psi}\Psi$}

We close this section with a short discussion about the lattice counterpart of the operator $\bar{\Psi}\Psi$ 
in the MT model. A simple Jordan-Wigner transformation (\ref{JWtf}) shows, that certain bilinears of the 
lattice Fermi operators are simple expressions of the lattice spin operators:
 \begin{equation} \label{oplat1}
\begin{split}
\psi^+_n \psi_{n+1}&=\sigma_{n}^+ \, \sigma_{n+1}^-, \\
\psi^+_{n+1} \psi_{n}&=\sigma_{n}^- \, \sigma_{n+1}^+,
\end{split}
\end{equation}
where $\sigma^{\pm}_n$  are the usual spin creation and annihilation operators corresponding to the $n$th site of the 
lattice, while $\psi_n$ and $\psi^+_n$ are the lattice Fermi operators defined by (\ref{ferm1}) and (\ref{ferm2}).

Using the representation (\ref{gammamus}) for $\gamma^0,$ the following identification can be made for the unrenormalized bare operators on the lattice: 
\begin{equation} \label{PPbare}
\begin{split}
\bar{\Psi}\Psi(x)\big|_{x=n \, a}=\Psi^+_R(x)\Psi_L(x)\big|_{x=n \, a}+\Psi^+_L(x)\Psi_R(x)\big|_{x=n \, a}\rightarrow
\tfrac{1}{a}\psi^+_{R,n}\psi_{L,n}+\tfrac{1}{a}\psi^+_{L,n}\psi_{R,n}=\\
=\tfrac{1}{a}\psi^+_{2n} \psi_{2n+1}+\tfrac{1}{a}\psi^+_{2n+1} \psi_{2n}
=\tfrac{1}{a}\sigma_{2n}^+ \, \sigma_{2n+1}^-+\tfrac{1}{a}\sigma_{2n}^- \, \sigma_{2n+1}^+,
\end{split}
\end{equation}
where the term $\tfrac{1}{a}$ is introduced to account for the correct bare dimension of the continuum Fermi field.

A similar computation shows that the pseudo-scalar combination of the Fermi operators correspond to the 
antisymmetric combination of the spin operators:
\begin{equation} \label{PP5bare}
\begin{split}
\bar{\Psi}\gamma^5 \Psi(x)\big|_{x=n \, a}\rightarrow \tfrac{1}{a}\psi^{+}_{2n} \psi_{2n+1}-\tfrac{1}{a}\psi^{+}_{2n+1} \psi^{}_{2n}
=\tfrac{1}{a}\sigma_{2n}^- \, \sigma_{2n+1}^+-\tfrac{1}{a}\sigma_{2n}^+ \, \sigma_{2n+1}^-.
\end{split}
\end{equation}

Thus (\ref{PPbare}) and (\ref{PP5bare}) implies, that the determination of the expectation values 
of the bare scalar- and pseudo-scalar fermion bilinears is equivalent to computing the two-point functions 
of neighboring spin operators. This task is completed in the rest of the paper via the QISM \cite{FST79}-\cite{LukTer}. 

We note that beyond the computation of 2-point functions $\langle \sigma^{\pm}_n \sigma^{\mp}_{n+1}\rangle$
the 2-point function $\langle e_{n} \, e_{n+1} \rangle$ with $e_n=\tfrac{1}{2}(1_n-\sigma^z_n)$ can also be computed 
with the techniques presented in this paper.
This latter 2-point function contains a combination a 4-fermion term, as well: 
$e_{n} \, e_{n+1}=(\psi^{+}_n\psi_n-\tfrac{1}{2})(\psi^{+}_{n+1}\psi_{n+1}-\tfrac{1}{2}).$
A usual argument based on the bare dimensions of the operators implies that 
this operator has a nontrivial 
mixing under renormalization. This means that the correct 
implementation of the renormalization process requires 
the computation of the expectation values of further operators. This investigation is left for future work.


\section{Computation of lattice correlators}

The strategy of computing the fermionic (solitonic) expectation values of the operators $\bar{\Psi}\Psi$ and
 $\bar{\Psi}\gamma^5 \Psi$
consists of three main steps. First, one has to compute the expectation values of the lattice operators 
$\sigma^{\pm}_n \sigma^{\mp}_{n+1}$ in pure hole states. The second step is to consider the symmetric (\ref{PPbare}) and 
anti-symmetric (\ref{PP5bare}) combinations of these expectation values in order to describe the diagonal 
form-factors of the operators $\bar{\Psi}\Psi$ and $\bar{\Psi}\gamma^5 \Psi,$ respectively.
Finally, one has to take the continuum limit of the lattice results by 
sending the number of lattice sites $N$ to infinity such that the inhomogeneity parameter $\rho_0$ is tuned 
according to (\ref{rho}).  
In \cite{En} the efficiency of this 
method has been demonstrated via the computation of the solitonic (fermionic) expectation values the $U(1)$ current of the model.
In this section we describe in detail the lattice part of the computations.

Consider a vector of the Hilbert-space obtained by successive actions of creation operators on the 
bare vacuum:
\begin{equation} \label{bra}
|\vec{\lambda} \rangle=B(\lambda_1)\, B(\lambda_2)...B(\lambda_m) \, |0 \rangle.
\end{equation}
Such a state is called Bethe-state if the numbers $\lambda_j$ are arbitrary and it is called Bethe-eigenstate
if  the set $\{\lambda_j\}_{j=1,..,m}$ is equal to the set of roots of the Bethe equations (\ref{BAE}).
The corresponding "bra" vector can be defined by acting from the right with the annihilation operators on the 
"bra" bare vacuum:
\begin{equation} \label{ket}
\langle \vec{\lambda}|=\langle 0|\, C(\lambda_m)... C(\lambda_2) \,C(\lambda_1).
\end{equation}

The determination of the 
diagonal form-factors of $\bar{\Psi}\Psi$ and $\bar{\Psi}\gamma^5\Psi$ requires the computation
of the following two special 2-point functions:
\begin{equation} \label{sspm}
\begin{split}
\langle \sigma^\pm_n \sigma^{\mp}_{n+1} \rangle_\lambda=\frac{\langle \vec{\lambda}|\sigma^\pm_n \sigma^{\mp}_{n+1}|\vec{\lambda}\rangle}
{\langle \vec{\lambda} |\vec{\lambda}\rangle },
\end{split}
\end{equation}
where here $|\vec{\lambda}\rangle$ denotes a Bethe-eigenstate. 

The determination of these 2-point functions can be achieved in a purely algebraic way \cite{KMT99,KMT00}
within the framework of the QISM \cite{FST79}, such that only the Yang-Baxter algebra relations 
and the expression of local spin operators in terms of the elements of the monodromy matrix (\ref{monodromy}) of the model are used 
\cite{KMT99}.  
 
The core of the algebraic computations is the relation between the local spin operators and the 
elements of the Yang-Baxter algebra \cite{KMT99}:
\begin{equation} \label{ISM}
E^{ab}_n=\prod\limits_{i=1}^{n-1} \, (A+D)(\xi_i) \, \, T_{ab}(\xi_n) \, \prod\limits_{i=n+1}^{N} \, 
(A+D)(\xi_i), \qquad a,b=1,2,
\end{equation}
where the operator $E_n$ is given in terms of local spin operators as follows:
\begin{equation} \label{En}
E^{11}_n=\frac{1}{2}({1}_n+\sigma^z_n), \quad E_n^{12}=\sigma^-_n, \quad E_n^{21}=\sigma^+_n, \quad E^{22}_n=\frac{1}{2}({1}_n-\sigma^z_n).
\end{equation}
The formulas (\ref{ISM}) and (\ref{En}) imply the following representation for the 
2-point correlators (\ref{sspm}) of our interest:
\begin{equation} \label{sigma_mp0}
\begin{split}
\langle \sigma_n^- \, \sigma_{n+1}^+ \rangle_\lambda=
\frac{\langle \vec{\lambda}|\sigma_n^- \, \sigma_{n+1}^+ | \vec{\lambda}\rangle }{\langle \vec{\lambda}|\vec{\lambda} \rangle}=
\frac{1}{{\cal T}_{\vec{\lambda}}(\xi_n|\vec{\xi}) \,{\cal T}_{\vec{\lambda}}(\xi_{n+1}|\vec{\xi})} \, \frac{\langle \vec{\lambda}|B(\xi_n) \, C(\xi_{n+1})| \vec{\lambda}\rangle}{\langle
\vec{\lambda}|\vec{\lambda} \rangle},
\end{split}
\end{equation}
\begin{equation} \label{sigma_pm0}
\begin{split}
\langle \sigma_n^+ \, \sigma_{n+1}^- \rangle_\lambda=
\frac{\langle \vec{\lambda}|\sigma_n^+ \, \sigma_{n+1}^- | \vec{\lambda}\rangle }{\langle \vec{\lambda}|\vec{\lambda} \rangle}=
\frac{1}{{\cal T}_{\vec{\lambda}}(\xi_n|\vec{\xi}) \,{\cal T}_{\vec{\lambda}}(\xi_{n+1}|\vec{\xi})} \, \frac{\langle \vec{\lambda}|C(\xi_n) \, B(\xi_{n+1})| \vec{\lambda}\rangle}
{\langle \vec{\lambda}|\vec{\lambda} \rangle},
\end{split}
\end{equation}
where here $|\vec{\lambda} \rangle$ denotes a Bethe-eigenstate, ${\cal T}_{\vec{\lambda}}(\lambda|\vec{\xi})$ 
denotes the eigenvalue of the transfer matrix (\ref{trans}) on the state $| \vec{\lambda}\rangle$ and $\xi_n$ is 
the inhomogeneity parameter belonging to the $n$th site of the vertex-model.


To compute (\ref{sigma_mp0}) and (\ref{sigma_pm0}), 
we need to know, how an operator $B(\xi_n)$ with $\xi_n$ being an inhomogeneity of the vertex model, acts on a "bra"-vector (\ref{ket}).
This is given by the formula \cite{KMT00}:
\begin{equation} \label{Bact}
\begin{split}
\langle 0|\! \!\!  \prod\limits_{k=1}^M \! \! C(\lambda_k)  B(\xi_n)\!\! = \! \! \sum\limits_{a=1}^M  \!
{\frak f}_M^{(0)}(\lambda_a|\xi_n)  \!\! \left\{ \!
{\frak f}_M^{(1)}(\lambda_a|\xi_n)  \langle 0|\!\! \prod\limits_{ k=1 \atop k \neq a
 }^M  \!\!\! C(\lambda_k) \!\! + \! \!\!
\sum\limits_{b=1 \atop b \neq a}^M  {\frak f}_M^{(2)}(\lambda_a,\lambda_b|\xi_n) 
\langle 0|\!\!\! \prod\limits_{ k=1 \atop k \neq a,b
 }^M  \! \!\! \! C(\lambda_k)  C(\xi_n) \!\!
\right\}\!\!.
\end{split} 
\end{equation}
Here the functions ${\frak f}_M^{(0)},{\frak f}_M^{(1)} \text{and }  {\frak f}_M^{(2)}$ are of the form:
\begin{eqnarray} \label{f012}
{\frak f}_M^{(0)}(\lambda_a|\xi)&=&\frac{1}{r(\lambda_a) \, \sinh(\lambda_a-\xi)} \, 
\frac{\prod\limits_{k=1}^M \sinh(\lambda_a-\lambda_k-i \, \gamma)}{ \prod\limits_{ k=1 \atop k \neq a
 }^M  \sinh(\lambda_a-\lambda_k) }, \nonumber \\ 
{\frak f}_M^{(1)}(\lambda_a|\xi)&=&\frac{\sinh(\lambda_a-\xi)}{\sinh(\lambda_a-\xi-i \, \gamma)} \,
\prod\limits_{j=1}^M \, \frac{\sinh(\lambda_j-\xi-i \, \gamma)}{\sinh(\lambda_j-\xi)},
 \\
{\frak f}_M^{(2)}(\lambda_a,\lambda_b|\xi)&=&\frac{1}{\sinh(\lambda_a-\lambda_b-i \, \gamma) \, \sinh(\xi-\lambda_b)} \, 
\frac{\prod\limits_{j=1}^M \sinh(\lambda_j-\lambda_b-i \, \gamma)}{ \prod\limits_{ j=1 \atop j \neq a,b
 }^M  \sinh(\lambda_j-\lambda_b)},
\nonumber 
\end{eqnarray}
where 
\begin{equation} \label{rdef}
r(\lambda)=\prod\limits_{j=1}^N \, \frac{\sinh(\lambda-\xi_j-i \,\gamma)}{\sinh(\lambda-\xi_j)}.
\end{equation}
From its definition it follows that its inverse becomes zero at the positions of the inhomogeneities 
of the lattice model:
\begin{equation} \label{rid1}
 \frac{1}{r(\xi_j)}=0, \quad j=1,...,N.
\end{equation}
We note that in general, in (\ref{Bact}) and (\ref{f012}), $\lambda_k$s can take any complex value
and they do not need to be solutions of the Bethe-equations (\ref{BAE}).
On the other hand it follows from (\ref{rdef}) and (\ref{BAE}) that if $\lambda_k$s are solutions 
of (\ref{BAE}), then  $r(\lambda)$ satisfies the  identity:
\begin{equation} \label{rid2}
\prod\limits_{k=1}^m r(\lambda_k)=1.
\end{equation}
Straightforward application of (\ref{Bact}) to (\ref{sigma_mp0}) and (\ref{sigma_pm0}) lead to the 
following formulas:
\begin{equation}\label{muBC}
\begin{split}
\langle \sigma_n^- \, \sigma_{n+1}^+ \rangle_\lambda=\frac{1}{{\cal T}_{\vec{\lambda}}(\xi_n|\vec{\xi}) \,{\cal T}_{\vec{\lambda}}(\xi_{n+1}|\vec{\xi})} \, \Bigg \{
\sum\limits_{a=1}^m {\frak f}_{m}^{(0)}(\lambda_a|\xi_n) \,
{\frak f}_{m}^{(1)}(\lambda_a|\xi_n) \, \frac{\langle \vec{\mu}^{(a)}(\xi_{n+1})|\vec{\lambda}\rangle}{\langle \vec{\lambda}|\vec{\lambda} \rangle}+ \\
+\sum\limits_{a=1}^{m} {\frak f}_{m}^{(0)}(\lambda_a|\xi_n) \sum\limits_{b=1 \atop b \neq a}^m 
{\frak f}_{m}^{(2)}(\lambda_a,\lambda_b|\xi_n) \, \frac{\langle \vec{\mu}^{(a,b)}(\xi_n,\xi_{n+1})|\vec{\lambda}\rangle}{\langle \vec{\lambda}|\vec{\lambda}\rangle} 
\Bigg \},
\end{split}
\end{equation}
\begin{equation}\label{muCB}
\begin{split}
& \langle \sigma_n^+ \, \sigma_{n+1}^- \rangle_\lambda=\frac{1}{{\cal T}_{\vec{\lambda}}(\xi_n|\vec{\xi}) \,{\cal T}_{\vec{\lambda}}(\xi_{n+1}|\vec{\xi})} \, \Bigg\{ \sum\limits_{a=1}^{m} {\frak f}_{m+1}^{(0)}(\lambda_a|\xi_{n+1}) \sum\limits_{b=1 \atop b \neq a}^m 
{\frak f}_{m+1}^{(2)}(\lambda_a,\lambda_b|\xi_{n+1}) \, \frac{\langle \vec{\mu}^{(a,b)}(\xi_n,\xi_{n+1})|\vec{\lambda}\rangle}{\langle \vec{\lambda}|\vec{\lambda}\rangle}+ \\
&+\sum\limits_{a=1}^{m}
{\frak f}_{m+1}^{(0)}(\lambda_a|\xi_{n+1}) \,\left[
 {\frak f}_{m+1}^{(1)}(\lambda_a|\xi_{n+1})
 \frac{\langle \vec{\mu}^{(a)}(\xi_{n})|\vec{\lambda}\rangle}{\langle \vec{\lambda}|\vec{\lambda}\rangle}+
%
\, {\frak f}_{m+1}^{(2)}(\lambda_a,\xi_n|\xi_{n+1})
 \frac{\langle \vec{\mu}^{(a)}(\xi_{n+1})|\vec{\lambda}\rangle}{\langle \vec{\lambda}|\vec{\lambda}\rangle} 
\right]
\Bigg \},
\end{split}
\end{equation}
where in ${\frak f}_{m+1}^{(0,1,2)}$ defined in (\ref{f012}) with $\lambda_{m+1}\equiv \xi_n,$ and we introduced 
the following notation for the states entering the scalar products in (\ref{muBC}) and (\ref{muCB}):
\begin{itemize}
\item $|\vec{\lambda}\rangle$ denotes a Bethe-eigenstate (\ref{bra}) characterized by $m$ Bethe-roots. 
\item $|\vec{\mu}^{(a)}(\xi)\rangle$ denotes a Bethe-state, the difference of which 
from $|\vec{\lambda}\rangle$ is that a single $\lambda_a \to \xi$ replacement should be done in (\ref{bra}).
\item $| \vec{\mu}^{(a,b)}(\xi,\xi')\rangle$ denotes a Bethe-state, 
which differs from $|\vec{\lambda}\rangle$ by the $\lambda_a \to \xi$ and  $\lambda_b \to \xi'$
 replacements in (\ref{bra}).
\end{itemize}
We note that in (\ref{muBC}) and (\ref{muCB}), we exploited that ${\frak f}_{M}^{(1)}(\xi_n|\xi_{n'})=0$ 
for any values of $M$ and of $n, n' \in \{1,...,N\}.$


Formulas in (\ref{muBC}) and (\ref{muCB}) imply, that in order to carry out 
the computation of the necessary special 2-point functions, one needs to know
 the scalar product of a Bethe-state and a Bethe-eigenstate. This is given by
 Slavnov's determinant formula \cite{Sla89}.
Let $|\vec{\mu} \rangle$ an arbitrary Bethe-state in the sense of (\ref{bra}) and $|\vec{\lambda} \rangle$ be a Bethe-eigenstate.
Then their scalar product can be determined with the help of the formula \cite{Sla89}:
\begin{equation} \label{skalarH}
\langle \vec{\mu}|\vec{\lambda} \rangle=\langle \vec{\lambda} |\vec{\mu} \rangle=
\prod\limits_{l=1}^m \, \frac{1}{ r(\mu_l)} \cdot
\frac{\text{det} \, H(\vec{\mu}|\vec{\lambda})}{ \prod\limits_{j>k} \sinh(\mu_k-\mu_j) \, \sinh(\lambda_j-\lambda_k)},
\end{equation}
where $H(\vec{\mu}|\vec{\lambda})$ is an $m \times m$ matrix with entries:
\begin{equation} \label{Hdef}
H_{ab}(\vec{\mu}|\vec{\lambda})=\frac{\sinh(-i \, \gamma)}{\sinh(\lambda_a-\mu_b)} 
\left( r(\mu_b) \, \frac{\prod\limits_{k=1}^m  \sinh(\lambda_k-\mu_b-i \, \gamma)}{\sinh(\lambda_a-\mu_b-i \, \gamma)} 
-\frac{\prod\limits_{k=1}^m  \sinh(\lambda_k-\mu_b+i \, \gamma)}{\sinh(\lambda_a-\mu_b+i \, \gamma)}
\right).
\end{equation}
An important special case of (\ref{skalarH}), when the scalar product of two identical Bethe-eigenstates are considered.
This is given by the Gaudin formula \cite{Gaudin0,Gaudin1,Korepin}:
\begin{equation} \label{Gaudin}
\langle \vec{\lambda} |\vec{\lambda} \rangle  =
\frac{\prod\limits_{j=1}^{m} \prod\limits_{k=1}^{m} \sinh(\lambda_j-\lambda_k-i \, \gamma)
 }{ \prod\limits_{j>k} \sinh(\lambda_k-\lambda_j) \, \sinh(\lambda_j-\lambda_k)} \cdot \text{det} \, \Phi(\vec{\lambda}),
\end{equation}
where  
 $\Phi(\vec{\lambda})$ is the Gaudin-matrix, which is related to the counting-function (\ref{countfv}) by the formula:
\begin{equation} \label{Phi}
\Phi_{ab}(\vec{\lambda})=-i \,\frac{\partial}{\partial \lambda_b} \, Z_{\lambda}(\lambda_a|\vec{\lambda}), \quad a,b=1,..,m.
\end{equation}
Using the actual form of the counting function (\ref{countfv}), from (\ref{Phi}) one obtains the following form 
for the matrix elements of $\Phi(\vec{\lambda})$:
\begin{equation} \label{Phiab}
\Phi_{ab}(\vec{\lambda})=-i \, Z^{\prime}_{\lambda}(\lambda_a) \, \delta_{ab}-2 \pi \, i \, K(\lambda_a-\lambda_b|\gamma), \quad a,b=1,..m,
\end{equation}
where
\begin{equation} \label{K1}
K(\lambda|\gamma)=\frac{1}{2 \pi} \frac{\sin(2 \, \gamma)}{\sinh(\lambda-i \, \gamma)\, \sinh(\lambda+i \, \gamma)}.
\end{equation}
As it can be seen from (\ref{muBC}) and (\ref{muCB}), 
during the computation of the special 2-point functions considered in this work, such scalar 
products arise, in which the components of the vector $\vec{\mu}$ take values either from the set of Bethe-roots 
$\{\lambda_j\}_{j=1,..m}$ or from the set of inhomogeneities $\{\xi_k\}_{k=1,..N}$ of the model.
In these cases the matrix elements of $H(\vec{\mu}|\vec{\lambda})$ remarkably simplify:
\begin{equation} \label{HPhi}
H_{ab}(\vec{\mu}|\vec{\lambda})\big|_{\mu_b \to \lambda_c}=(-1)^{m-1} \, \prod\limits_{j=1}^m \, 
\sinh(\lambda_c-\lambda_j-i \, \gamma) \, \Phi_{ac}(\vec{\lambda}), \quad a,b,c=1,..,m.
\end{equation}
 \begin{equation} \label{Hxi}
\frac{1}{r(\mu_b)} H_{ab}(\vec{\mu}|\vec{\lambda})\big|_{\mu_b \to \xi_c}\!\!=\!\!
\frac{(-1)^{m} \,\sinh(-i\, \gamma) \prod\limits_{j=1}^m  \sinh(\xi_c-\lambda_j+i \, \gamma)}
{\sinh(\lambda_a-\xi_c)  \sinh(\lambda_a-\xi_c-i \, \gamma)}, \quad a,b=1,..,m, \quad c=1,..,N.
\end{equation}
These simplifications allow{\footnote{The main technical steps of the computations are 
the same as those given in sections 4. of \cite{KMT00} and of \cite{En}.}} 
one to compute a typical scalar product arising in the computation of 
diagonal form factors:
\begin{equation} \label{mulambda}
\begin{split}
\frac{\langle \vec{\mu}^{(a_1,...,a_K)}(\xi_{\alpha_1},..,\xi_{\alpha_K}) | \vec{\lambda} \rangle}{\langle \vec{\lambda}| \vec{\lambda} \rangle}&=
\text{det} \, Y \, \prod\limits_{k=1}^K \prod\limits_{j=1}^m \frac{\sinh(\xi_{\alpha_k}-\lambda_j+i \, \gamma)}{\sinh(\lambda_{a_k}-\lambda_j-i \, \gamma)}
\, \prod\limits_{k>j}^K \frac{\sinh(\lambda_{a_k}-\lambda_{a_j})}{\sinh(\xi_{\alpha_k}-\xi_{\alpha_j})} \times \\
&\times \prod\limits_{k=1}^K \prod\limits_{j=1 \atop j \neq a_1,..a_K}^m \frac{\sinh(\lambda_{a_k}-\lambda_j)}{\sinh(\xi_{\alpha_k}-\xi_{\alpha_j})}, 
\qquad \qquad K\leq m,
\end{split}
\end{equation}
where $\langle \vec{\mu}^{(a_1,...,a_K)}(\xi_{\alpha_1},..,\xi_{\alpha_K}) |$ denotes a state, which is obtained from $\langle \vec{\lambda}|=\langle \lambda_1,...\lambda_m|$ 
by replacing $K$ pieces of $\lambda_j$ to certain inhomogeneities of the lattice model:
\begin{equation} \label{replacement}
\begin{split}
\lambda_{a_k} \rightarrow \xi_{\alpha_k}, \qquad a_k \in \{1,.., m\}, \qquad \alpha_k\in\{1,...,N \}, \qquad 1\leq k\leq K\leq m,
\end{split}
\end{equation}
such that both sets \{$a_k\}$ and $\{\alpha_k\}$ contain distinct numbers. 
In (\ref{mulambda}) $Y$ denotes a $K \times K$ matrix with entries as follows:
\begin{equation} \label{Yentry}
\begin{split}
Y_{ij}=r(\lambda_{a_i}) \, X_{a_i}(\xi_{\alpha_j}),  \qquad i,j\in \{1,...,K\},
\end{split}
\end{equation}
where the $m$-component vector $X_{b}(\xi)$ is the solution of a set of linear equations:
\begin{equation} \label{Xlin}
\begin{split}
\sum\limits_{b=1}^m \Phi_{ab}(\vec{\lambda}) \, X_{b}(\xi)={\cal V}_a(\xi), \qquad a=1,..,m,
\end{split}
\end{equation}
with 
\begin{equation} \label{NUa}
\begin{split}
{\cal V}_a(\xi)=\frac{-\sinh(-i \, \gamma)}{\sinh(\lambda_a-\xi) \, \sinh(\lambda_a-\xi-i \,\gamma)}, \qquad a \in \{1,...,m \}.
\end{split}
\end{equation}
In subsection 4.1 of \cite{En} it has been shown, that the discrete linear problem (\ref{Xlin}) can be transformed
 into a set of linear integral equations. Latter formulation proves to be very convenient, 
when the continuum limit is taken. For the paper to be self-contained, we recall the derivation of the 
transformation of (\ref{Xlin}) into linear integral equations.

The actual forms (\ref{Phiab}) of the Gaudin-matrix and the source vector (\ref{NUa}) suggest, the following 
Ansatz for the $m$-component solution vector $X_a(\xi)$ of the linear equations (\ref{Xlin}):
\begin{equation}\label{Xansatz}
X_a(\xi)=X(\lambda_a|\xi), \qquad a=1,..,m,
\end{equation} 
where $X(\lambda|\xi)$ is supposed to be a meromorphic function in $\lambda$ on the complex plane, such that 
it is analytic in a small neighborhood of the real axis. Thus our task is the determine the functional form of
 $X(\lambda|\xi).$ Then using (\ref{Xansatz}), (\ref{Phiab}) 
and (\ref{K1}) the linear equations (\ref{Xlin}) take the form:
\begin{equation} \label{lineq1}
-i \, Z^{\prime}_\lambda(\lambda_a) \, X(\lambda_a|\xi)-2 \pi \, i \sum\limits_{b=1}^m K(\lambda_a-\lambda_b|\gamma) \, X(\lambda_b|\xi)=2 \pi \, i \, K(\lambda_a-\rho(\xi)|\tfrac{\gamma}{2}), \quad a=1,..,m,
\end{equation}
where $\rho(\xi)=\xi+i \tfrac{\gamma}{2}.$ In our computations $\xi$ takes value from the set of 
inhomogeneities of the vertex-model. The actual choice for the inhomogeneities given by 
(\ref{xis}) and (\ref{rho}) imply that we can restrict our investigations to the case, when
 $\rho(\xi) \in \mathbb{R}.$ Thus, in the sequel we will assume $\rho(\xi)$ to be real.

To transform (\ref{lineq1}) into integral equations one needs to use the lemma as follows \cite{ddv95,ddv97}:
\newline {\bf Lemma:}
{ \it Let $\{\lambda_j\}_{j=1,..,m}$ solutions of the Bethe-equations (\ref{BAE})
 and let $f(\lambda)$ a meromorphic function, which is continuous and bounded on the real axis. 
Denote $p^{(f)}$ its pole located the closest to the real axis.
Then for $|\text{Im}\,\mu|<|\text{Im} \, p^{(f)}|$ the following equation
 holds:
\begin{equation} \label{sumrule}
\begin{split}
\sum\limits_{j=1}^{m} f(\mu-\lambda_j)&=\sum\limits_{j=1}^{m_C}f(\mu-c_j)-\sum\limits_{j=1}^{m_H} f(\mu-h_j)+
\int\limits_{-\infty}^{\infty} \!  \frac{d \lambda}{2 \pi} \, f(\mu-\lambda) \, Z'_\lambda(\lambda) \\
&-\sum\limits_{\alpha=\pm} \int\limits_{-\infty}^{\infty} \! \frac{d \lambda}{2 \pi}  f(\mu-\lambda-i \, \alpha \, \eta) \,Z'_\lambda(\lambda+i \, \alpha \, \eta) \, 
{\cal F}^{(\lambda)}_{\alpha}(\lambda+i \, \alpha \, \eta),
\end{split}
\end{equation}
where $h_j$ and $c_j$ denote the positions of holes and complex Bethe-roots, respectively and ${\cal F}^{(\lambda)}_{\pm}(\lambda)$ is given by 
\begin{equation} \label{calFl}
{\cal F}^{(\lambda)}_{\pm}(\lambda)=\frac{(-1)^\delta \, e^{\pm i \, Z_\lambda(\lambda)}}
{1+(-1)^\delta \, e^{\pm i \, Z_\lambda(\lambda)}},
\end{equation}
 $\eta$ is a 
small positive contour-integral parameter 
which should satisfy the inequalities:
\begin{equation}\label{ineqlemma}
0<\eta<\text{min} \{|\text{Im} \, p^{\pm}_\lambda| \}, \qquad |\text{Im}\,\mu \pm \eta|<|\text{Im} \, p^{(f)}|,
\end{equation}
where $p^{\pm}_\lambda$ denotes those complex poles of ${\cal F}^{(\lambda)}_{\pm}(\lambda)$, which are 
located the closest to the real axis.} The validity of formula (\ref{sumrule}) in $\mu$ can be extended to the whole complex plane by an appropriate analytical continuation method \cite{ddv97}.

Then one has to apply (\ref{sumrule}) to the discrete sum arising in (\ref{lineq1}). From the actual form of 
(\ref{sumrule}) one can recognize, that under the integrations a factor $Z^{\prime}_\lambda(\lambda)$ always arises. 
To eliminate this factor from the equations, it is worth to formulate the integral form of the discrete 
set of equations (\ref{lineq1}) in terms of the function: 
\begin{equation} \label{Xrepr}
\begin{split}
{\cal G}(\lambda|\xi)=Z^{\prime}_\lambda(\lambda) \, X(\lambda|\xi).
\end{split}
\end{equation}
In the language of this function, 
 the discrete set of linear equations (\ref{lineq1}) take the form 
 of a set of linear integral equations as follows: 
\begin{equation} \label{llie1}
\begin{split}
 {\cal G}(\lambda|\xi)&+\! \!\! \int\limits_{-\infty}^{\infty}  d\lambda' \, K(\lambda-\lambda'|\gamma) \, {\cal G}(\lambda'|\xi) - \\
&-\! \! \sum\limits_{\alpha=\pm} \int\limits_{-\infty}^{\infty} d\lambda' \,  K(\lambda-\lambda'-i \, \alpha \, \eta|\gamma) \, {\cal G}(\lambda'+i \, \alpha \, \eta|\xi) \, {\cal F}^{(\lambda)}_{\alpha}(\lambda'+i \, \alpha \, \eta)=\\
&=-2 \pi \, K(\lambda-\rho(\xi)|\tfrac{\gamma}{2})+\sum\limits_{j=1}^{m_H} \, K(\lambda-h_j|\gamma) \, X(h_j|\xi),
\end{split}
\end{equation}
where as a consequence of (\ref{Xrepr}) the "discrete degrees of freedom" satisfy the equations:
\begin{equation} \label{XG}
X(h_j|\xi)=\frac{{\cal G}(h_j|\xi)}{Z'_\lambda(h_j)}, \qquad j=1,..., m_H.
\end{equation}
To be more precise (\ref{lineq1}) and (\ref{sumrule}) implies that (\ref{llie1}) holds only at the positions 
$\{\lambda_a\}_{a=1...m}.$ For the pure hole states of our interest all Bethe-roots are real, $\lambda_a \in \mathbb{R},$  and we need the functional form of ${\cal G}(\lambda|\xi)$ or equivalently of $X(\lambda|\xi)$ to 
solve (\ref{lineq1}). It follows that if (\ref{llie1}) is fulfilled everywhere in an appropriate neighborhood of the 
real axis, then it will be satisfied at the discrete points $\{\lambda_a\}_{a=1...m},$ as well. 
Thus, we require ${\cal G}(\lambda|\xi)$ to be the solution of the set of linear 
integral equations (\ref{llie1}). Finally acting{\footnote{This action is necessary in order for the final equations to have well defined continuum limit.}} 
on (\ref{llie1}) with the inverse of the integral
 operator{\footnote{To be more precise the kernel of the integral operator in "lambda" space is given by 
 $\delta(\lambda-\lambda')+K(\lambda-\lambda'|\gamma)$
with $\delta(\lambda)$ being the Dirac-delta function.}} 
$1+K,$ one obtains the final form \cite{En} of the linear integral equations satisfied by ${\cal G}(\lambda|\xi)$, when only pure hole excitations above the anti-ferromagnetic vacuum are considered:
\begin{equation} \label{llie}
\begin{split}
 {\cal G}(\lambda|\xi)&- \sum\limits_{\alpha=\pm} \int\limits_{-\infty}^{\infty} d\lambda' \,  G_{\lambda}(\lambda-\lambda'-i \, \alpha \, \eta) \, {\cal G}(\lambda'+i \, \alpha \, \eta|\xi) \, {\cal F}^{(\lambda)}_{\alpha}(\lambda'+i \, \alpha \, \eta)=\\
&=S_0(\lambda|\xi)+\sum\limits_{j=1}^{m_H} \,
2 \pi \, G_\lambda(\lambda-h_j) \, X(h_j|\xi),  
\end{split}
\end{equation}
with
\begin{equation}\label{ess0}
S_0(\lambda|\xi)=-\frac{\pi}{\gamma} \, \frac{1}{\cosh\left(\tfrac{\pi}{\gamma}(\lambda-\rho(\xi))\right)}, 
\qquad \rho(\xi)=\xi+i \, \tfrac{\gamma}{2} \in \mathbb{R},
\end{equation}
where $h_j$s denote the positions of the holes, $\eta$ is a 
small positive contour-integral parameter,  
and $G_\lambda(\lambda)$ is related to the kernel of NLIE equations (\ref{G}) by:
\begin{equation} \label{Gl}
G_\lambda(\lambda)=\frac{1}{2 \gamma} \, G\left(\tfrac{\pi}{\gamma} \lambda \right), \quad \text{with} \quad \gamma=\tfrac{\pi}{p+1}.
\end{equation}

Now, we can return to the computation of the 2-point functions of our interest.
Inserting (\ref{transsajat}) and (\ref{mulambda}) into (\ref{muBC}) and (\ref{muCB}), 
after some simplification one ends up with the formulas as follows:
\begin{equation} \label{sigma_mp1}
\begin{split}
\langle \sigma_n^- \, \sigma_{n+1}^+ \rangle_\lambda =-\! \! \sum\limits_{a=1}^m  &\frac{\sinh(\lambda_a-\xi_{n+1})}{\sinh(\lambda_a-\xi_n-i \, \gamma)}
 X(\lambda_a|\xi_{n+1})+\! \! \! \sum\limits_{a,b=1}^m  \frac{\sinh(\lambda_a-\xi_{n+1})  \sinh(\lambda_b-\xi_{n+1})}{\sinh(\lambda_a-\lambda_b-i \, \gamma)} \times \\
&\times \, \frac{X(\lambda_a|\xi_n)\, X(\lambda_b|\xi_{n+1})-X(\lambda_a|\xi_{n+1})\, X(\lambda_b|\xi_{n})}{\sinh(\xi_n-\xi_{n+1})},
\end{split}
\end{equation}
\begin{equation} \label{sigma_pm1}
\begin{split}
\langle & \sigma_n^+ \, \sigma_{n+1}^- \rangle_\lambda=-\frac{\sinh(\xi_n-\xi_{n+1}-i \, \gamma)}{\sinh(\xi_n-\xi_{n+1})} \, \sum\limits_{a=1}^m \frac{\sinh(\lambda_a-\xi_{n}-i \, \gamma)}{\sinh(\lambda_a-\xi_{n+1}-i \, \gamma)} \,
X(\lambda_a|\xi_n) \\
&+\sum\limits_{a,b=1}^m  \frac{\sinh(\lambda_a-\xi_{n}-i \, \gamma)  \sinh(\lambda_b-\xi_{n}+i \, \gamma)}{\sinh(\lambda_a-\lambda_b-i \, \gamma)}
 \, \frac{X(\lambda_a|\xi_n)\, X(\lambda_b|\xi_{n+1})-X(\lambda_a|\xi_{n+1})\, X(\lambda_b|\xi_{n})}{\sinh(\xi_n-\xi_{n+1})} \\
&+ \frac{\sinh(-i \, \gamma)}{\sinh(\xi_n-\xi_{n+1})} \, \sum\limits_{a=1}^m X(\lambda_a|\xi_{n+1}).
\end{split}
\end{equation}
In (\ref{sigma_mp1}) and (\ref{sigma_pm1}) we preserved the determinant structure implied by (\ref{mulambda}). Nevertheless for future computations it is better to 
shift the anti-symmetrization to the coefficient of the quadratic expression of $X.$  Then the quadratic in $X$ parts of (\ref{sigma_mp1}) and (\ref{sigma_pm1}) can be 
written as follows:
\begin{equation} \label{mpkvad}
\begin{split}
\langle \sigma_n^- \, \sigma_{n+1}^+ \rangle_\lambda^{quad} =\frac{1}{\sinh(\xi_n-\xi_{n+1})} \sum\limits_{a,b=1}^m {\frak f}(\lambda_a,\lambda_b|\xi_{n+1}) \,
X(\lambda_a|\xi_n)\, X(\lambda_b|\xi_{n+1}),
\end{split}
\end{equation}
\begin{equation} \label{pmkvad}
\begin{split}
\langle \sigma_{n}^+ \, \sigma_{n+1}^- \rangle_\lambda^{quad} =\frac{1}{\sinh(\xi_n-\xi_{n+1})} \sum\limits_{a,b=1}^m {\frak f}(\lambda_a,\lambda_b|\xi_{n}) \,
X(\lambda_a|\xi_n)\, X(\lambda_b|\xi_{n+1}),
\end{split}
\end{equation}
where ${\frak f}(\lambda,\lambda'|\xi)$ is an antisymmetric function given by the formula: 
\begin{equation} \label{fplusminus}
\begin{split}
{\frak f}(\lambda,\lambda'|\xi)=\cos(\gamma) \, \frac{\sinh(2(\lambda-\xi))-\sinh(2(\lambda'-\xi))-\sinh(2(\lambda-\lambda'))}{\cosh(2(\lambda-\lambda'))-\cos(2 \gamma)}.
\end{split}
\end{equation}
In (\ref{mpkvad}) and (\ref{pmkvad}) the coefficient function ${\frak f}$ has better large $\lambda$ and $\lambda'$ asymptotics, than the coefficients of the quadratic terms 
of (\ref{sigma_mp1}) and (\ref{sigma_pm1}). This property proves to be very useful, when the discrete sums in (\ref{sigma_mp1}) and (\ref{sigma_pm1}) are transformed into integral expressions.

The typical sums arising in (\ref{sigma_mp1}), (\ref{sigma_pm1}), (\ref{mpkvad}) and (\ref{pmkvad}) are of the form:
\begin{eqnarray} \label{SUM1}
\Sigma_\lambda^{(1)}[f](\xi)&=&\sum\limits_{a=1}^m \, f(\lambda_a) \, X(\lambda_a|\xi), \\
\label{SUM2}
\Sigma_\lambda^{(2)}[f](\xi,\xi')&=&\sum\limits_{a,b=1}^m \, f(\lambda_a,\lambda_b) \, X(\lambda_a|\xi)
\, X(\lambda_b|\xi'),
\end{eqnarray}
where in (\ref{SUM2}) $f(\lambda,\lambda')$ is meant to be an antisymmetric function.
Formulas (\ref{SUM1}) and (\ref{SUM2}) are not appropriate to take the continuum limit, this is 
why it is worth to transform these sums into integral expressions with the help of lemma (\ref{sumrule}).
The transformation procedure together with the integral representations of (\ref{SUM1}) and (\ref{SUM2}) 
can be found in appendix \ref{appSUM}. The final integral representations for (\ref{SUM1}) and (\ref{SUM2}) 
are given by (\ref{SUM1form}) and (\ref{SUM2form}) together with the related definitions.


To describe the scalar and pseudo-scalar fermion bilinears, it is worth to introduce the 
lattice operators $O^+_{2n}$ and $O^-_{2n}$ with the definitions:
\begin{equation} \label{Opm}
\begin{split}
O^{\pm}_{2n}=\sigma_{2n}^- \, \sigma_{2n+1}^+ \pm \sigma_{2n}^+ \, \sigma_{2n+1}^-.
\end{split}
\end{equation}
According to (\ref{PPbare}) and (\ref{PP5bare}) they correspond to the lattice counterparts of
 the bare fermion bilinears $\bar{\Psi}\Psi$ and $\bar{\Psi}\gamma^5 \Psi,$ respectively.
Using the formulas (\ref{sigma_mp1}), (\ref{sigma_pm1}), (\ref{mpkvad}), (\ref{pmkvad}), (\ref{SUM1}) and (\ref{SUM2}) 
the following formal representation can be given for the expectation values of $O^\pm_{2n}:$
\begin{equation} \label{Opmexp1}
\begin{split}
\langle O_{2n}^{\pm} \rangle_\lambda&=-\Sigma_\lambda^{(1)}[f_3](\xi_-)\pm \frac{\sinh(-i \, \gamma)}{\sinh(2 \rho_0)}
\, \Sigma_\lambda^{(1)}[f_1](\xi_-) \mp \frac{\sinh(2 \rho_0-i \, \gamma)}{\sinh(2 \rho_0)}
\, \Sigma_\lambda^{(1)}[f_2](\xi_+) \\
&+\frac{1}{\sinh(2\rho_0)}\Sigma_\lambda^{(2)}[f_\pm](\xi_+,\xi_-), \qquad \qquad \xi_{\pm}=\pm \rho_0-i\, \tfrac{\gamma}{2},
\end{split}
\end{equation}
where the functions $f_1,f_2,f_3$ and $f_{\pm}$ are of the form:
\begin{eqnarray}\label{fv1}
f_1(\lambda)&=&1, \\
f_2(\lambda)&=&\frac{\sinh(\lambda-\rho_0-i \, \tfrac{\gamma}{2})}{\sinh(\lambda+\rho_0-i \, \tfrac{\gamma}{2})},
\label{fv2} \\
f_3(\lambda)&=&\frac{\sinh(\lambda+\rho_0+i \, \tfrac{\gamma}{2})}{\sinh(\lambda-\rho_0-i \, \tfrac{\gamma}{2})},
\label{fv3} \\
f_+(\lambda,\lambda')&=& 2 \cos(\gamma) \,
 \frac{\cosh(2\rho_0) \left[\sinh(2\lambda+i \, \gamma)-\sinh(2\lambda'+i \, \gamma)  \right]-\sinh(2(\lambda-\lambda'))}{\cosh(2(\lambda-\lambda'))-\cos(2 \gamma)},
\label{f_+} \\
f_-(\lambda,\lambda')&=& 2 \cos(\gamma) \,\sinh(2\rho_0)
 \frac{\cosh(2\lambda+i \, \gamma)-\cosh(2\lambda'+i \, \gamma)}{\cosh(2(\lambda-\lambda'))-\cos(2 \gamma)},
\label{f_-}
\end{eqnarray}
and we exploited the concrete inhomogeneity structure of the model, namely that $\xi_{2n}=\rho_0-i \, \tfrac{\gamma}{2}$ and $\xi_{2n+1}=-\rho_0-i \, \tfrac{\gamma}{2},$ with $\rho_0$ given by (\ref{rho}).

Using the integral representations (\ref{SUM1form}) and (\ref{SUM2form}) for $\Sigma_\lambda^{(1)}[f](\xi)$ and $\Sigma_\lambda^{(2)}[f](\xi,\xi'),$ respectively, the formula (\ref{Opmexp1}) can be rephrased as follows:
\begin{equation} \label{Opm0XG}
\langle O_{2n}^{\pm} \rangle_\lambda={\cal O}^{\pm}_0+{\cal O}^{\pm}_X+{\cal O}^{\pm}_{\cal G}+{\cal O}^{\pm}_{XX}
+{\cal O}^{\pm}_{{\cal G}{\cal G}}
+{\cal O}^{\pm}_{X{\cal G}},
\end{equation}
where
\begin{eqnarray}
{\cal O}^{\pm}_0&=&-{\cal J}_0[f_3](\xi_-)\pm \frac{\sinh(-i \, \gamma)}{\sinh(2 \rho_0)}
\, {\cal J}_0[f_1](\xi_-) \mp \frac{\sinh(2 \rho_0-i \, \gamma)}{\sinh(2 \rho_0)}
\, {\cal J}_0[f_2](\xi_+), \label{O0} \\
{\cal O}^{\pm}_X&=&-{\cal S}_X[f_3](\xi_-)\pm \frac{\sinh(-i \, \gamma)}{\sinh(2 \rho_0)}
\, {\cal S}_X[f_1](\xi_-) \mp \frac{\sinh(2 \rho_0-i \, \gamma)}{\sinh(2 \rho_0)}
\, {\cal S}_X[f_2](\xi_+) \nonumber \\
&+&\frac{\Sigma_X[f_\pm](\xi_+,\xi_-)}{\sinh(2\rho_0)},  \label{OX} \\
{\cal O}^{\pm}_{\cal G}&=&-{\cal S}_{\cal G}[f_3](\xi_-)\pm \frac{\sinh(-i \, \gamma)}{\sinh(2 \rho_0)}
\, {\cal S}_{\cal G}[f_1](\xi_-) \mp \frac{\sinh(2 \rho_0-i \, \gamma)}{\sinh(2 \rho_0)}
\, {\cal S}_{\cal G}[f_2](\xi_+) \nonumber \\
&+&\frac{\Sigma_{\cal G}[f_\pm](\xi_+,\xi_-)}{\sinh(2\rho_0)},  \label{OG} \\
{\cal O}^{\pm}_{XX}&=&\frac{\Sigma_{XX}[f_\pm](\xi_+,\xi_-)}{\sinh(2\rho_0)},  \label{OXX} \\
{\cal O}^{\pm}_{X{\cal G}}&=&\frac{\Sigma_{X{\cal G}}[f_\pm](\xi_+,\xi_-)}{\sinh(2\rho_0)}, \label{OXG} \\
{\cal O}^{\pm}_{{\cal G}{\cal G}}&=&\frac{\Sigma_{{\cal G}{\cal G}}[f_\pm](\xi_+,\xi_-)}{\sinh(2\rho_0)}, 
\label{OGG}
\end{eqnarray}
such that the functionals ${\cal J}_0,$ ${\cal S}_X,$ ${\cal S}_{\cal G},$ $\Sigma_{0},$
$\Sigma_{X},$ $\Sigma_{\cal G},$  $\Sigma_{X{\cal G}},$ $\Sigma_{XX}$ and $\Sigma_{{\cal G}{\cal G}},$ are given by 
the formulas (\ref{calJG}), (\ref{SX}), (\ref{ScalG}), (\ref{SUM0}), (\ref{SUMX}), (\ref{SUMF}), 
(\ref{SUMXF}), (\ref{SUMXX}), (\ref{SUMFF}), respectively.
 
The lower index of the terms in the right hand side of (\ref{Opm0XG}) carry information about how these 
terms depend on the dynamical variables $X(\lambda|\xi)$ and ${\cal G}(\lambda|\xi).$ Namely:
\begin{itemize}
\item ${\cal O}^{\pm}_0$ stands for the "bulk" term, which is independent of $X$ and ${\cal G},$ 
\item ${\cal O}^{\pm}_X$ linear in $X$ and independent of ${\cal G},$ 
\item ${\cal O}^{\pm}_{\cal G}$ is linear in ${\cal G}$ and independent of $X,$
\item ${\cal O}^{\pm}_{X{\cal G}}$ is linear in both ${\cal G}$ and $X,$
\item ${\cal O}^{\pm}_{XX}$ is quadratic in $X$ and independent of ${\cal G},$ and finally
\item  ${\cal O}^{\pm}_{{\cal G}{\cal G}}$ is quadratic in ${\cal G}$ and independent of $X.$
\end{itemize}

Formula (\ref{Opm0XG}) together with (\ref{O0})-(\ref{OGG}) and the integral
 representations (\ref{SX})-(\ref{J_GG}) given in appendix \ref{appSUM} constitutes our final result  
for the lattice expectation values of the scalar and pseudo-scalar fermion 
bilinears. To get the expectation values of these operators in the continuum theory,
 the lattice formula (\ref{Opm0XG}) should be evaluated in the continuum limit. 
This will be discussed in the next section.

\section{Continuum limit}

In this section the expectation value formulas (\ref{Opm0XG}) are evaluated at the continuum limit. 
This task reduces to the 
evaluation of the sums and integrals entering (\ref{Opm0XG}) in the large $\rho_0$ 
limit{\footnote{In the rest of the paper, the terms "large $\rho_0$ limit" and "large $N$ 
limit" will be equivalently used for the continuum limit procedure.}}. 
The $\rho_0$ dependence of these terms is determined by the $\rho_0$ dependence of 
the functions $f_1,f_2,f_3,f_\pm$ given in 
(\ref{fv1})-(\ref{f_-}), and the $\rho_0$ dependence of $X(h_j|\xi_\pm)$ and ${\cal G}(\lambda|\xi_\pm).$
 Latter is governed by the linear integral equation (\ref{llie}).
First, it is worth to discuss the continuum limit of the variables $X(h_j|\xi_\pm)$ and ${\cal G}(\lambda|\xi_\pm).$
They are solutions of the set of equations (\ref{XG})-(\ref{ess0}). 
The continuum limit means, that 
one has to take the number of lattice sites $N$ to infinity, such that the inhomogeneity 
parameter $\rho_0$ is tuned with $N$ according to the formula (\ref{rho}). This means, that in the continuum limit 
procedure $\rho_0$ also tends to infinity, but logarithmically in $N$ or $a.$

In this limit the counting-function $Z_\lambda(\lambda)$ (\ref{countfv})and its nonlinear combinations ${\cal F}^{(\lambda)}_\pm(\lambda)$ (\ref{calFl}) tend to their (finite) continuum counterparts:
\begin{equation}
\begin{split}
Z_\lambda(\lambda) \to Z_{\lambda,c}(\lambda)=Z(\tfrac{\pi}{\gamma}\lambda), \label{Zlcont} \\
{\cal F}^{(\lambda)}_\pm(\lambda) \to {\cal F}^{(\lambda)}_{\pm,c}(\lambda)=
{\cal F}_\pm(\tfrac{\pi}{\gamma}\lambda), 
\end{split}
\end{equation}
where $Z(\theta)$ is solution of the continuum NLIE (\ref{DDVcont}) and ${\cal F}_\pm(\theta)$ is
 given by (\ref{calF}). This implies that in the leading order in $\tfrac{1}{N}$ computations 
$Z_\lambda(\lambda)$ and ${\cal F}^{(\lambda)}_\pm(\lambda)$ can be replaced by their continuum 
counterparts given by (\ref{Zlcont}).

Then equation (\ref{llie}) implies, that
the large $N$ limit of ${\cal G}(\lambda|\xi_\pm)$ is governed by the large $\rho_0$ expansion of $S_0(\lambda|\xi_\pm):$  
\begin{equation} \label{S0kifejt}
\begin{split}
S_0(\lambda|\xi_\pm)=-\tfrac{2 \pi}{\gamma} \, e^{\pm \tfrac{\pi}{\gamma}\lambda}\, e^{- \tfrac{\pi}{\gamma}\rho_0}+O(e^{- 2 \,\tfrac{\pi}{\gamma}\rho_0}).
\end{split}
\end{equation}
According to (\ref{rho}) $e^{- \tfrac{\pi}{\gamma} \rho_0} \sim \tfrac{1}{N}\sim a.$ 
Since apart from $S_0$, all terms in the equations (\ref{XG})-(\ref{ess0}) 
are proportional to ${\cal G}(\lambda|\xi)$ and $X(h_j|\xi),$ (\ref{S0kifejt}) implies that:
\begin{equation} \label{GXNa}
\begin{split}
{\cal G}(\lambda|\xi_\pm)\sim \tfrac{1}{N}+O(\tfrac{1}{N^2}) \sim a+O(a^2), \\
X(\lambda|\xi_\pm)\sim \tfrac{1}{N}+O(\tfrac{1}{N^2}) \sim a+O(a^2).
\end{split}
\end{equation}
With the help of (\ref{S0kifejt}) and (\ref{GXNa}), one can immediately give a rough estimate for the 
large $N$ magnitude of the different terms arising in the right hand side of (\ref{Opm0XG}). 
This is implied by their $S_0,$ $X,$ and ${\cal G}$ content:
\begin{eqnarray}
{\cal O}^{\pm}_0 &\sim& S_0 \sim \tfrac{1}{N} \sim e^{-(1+p) \rho_0} \qquad \qquad
{\cal O}^{\pm}_{XX} \sim X \, X \sim \tfrac{1}{N^2} \sim e^{-2(1+p) \rho_0} \nonumber \\
{\cal O}^{\pm}_X &\sim& X \sim \tfrac{1}{N} \sim e^{-(1+p) \rho_0} \qquad \qquad
{\cal O}^{\pm}_{X{\cal G}} \sim X \, {\cal G} \sim \tfrac{1}{N^2} \sim e^{-2(1+p) \rho_0} \label{rough} \\
{\cal O}^{\pm}_{\cal G} &\sim& {\cal G} \sim \tfrac{1}{N} \sim e^{-(1+p) \rho_0} \qquad \qquad
{\cal O}^{\pm}_{{\cal G}{\cal G}} \sim {\cal G} \, {\cal G} \sim \tfrac{1}{N^2} \sim e^{-2(1+p) \rho_0}. \nonumber 
\end{eqnarray}
This rough estimate implies, that in the continuum limit the terms in (\ref{Opm0XG}) being quadratic or 
multilinear
 in ${\cal G}$ and $X,$ (i.e. $O^{\pm}_{XX},{\cal O}^{\pm}_{{\cal G}{\cal G}},{\cal O}^{\pm}_{{X}{\cal G}}$) 
become negligible with respect to the constant (${\cal O}^{\pm}_0$) and linear terms (${\cal O}^{\pm}_{\cal G},{\cal O}^{\pm}_{X}$). Thus only the constant and purely linear terms determine the leading order behavior of 
$\langle O^{\pm}_{2n} \rangle_\lambda$ in the large{\footnote{By large $N$ limit, we mean the continuum 
limit procedure, which means that we consider the $N \to \infty$ limit, such that at the same time 
$\rho_0$ is also tuned with $N$ according to the formula (\ref{rho}).}} $N$ limit.

Though we would like to emphasize, that (\ref{rough}) is only a rough and not the exact estimate for the large $N$ behavior for the quantities entering the rhs. of (\ref{Opm0XG}). Its purpose is to give a fast intuitive 
argument, why the multilinear and quadratic in $X$ and ${\cal G}$ terms become negligible in the continuum limit. 

The rough estimate (\ref{rough}) was derived by neglecting the $\rho_0$ dependence of the functions $f_1,f_2,f_3,f_\pm$ given by (\ref{fv1})-(\ref{f_-}). 
For presentational purposes we anticipate the exact result. 
The careful computations presented in the 
rest of this section and in appendix \ref{appArg} show, that 
the exact leading order large $N$ or equivalently $\rho_0$ behavior of the sums and integrals entering
 (\ref{Opm0XG}) is given by the formula:
\begin{eqnarray}
{\cal O}^{\pm}_0 &\sim& {\cal O}^{\pm}_X \sim {\cal O}^{\pm}_{\cal G} \sim 
\sim e^{-(1+p) \rho_0} \sim e^{(1-p) \rho_0}  \sim  e^{-2 \,p \, \rho_0}, \nonumber \\
{\cal O}^{\pm}_{{\cal G}{\cal G}} &\sim& {\cal O}^{\pm}_{{X}{\cal G}} \sim {\cal O}^{\pm}_{{X}{X}} \sim
e^{-2(1+p) \rho_0} \sim \tfrac{1}{N^2} \sim a^2. \label{roughexact}
\end{eqnarray}
This formula also implies that multilinear and quadratic in $X$ and ${\cal G}$ terms are 
negligible in the continuum limit. This statement is shown in appendix \ref{multilinQ}. 
By comparing (\ref{roughexact}) and (\ref{rough}) it can also be seen 
that the rough estimate came from a simplified train of thoughts is exact at the $p=1$ free fermion point.

Now, our goal is to compute the leading order in $N$ term of $\langle O^\pm_{2n} \rangle_\lambda$ in the 
continuum limit.
To carry out this purpose, it is worth to formulate the problem in terms of the finite parts of the 
leading order in $N$ terms of ${\cal G}(\lambda|\xi_\pm)$ and $X(h_j|\xi_\pm).$ These finite parts are defined by the 
following large $N$ (or equivalently $\rho_0$) expansions of ${\cal G}(\lambda|\xi_\pm)$ and $X(h_j|\xi_\pm):$
\begin{equation} \label{GXlead}
\begin{split}
{\cal G}(\lambda|\xi_\pm)&=-\tfrac{2 \pi}{\gamma} \, e^{- \tfrac{\pi}{\gamma}\rho_0} \, {\cal G}^{(\pm)}(\lambda)+O(\tfrac{1}{N^2}), \\
X(h_j|\xi_\pm)&=-\tfrac{2 \pi}{\gamma} \, e^{- \tfrac{\pi}{\gamma}\rho_0} \, X^{(\pm)}_j+O(\tfrac{1}{N^2}), \qquad j=1,...,m_H.
\end{split}
\end{equation}
From (\ref{llie}), (\ref{XG}) and (\ref{Zlcont}) it follows, 
that the finite parts ${\cal G}_\pm$ and $X^{(\pm)}_j$ satisfy the equations:
\begin{equation} \label{GPMllie}
\begin{split}
 &{\cal G}^{(\pm)}(\lambda)- \sum\limits_{\alpha=\pm} \int\limits_{-\infty}^{\infty} d\lambda' \,  G_{\lambda}(\lambda-\lambda'-i \, \alpha \, \eta) 
 \, {\cal G}^{(\pm)}(\lambda'+i \, \alpha \, \eta) \, {\cal F}^{(\lambda)}_{\alpha,c}(\lambda'+i \, \alpha \, \eta)=\\
&=e^{\pm \tfrac{\pi}{\gamma}\lambda}+\sum\limits_{j=1}^{m_H} \,
2 \pi \, G_\lambda(\lambda-h_j) \, X^{(\pm)}_j,  \qquad X^{(\pm)}_j=\frac{{\cal G}^{(\pm)}(h_j)}{Z'_{\lambda,c}(h_j)}, \qquad j=1,...,m_H.
\end{split}
\end{equation}
We note, that everywhere in (\ref{GPMllie}) the finite continuum limit of the counting-function arises. 


Now, we are in the position to determine the leading order term of the expectation value 
$\langle O^{\pm}_{2n}\rangle_\lambda$ in the large $N$ limit. As (\ref{roughexact}) implies, only the first three terms from the 
rhs. of (\ref{Opm0XG}) will contribute at leading order. Namely, 
$\langle O^{\pm}_{2n}\rangle_\lambda={\cal O}^\pm_0+{\cal O}^\pm_X+{\cal O}^\pm_{\cal G}+\text{\em "next to leading order terms".}$
  
The term ${\cal O}^\pm_0$ is independent of the positions of the holes. This means that this contribution is 
independent of the matrix element of the operator. This is why we will call it the bulk term: $\langle O^\pm_{2n} \rangle^{bulk}_{\lambda}\equiv{\cal O}^\pm_{0}.$  Formula  (\ref{O0}) implies, that 
this bulk term can be represented as follows:
\begin{equation}
\begin{split} \label{Cbulk}
\langle O^\pm_{2n} \rangle^{bulk}_{\lambda}=-{\cal J}_0[f_3](\xi_-)\! \pm \! \frac{\sinh(-i \, \gamma)}{\sinh(2 \rho_0)}
\, {\cal J}_0[f_1](\xi_-) \mp \frac{\sinh(2 \rho_0-i \, \gamma)}{\sinh(2 \rho_0)}
\, {\cal J}_0[f_2](\xi_+)\!+\!\frac{\Sigma_0[f_\pm](\xi_+,\xi_-)}{\sinh(2 \rho_0)},
\end{split}
\end{equation}
where ${\cal J}_0$ and $\Sigma_0$ are given by (\ref{calJG}) and (\ref{SUM0}), respectively.

On the other hand, with the help of (\ref{OX}), (\ref{OG}), (\ref{SX})-(\ref{calJG}), 
(\ref{SUMX}), (\ref{SUMF}), (\ref{FX}) and (\ref{J_S}), (\ref{J_SG}) the sum
 ${\cal O}^\pm_X+{\cal O}^\pm_{\cal G}$ can be written as follows:
\begin{equation} \label{Ox+gV1}
\begin{split}
{\cal O}^\pm_X+{\cal O}^\pm_{\cal G}&=
\sum\limits_{j=1}^{m_H} \left[ C^{(\pm)}_+(h_j) \, X(h_j|\xi_+) +C^{(\pm)}_-(h_j) \, X(h_j|\xi_-)  \right]+          \\
+\sum\limits_{\alpha=\pm}\int\limits_{-\infty}^{\infty}  
\frac{d \lambda}{2 \pi} \,&{\cal F}_{\alpha}^{(\lambda)}(\lambda+i\,  \alpha \,  \eta) \,
\left[
{\cal G}(\lambda+i \, \alpha \,  \eta|\xi_+)\, C^{(\pm)}_+(\lambda+i\,  \alpha) +
{\cal G}(\lambda+i \, \alpha \,  \eta|\xi_-)\, C^{(\pm)}_-(\lambda+i\,  \alpha) \right],
\end{split}
\end{equation}
where 
\begin{equation} \label{Cpmp}
\begin{split}
C^{(\pm)}_+(\lambda)=\mp \frac{\sinh(2 \rho_0-i \, \gamma)}{\sinh(2 \rho_0)} \, {\cal J}_G[f_2](\lambda)+\frac{1}{\sinh(2 \rho_0)} 
\left[ {J}_S[f_\pm](\lambda|\xi_-)-{J}_{SG}[f_\pm](\lambda|\xi_-)\right],
\end{split}
\end{equation}
\begin{equation} \label{Cpmm}
\begin{split}
C^{(\pm)}_-(\lambda)=\!-{\cal J}_G[f_3](\lambda)\! \pm \!\frac{\sinh(-i \, \gamma)}{\sinh(2 \rho_0)} \, {\cal J}_G[f_1](\lambda)\!\!-\!\!\frac{1}{\sinh(2 \rho_0)} 
\left[ {J}_S[f_\pm](\lambda|\xi_+)\!\!-\!\!{J}_{SG}[f_\pm](\lambda|\xi_+)\right],
\end{split}
\end{equation}
with ${\cal J}_G,$ ${J}_S$ and ${J}_{SG}$ given in (\ref{calJG}), (\ref{J_S}) and (\ref{J_SG}), respectively.
We are interested in the leading order large $N$ expression for ${\cal O}^\pm_X+{\cal O}^\pm_{\cal G},$ 
this is why from (\ref{GXlead}) the leading order expressions of $X(h_j|\xi_\pm)$ and
 of ${\cal G}(\lambda|\xi_\pm)$ can be replaced into (\ref{Ox+gV1}). 
Similarly, the ${\cal F}_{\alpha}^{(\lambda)} (\lambda) \to {\cal F}_{\alpha,c}^{(\lambda)}(\lambda)$ replacement can also be 
done at leading order. 
 As a result one obtains:
 \begin{equation}
\begin{split} \label{OX+OG}
{\cal O}^\pm_X+{\cal O}^\pm_{\cal G}&= -\frac{2 \pi}{\gamma} \, e^{-(1+p) \rho_0} \, \Bigg\{
\sum\limits_{j=1}^{m_H} \left[ C^{(\pm)}_+(h_j) \, X_j^{(+)} +C^{(\pm)}_-(h_j) \, X_j^{(-)}  \right]+          \\
+\!\sum\limits_{\alpha=\pm}\int\limits_{-\infty}^{\infty}  \!\!\!
\frac{d \lambda}{2 \pi} \,&{\cal F}_{\alpha,c}^{(\lambda)}(\lambda+i\,  \alpha \,  \eta) \!\!
\left[
{\cal G}^{(+)}(\lambda+i \, \alpha \,  \eta)\, C^{(\pm)}_+(\lambda+i\,  \alpha) \!\!+ \!\!
{\cal G}^{(-)}(\lambda+i \, \alpha \,  \eta)\, C^{(\pm)}_-(\lambda+i\,  \alpha) \right]\!\!\! \Bigg\}\!\!+\!...,
\end{split}
\end{equation}
where the dots stand for subleading terms in the large $N$ limit.

The careful evaluation of the functionals (\ref{Cbulk}), (\ref{Cpmp}) and (\ref{Cpmm}) 
in the large $\rho_0$ limit, 
which is presented in appendix \ref{appArg},   
leads to the following large $\rho_0$ asymptotics 
for the bulk term and for $C^{(\pm)}_\pm(\lambda):$
\begin{equation} \label{Opmbulk}
\begin{split}
\langle O^{\pm}_{2n}\rangle^{bulk}_\lambda=\left\{ 
\begin{array}{r}
-\tfrac{p+1}{\sin \gamma} \, \tan(\tfrac{p \, \pi}{2})\, e^{-2 p \rho_0}
+O(e^{-2 \rho_0}) , \qquad  \text{case:} \, +, \\
O(e^{-2 \rho_0}), \qquad \text{case:} \, -.
\end{array}\right.
\end{split}
\end{equation}
\begin{equation}
\begin{split}\label{CcalK}
 C^{(\pm)}_+(\lambda)& \stackrel{\rho_0 \to \infty}{=} \mp {\cal K}_+(\lambda|\rho_0)+...,\\
 C^{(\pm)}_-(\lambda)& \stackrel{\rho_0 \to \infty}{=} -{\cal K}_-(\lambda|\rho_0)+...,
\end{split}
\end{equation}
where
\begin{equation} \label{calKpm}
\begin{split}
{\cal K}_\pm(\lambda|\rho_0)=\tfrac{p+1}{2 \sin \gamma} \, e^{\mp (p+1) \lambda} \, e^{(1-p) \rho_0}+O(e^{-2 \rho_0}), 
\end{split}
\end{equation}
and the dots stand for terms tending to zero, when $\rho_0 \to \infty$ and $p<1.$ 

Putting the results (\ref{Cbulk}), (\ref{Opmbulk}), (\ref{OX+OG}), (\ref{CcalK}) together, 
one obtains the following leading order result for $\langle O^{\pm}_{2n}\rangle_\lambda$ in the attractive regime:
\begin{equation} \label{Opmlarge}
\begin{split}
\langle O^{\pm}_{2n}\rangle_\lambda=\langle O^{\pm}_{2n}\rangle^{bulk}_\lambda + \tfrac{2 \pi}{\gamma} \, e^{- (p+1) \rho_0} 
\Bigg\{  \sum\limits_{j=1}^{m_H} {\cal K}_-(h_j|\rho_0) \, X_j^{(-)}\pm  \sum\limits_{j=1}^{m_H}  {\cal K}_+(h_j|\rho_0) \, X_j^{(+)}+         \\
+\sum\limits_{\alpha=\pm}\int\limits_{-\infty}^{\infty}  
\frac{d \lambda}{2 \pi} \,{\cal F}_{\alpha,c}^{(\lambda)}(\lambda+i\,  \alpha \,  \eta) \,
\left[
{\cal G}^{(-)}(\lambda+i \, \alpha \,  \eta)\, {\cal K}_-(\lambda+i\,  \alpha) \pm
{\cal G}^{(+)}(\lambda+i \, \alpha \,  \eta)\, {\cal K}_+(\lambda+i\,  \alpha) \right] \Bigg\}.
\end{split}
\end{equation}
Here as a consequence of (\ref{roughexact}), the contributions coming from 
${\cal O}^{\pm}_{XX},$ ${\cal O}^{\pm}_{X{\cal G}}$ and ${\cal O}^{\pm}_{{\cal G}{\cal G}}$ 
were neglected. 

It turns out, the leading order expression (\ref{Opmlarge}) for $\langle O^{\pm}_{2n}\rangle_\lambda$ 
can be rephrased in terms of the variables ${\cal G}_d(\theta),X^{(d)}_j$ of (\ref{llie_d}) and ${\cal G}_\ell(\theta), \, X^{(\ell)}_j$ of (\ref{llie_l}). The reason for this is that  ${\cal G}^{(\pm)}(\lambda)$ and $X^{(\pm)}_j$ of  (\ref{GPMllie}) can be simply expressed in terms of ${\cal G}_d(\theta),X^{(d)}_j$ and ${\cal G}_d(\theta),X^{(d)}_j.$

Consider the following linear combinations of ${\cal G}^{(\pm)}(\lambda)$ and $X^{(\pm)}_j$: 
\begin{equation} \label{LTchange}
\begin{split}
\hat{\cal G}_d(\theta)=\frac{\ell}{2} \left[{\cal G}^{(+)}(\tfrac{\gamma}{\pi}\theta)+{\cal G}^{(-)}(\tfrac{\gamma}{\pi}\theta) \right], \qquad
\hat{X}^{(d)}_j=\frac{\ell}{2} \, \frac{\pi}{\gamma} \, \left( X^{(+)}_j+X^{(-)}_j\right), \\
\hat{\cal G}_\ell(\theta)=\frac{1}{2} \left[{\cal G}^{(+)}(\tfrac{\gamma}{\pi}\theta)-{\cal G}^{(-)}(\tfrac{\gamma}{\pi}\theta) \right], \qquad
\hat{X}^{(\ell)}_j=\frac{1}{2} \, \frac{\pi}{\gamma} \, \left( X^{(+)}_j-X^{(-)}_j\right).
\end{split}
\end{equation}
As a consequence of the linearity of (\ref{GPMllie}), it can be shown that the new variables 
$\hat{\cal G}_d(\theta),\hat{\cal G}_\ell(\theta)$, $\hat{X}^{(d)}_j, \hat{X}^{(\ell)}_j$ satisfy the 
linear integral equations as follows: 
\begin{equation} \label{llie_dhat}
\begin{split}
\hat{\cal G}_{d}(\theta)&-\sum\limits_{\alpha =\pm} \int\limits_{-\infty}^{\infty} \! \frac{d\theta'}{2 \pi} G(\theta-\theta'-i \, \alpha \, \eta)
\hat{\cal G}_{d}(\theta'+i \, \alpha \, \eta) \, {\cal F}_{\alpha} (\theta'+i \, \alpha \, \eta)
=\\
&=\ell \, \cosh(\theta)+\sum\limits_{j=1}^{m_H} \, G(\theta-H_j) \, \hat{X}_j^{(d)}, \qquad
\hat{X}_j^{(d)}=\frac{\hat{\cal G}_{d}(H_j)}{Z'(H_j)}, \qquad j=1,...,m_H. 
\end{split}
\end{equation}
\begin{equation} \label{llie_lhat}
\begin{split}
\hat{\cal G}_{\ell}(\theta)&-\sum\limits_{\alpha =\pm} \int\limits_{-\infty}^{\infty} \! \frac{d\theta'}{2 \pi} G(\theta-\theta'-i \, \alpha \, \eta)
\hat{\cal G}_{\ell}(\theta'+i \, \alpha \, \eta) \, {\cal F}_{\alpha} (\theta'+i \, \alpha \, \eta)
=\\
&= \sinh(\theta)+\sum\limits_{j=1}^{m_H} \, G(\theta-H_j) \, \hat{X}_j^{(\ell)}, \qquad
\hat{X}_j^{(\ell)}=\frac{\hat{\cal G}_{\ell}(H_j)}{Z'(H_j)}, \qquad j=1,...,m_H, 
\end{split}
\end{equation}
where $Z(\theta)$ and $H_j$ are the counting function  and the positions of the holes in rapidity convention.
They are solutions of the equations (\ref{DDVcont}) and (\ref{contQHk}).

Comparing (\ref{llie_dhat}) and (\ref{llie_lhat}) to (\ref{llie_d}) and (\ref{llie_l}) one can recognize that 
\begin{equation}\label{GXcompare}
\begin{split}
\hat{\cal G}_{d}(\theta)&={\cal G}_{d}(\theta), \qquad  \hat{X}_j^{(d)}={X}_j^{(d)}, \qquad j=1,...,m_H, \\
\hat{\cal G}_{\ell}(\theta)&={\cal G}_{\ell}(\theta), \qquad \hat{X}_j^{(\ell)}={X}_j^{(\ell)}, \qquad j=1,...,m_H.
\end{split}
\end{equation}
Using (\ref{GXcompare}) and substituting 
the inverse relation of (\ref{LTchange}) into (\ref{Opmlarge}) together with a 
change of integrating variables from $\lambda$ to $\theta$ 
one ends up with the final result:
\begin{equation} \label{O+final}
\begin{split}
&\langle O^+_{2n} \rangle_\lambda= \frac{2 \, (p+1) \, e^{-2 \, p  \, \rho_0 }}{\sin \gamma}
\left\{ -\tfrac{1}{2} \, \tan(\tfrac{p \, \pi}{2})+
 \, \sum\limits_{k=1}^{m_H} \left\{ \, \cosh(H_k) \,
\frac{X_k^{(d)}}{\ell}-\sinh(H_k) \, X_k^{(\ell)} \right\}\right.+ \\
\,&\left.+  \sum\limits_{\alpha =\pm} \int\limits_{-\infty}^{\infty}\!
 \! d\theta \! \left[ \cosh(\theta+i \, \alpha \, \eta)
\, \frac{{\cal G}_d(\theta+i \, \alpha \, \eta)}{\ell}
-\sinh(\theta+i \, \alpha \, \eta)
\, {\cal G}_{\ell}(\theta+i \, \alpha \, \eta) \right]\!
 {\cal F}_{\alpha} (\theta+i \, \alpha \, \eta)\!+\!...\!\! \right\}\!,
\end{split}
\end{equation}
\begin{equation} \label{O-final}
\begin{split}
&\langle O^-_{2n} \rangle_\lambda= \frac{2 \, (p+1) \, e^{-2 \, p  \, \rho_0 }}{\sin \gamma}
\left\{ 
 \, \sum\limits_{k=1}^{m_H} \left( \, \sinh(H_k) \,
\frac{X_k^{(d)}}{\ell}-\cosh(H_k) \, X_k^{(\ell)} \right)\right.+ \\
\,&\left.+  \sum\limits_{\alpha =\pm} \int\limits_{-\infty}^{\infty}
 \!\! d\theta \! \left[ \sinh(\theta+i \, \alpha \, \eta)
\, \frac{{\cal G}_d(\theta+i \, \alpha \, \eta)}{\ell}
-\cosh(\theta+i \, \alpha \, \eta)
\, {\cal G}_{\ell}(\theta+i \, \alpha \, \eta) \right]\!
 {\cal F}_{\alpha} (\theta+i \, \alpha \, \eta)\!\!+\!\!... \!\!\right\}\!,
\end{split}
\end{equation}
where dots mean next to leading order terms in the $N$ tends to infinity limit of the attractive regime. 
Comparing (\ref{tetaTfinal}) and (\ref{O+final}) one can easily recognize the proportionality of
$\langle O^+_{2n} \rangle_\lambda$ and $\langle \Theta_T \rangle$:
\begin{equation} \label{O+Teta}
\begin{split}
\langle O^+_{2n} \rangle_\lambda= \frac{2 \, (p+1) \, e^{-2 \, p  \, \rho_0 }}{\sin \gamma} \frac{\langle \Theta_T \rangle}{2 \pi {\cal M}^2}+....
\end{split}
\end{equation}
According to (\ref{PPbare}) the expectation value for the bare fermion bilinear is given by:
\begin{equation} \label{PPexp}
\begin{split}
\langle \bar{\Psi} \Psi \rangle=\tfrac{1}{a}\, \langle O^+_{2n} \rangle_\lambda=\frac{{\cal M} \, (p+1)}{2 \sin \gamma} \, \left( \tfrac{4}{{\cal M} a}\right)^{\tfrac{1-p}{p+1}}
 \frac{\langle \Theta_T \rangle}{2 \pi {\cal M}^2}+...,
\end{split}
\end{equation}
where we exploited the relation (\ref{rho}) between the lattice constant $a$ and the inhomogeneity parameter $\rho_0.$
Using the relation between $p$ and $\beta$ in (\ref{csatrel1}), one can see that 
$\langle \bar{\Psi} \Psi \rangle$ is proportional to the expectation value of the stress energy tensor, and it scales as $a^{\beta^2/4 \pi-1}$ as
it is expected from (\ref{ppT}) obtained via 
purely field theoretical considerations. 


\section{Large volume expansion}

In this section we rephrase the leading order terms in the large $N$ expansions of (\ref{O+final}) and (\ref{O-final}) in the form of a 
systematic large volume series. To get rid of the unnecessary constants, we consider the following quantities:
\begin{equation} \label{calO+}
\begin{split}
{\cal O}^+&=\sum\limits_{\alpha =\pm} \int\limits_{-\infty}^{\infty}
 \! d\theta \,\left[ \cosh(\theta+i \, \alpha \, \eta)
\, \frac{{\cal G}_d(\theta+i \, \alpha \, \eta)}{\ell}
-\sinh(\theta+i \, \alpha \, \eta)
\, {\cal G}_{\ell}(\theta+i \, \alpha \, \eta) \right]\,
 {\cal F}_{\alpha} (\theta+i \, \alpha \, \eta)+  \\
&+\sum\limits_{k=1}^{m_H} \left\{ \, \cosh(H_k) \,
\frac{X_k^{(d)}}{\ell}-\sinh(H_k) \, X_k^{(\ell)} \right\},
\end{split}
\end{equation}
\begin{equation} \label{calO-}
\begin{split}
{\cal O}^-&=\sum\limits_{\alpha =\pm} \int\limits_{-\infty}^{\infty}
 \! d\theta \,\left[ \sinh(\theta+i \, \alpha \, \eta)
\, \frac{{\cal G}_d(\theta+i \, \alpha \, \eta)}{\ell}
-\cosh(\theta+i \, \alpha \, \eta)
\, {\cal G}_{\ell}(\theta+i \, \alpha \, \eta) \right]\,
 {\cal F}_{\alpha} (\theta+i \, \alpha \, \eta)+ \\
 \, &+\sum\limits_{k=1}^{m_H} \left\{ \, \sinh(H_k) \,
\frac{X_k^{(d)}}{\ell}-\cosh(H_k) \, X_k^{(\ell)} \right\},
\end{split}
\end{equation}
where ${\cal G}_d(\theta), \, X^{(d)}$ and ${\cal G}_\ell(\theta), \, X^{(\ell)}$ are defined by the equations 
(\ref{llie_d}) and (\ref{llie_l}) respectively.
From (\ref{tetaTfinal}) and (\ref{O+final}) it can be seen that ${\cal O}^+$ is simply related to the 
fermionic expectation value of the trace of the stress-energy tensor:
\begin{equation} \label{calO+teta}
\begin{split}
{\cal O}^+=\frac{\langle \Theta_T \rangle-\langle \Theta^\infty_T \rangle}{2 \pi \, {\cal M}^2}.
\end{split}
\end{equation}
On the other hand ${\cal O}^-$ is proportional to the fermionic expectation value of the 
renormalized pseudo-scalar fermion bilinear $\langle\bar{\Psi}\gamma^5\Psi\rangle.$
Here we do not care about the actual value of the proportionality factor, since it will turn out, that this 
expectation value is zero between multi-fermion states.

The process of the evaluation of (\ref{calO+}) and (\ref{calO-}) in the large volume limit is very similar to 
the method used for computing the diagonal matrix elements of the trace of the stress-energy tensor 
in purely elastic scattering theories \cite{PST14}. The reason for this is that formally the NLIE equations
 (\ref{DDVcont}) are very similar to the Thermodynamic Bethe Ansatz (TBA) equations
 of a purely diagonally scattering 
theory of two types of particles. This analogy, the actual form of the large volume series of the $U(1)$ current 
of the theory \cite{En} together with 
the all order large volume series conjectures for the
 diagonal form factors of purely elastic scattering theories \cite{LM99,saleur,Pozsg11,Pozsg13,PST14}, led to 
the following large volume series conjecture for the diagonal multi-fermion (soliton) expectation values of local 
operators in the MT (SG) models \cite{En}: 
\newline {\bf Conjecture:}
{\it
For any local operator ${\cal O}(x)$ in the MT (SG) model the expectation value in an $n$-fermion (soliton) state 
with rapidities $\{H_1,H_2,...,H_n\}$ 
can be written as:
\begin{equation} \label{Ps1}
\begin{split}
\langle H_1,...,H_n|{\cal O}(x)|H_1,...,H_n \rangle=&\frac{1}{\rho(H_1,..,H_n)} \\
& \times \sum\limits_{\{H_+\}\cup \{H_-\}} {\cal D}^{{\cal O}}(\{H_+\}) \, \rho(\{H_-\}|\{H_+\}),
\end{split}
\end{equation}
where $\rho(\vec{H})$ is the determinant of the exact Gaudin-matrix:
\begin{equation} \label{ro}
\rho(H_1,..,H_n)=\text{det} \, \hat{\Phi}(\vec{H}), \qquad \hat{\Phi}_{kj}(\vec{H})=\frac{d}{d H_j} Z(H_k|\vec{H}), \qquad j,k=1,..,m_H,
\end{equation}
the sum in (\ref{Ps1}) runs for all bipartite partitions of the rapidities of the sandwiching state:
$\{H_1,..,H_n \}=\{H_+\}\cup\{H_-\},$ such that
\begin{equation} \label{ropm}
\rho(\{H_+\}|\{H_-\}=\text{det} \, \hat{\Phi}_+(\vec{H}),
\end{equation}
with $\hat{\Phi}_+(\vec{H})$ being the sub-matrix of $\hat{\Phi}(\vec{H})$ corresponding to the subset $\{H_+\}.$
The quantity ${\cal D}^{\cal O}(\{H\})$ in (\ref{Ps1}) is called the dressed form-factor \cite{PST14} and
it is given by an infinite sum in terms of the connected diagonal form-factors of the theory:
\begin{equation} \label{ODdress}
\begin{split}
{\cal D}^{\cal O}(\{H_1,...,H_n\})=\sum\limits_{n_+=0}^\infty \sum\limits_{n_-=0}^\infty 
\frac{1}{n_+! \, n_-!} \! \int\limits_{-\infty}^{\infty} \! \prod\limits_{i=1}^{n_++n_-} \! \! \frac{d \theta_i}{2 \pi} \!
\prod\limits_{i=1}^{n_+} \! {\cal F}_+(\theta_i+i \, \eta) \! \! \! \prod\limits_{i=n_++1}^{n_++n_-} \! \!
{\cal F}_-(\theta_i-i \, \eta)\\
\times 
F^{\cal O}_c(H_1,H_2,...,H_n,\theta_1\!+\!i \, \eta,...,\theta_{n_+}\!+\!i \, \eta,\theta_{n_++1}\!-\!i \, \eta,...,\theta_{n_++n_-}\!-\!i \, \eta),
\end{split}
\end{equation}
where $F^{{\cal O}}_{c}$ denotes the connected diagonal form factors of $O(x)$ in pure fermion (soliton) states, $0<\eta<\text{min}(p \pi,\pi)$ is a small contour deformation parameter
 and ${\cal F}_\pm(\theta)$ are the nonlinear expressions of the counting function given by (\ref{calF}). 
}
We note that the structure (\ref{Ps1}) is the same for the purely elastic scattering theories and for the MT (SG) model. The difference arises in the concrete form of the exact Gaudin-matrix{\footnote{In general, the Gaudin-matrix
 is the derivative of the exact Bethe-equations with respect to the particle rapidities.}}
 and in the actual form of the dressed form factors. However up to exponentially small 
in volume corrections the formulas of purely elastic scattering theories are also 
appropriate to describe the multi-fermion (soliton) expectation values of local operators 
\cite{Palmai13},\cite{En}.

So far, conjecture (\ref{Ps1})-(\ref{ODdress}) has been checked against the diagonal fermionic (solitonic) form factors of the $U(1)$ current of the 
theory \cite{En} and now 
by rephrasing ${\cal O}^+$ as a large volume series, we will show that this conjecture remains valid in the 
case of the trace of the stress energy tensor, too.
Thus, our purpose is to bring ${\cal O}^\pm$ into the form of (\ref{Ps1}) and check whether the coefficients of 
$\rho(\{H_-\}|\{H_+\})$ agrees with ${\cal D}^{\cal O}(\{H_+\})$ given by (\ref{ODdress}). 

In \cite{PST14}, starting from the Thermodynamic Bethe Ansatz (TBA) equations, 
the analog formulas of (\ref{Ps1})-(\ref{ODdress}) were derived for the diagonal matrix elements of
the trace of the stress energy tensor in purely elastic scattering theories.
The computation we present below is an appropriate adaptation of the derivation given in section 3 of ref. \cite{PST14}.

The first step of the computation is to rewrite ${\cal G}_d (\theta), \, X^{(d)}$ and ${\cal G}_\ell(\theta) \, X^{(\ell)}$ 
in terms of the solutions of some "elementary" linear problems. 
For any function $f,$ let $f^{[\pm]}(\theta)=f(\theta\pm i \eta),$ then
 by definition an "elementary" solution indexed by $A$ 
satisfy the linear equations as follows:
\begin{equation} \label{lliegen}
\begin{split}
{\cal G}^{[\alpha]}_A(\theta)-\sum\limits_{\beta =\pm} \int\limits_{-\infty}^{\infty} \! \frac{d\theta'}{2 \pi} \psi_{\alpha \beta}(\theta-\theta') \, {\cal G}^{[\beta]}_A(\theta') \,{\cal F}_{\beta}^{[\beta]} (\theta')=f_A^{[\alpha]}(\theta), \qquad \alpha=\pm,
\end{split}
\end{equation}
where the symmetric kernel $\psi_{\alpha \beta}(\theta)$ is given by:
\begin{equation} \label{kernPSI}
\begin{split}
\psi_{\alpha \beta}(\theta)=G(\theta+i \, (\alpha-\beta) \, \eta), \qquad \alpha,\beta=\pm,
\end{split}
\end{equation}
and $f_A(\theta)$ is the source term specifying ${\cal G}^{[\alpha]}_A(\theta).$
 An "elementary" solution with unshifted argument satisfies the equations as follows:
\begin{equation} \label{lliegen0}
\begin{split}
{\cal G}_A(\theta)-\sum\limits_{\beta =\pm} \int\limits_{-\infty}^{\infty} \! \frac{d\theta'}{2 \pi} \, \psi_{\alpha \beta}(\theta-\! \theta'\!-i \, \alpha \, \eta) \, {\cal G}^{[\beta]}_A(\theta') \, {\cal F}_{\beta}^{[\beta]} (\theta')=f_A (\theta), \qquad \alpha=\pm.
\end{split}
\end{equation}
In our problem the index $A$ can take values from the set ${\frak I}=\{s,c,1,2,...,m_H\},$ such that the source functions
$f_A(\theta)$ in (\ref{lliegen}) take the form:
\begin{equation} \label{fAlist}
\begin{split}
f_s(\theta)=\sinh(\theta), \quad f_c(\theta)=\cosh(\theta), \quad f_j(\theta)=-G(\theta-H_j), \qquad j=1,...,m_H.
\end{split}
\end{equation}
From the defining linear equations (\ref{lliegen}) it can be shown, that  
the "elementary" solutions satisfy the following identities:
 \begin{equation} \label{idAB}
\begin{split}
\sum\limits_{\alpha=\pm} \int\limits_{-\infty}^{\infty} \frac{d \theta}{2 \pi} \, f_A^{[\alpha]}(\theta) \,
{\cal G}^{[\alpha]}_B(\theta) \, {\cal F}_{\alpha}^{[\alpha]}(\theta)=
\sum\limits_{\alpha=\pm} \int\limits_{-\infty}^{\infty} \frac{d \theta}{2 \pi} \, f_B^{[\alpha]}(\theta) \,
{\cal G}^{[\alpha]}_A(\theta) \, {\cal F}_{\alpha}^{[\alpha]}(\theta), \qquad A,B \in {\frak I},
\end{split}
\end{equation}
\begin{equation} \label{idAj}
\begin{split}
\sum\limits_{\alpha=\pm} \int\limits_{-\infty}^{\infty} \frac{d \theta}{2 \pi} \, f_A^{[\alpha]}(\theta)\,
{\cal G}^{[\alpha]}_j(\theta) \, {\cal F}_{\alpha}^{[\alpha]}(\theta)=f_A(H_j)-{\cal G}_A(H_j), 
\quad j=1,..,m_H, \qquad A \in {\frak I}.
\end{split}
\end{equation}
With the help of the linear equations (\ref{llie_d}), (\ref{llie_l}) and (\ref{lliegen}) with (\ref{fAlist}),
the pairs of quantities ${\cal G}_d(\theta), \, X^{(d)}$ and ${\cal G_\ell}(\theta), \, X^{(\ell)}$ can be expressed 
in terms of the elementary solutions as follows:
\begin{equation} \label{Gdelem}
\begin{split}
{\cal G}_d(\theta)=\ell \, {\cal G}_c(\theta)-\sum\limits_{j=1}^{m_H} {\cal G}_j(\theta) \, X_j^{(d)}, \qquad 
X_k^{(d)}=1, \qquad \qquad \qquad \qquad \quad k=1,...,m_H,
\end{split}
\end{equation}
\begin{equation} \label{Glelem}
\begin{split}
{\cal G}_\ell(\theta)={\cal G}_s(\theta)-\sum\limits_{j=1}^{m_H} {\cal G}_j(\theta) \, X_j^{(\ell)}, \qquad 
X_k^{(\ell)}=\sum\limits_{j=1}^{m_H} \hat{\Phi}^{-1}_{kj}(\vec{H})\,{\cal G}_{s}(H_j), \qquad k=1,...,m_H,
\end{split}
\end{equation}
where $\hat{\Phi}_{kj}(\vec{H})$ is the exact Gaudin-matrix defined by the formula:
\begin{equation} \label{exactGaudin}
\begin{split}
\hat{\Phi}_{kj}(\vec{H})=\frac{d}{d H_j} Z(H_k|\vec{H})=Z'(H_k) \, \delta_{jk}+{\cal G}_j(H_k), \qquad j,k=1,..,m_H.
\end{split}
\end{equation}
Using the formulas (\ref{idAB})-(\ref{exactGaudin}), $\hat{\Phi}_{kj}(\vec{H})$ and ${\cal O}^\pm$ can be expressed in terms of the 
elementary solutions of (\ref{lliegen}) as follows: 
\begin{equation} \label{Gaudinelem}
\begin{split}
\hat{\Phi}_{kj}(\vec{H})=\left( \ell \,{\cal G}_c(H_k)-\sum\limits_{k'=1}^{m_H} \, {\cal G}_{k'}(H_k)
\right) \delta_{kj}+{\cal G}_j(H_k), \qquad j,k=1,..,m_H,
\end{split}
\end{equation}
\begin{equation} \label{o+}
\begin{split}
{\cal O}^+\!\!=\!\!\! \sum\limits_{\alpha=\pm} \! \int\limits_{-\infty}^{\infty} \!\!\! 
\frac{d \theta}{2 \pi}  {\cal F}_{\alpha}^{[\alpha]}\!(\theta) \!\! \left[ \! \! 
\, f_c^{[\alpha]}(\theta)\,
{\cal G}^{[\alpha]}_c(\theta) \!- \!f_s^{[\alpha]}(\theta)\,
{\cal G}^{[\alpha]}_s(\theta) \! \right]\!\!+\!\! \frac{1}{\ell} \!\sum\limits_{j=1}^{m_H} \!{\cal G}_c(H_j)
\!-\!\!\!\!
\sum\limits_{j,k=1}^{m_H} \!\!{\cal G}_{s}(H_k) \,\hat{\Phi}^{-1}_{kj}(\vec{H})\,{\cal G}_{s}(H_j),
\end{split}
\end{equation}
\begin{equation} \label{o-}
\begin{split}
{\cal O}^-\!\!=\!\!\! \sum\limits_{\alpha=\pm} \! \int\limits_{-\infty}^{\infty} \!\!\! 
\frac{d \theta}{2 \pi}  {\cal F}_{\alpha}^{[\alpha]}\!(\theta) \!\! \left[ \! \! 
\, f_c^{[\alpha]}(\theta)\,
{\cal G}^{[\alpha]}_s(\theta) \!- \!f_s^{[\alpha]}(\theta)\,
{\cal G}^{[\alpha]}_c(\theta) \! \right]\!\!-\!\! \frac{1}{\ell} \!\sum\limits_{j=1}^{m_H} \!{\cal G}_s(H_j)
\!+\!\!\!\!
\sum\limits_{j,k=1}^{m_H} \!\!{\cal G}_{s}(H_k) \,\hat{\Phi}^{-1}_{kj}(\vec{H})\,{\cal G}_{c}(H_j).
\end{split}
\end{equation}
To bring (\ref{o+}) and (\ref{o-}) into the form of (\ref{Ps1}) one needs to use two theorems. 
\newline {\bf Theorem 1.} {\it The inverse of the Gaudin-matrix can be expressed 
in terms of its principal minors and sequences of its matrix elements \cite{PST50},\cite{PST14} by 
the formula as follows:
\begin{equation} \label{GC}
\begin{split}
\hat{\Phi}_{ij}^{-1}=\frac{{\cal C}_{ij}}{\text{det} \, \hat{\Phi}}, \qquad i,j=1,..,m_H,
\end{split}
\end{equation}
with ${\cal C}_{ij}$ being the co-factor matrix with entries:
\begin{equation} \label{calC}
\begin{split}
{\cal C}_{ij}=\left\{\begin{array}{c}
\text{det} \, \hat{\Phi}(\{i\}), \qquad \qquad \qquad \qquad \qquad \qquad \qquad \qquad \qquad \qquad \quad i=j, \\
\sum\limits_{n=0}^{m_H-2} \, \sum\limits_{\{\alpha \}} \,
(-1)^{n+1} \, \hat{\Phi}_{i \alpha_1} \, \hat{\Phi}_{\alpha_1 \alpha_2} \dots \hat{\Phi}_{\alpha_n j} \,
\, \text{det} \, \hat{\Phi}(\{j,i,\alpha_1,...,\alpha_n \}), \quad i \neq j,
\end{array}\right.
\end{split}
\end{equation}
where $\{\alpha \}=\{1,2,...,m_H\} \setminus \{i,j\}$ and $\hat{\Phi}(\{ {\cal I} \})$ denotes the matrix 
obtained by omitting from $\hat{\Phi}$ the rows and columns indexed by the set $\{{\cal I}\}.$
}
\newline {\bf Theorem 2.} {\it The determinant of 
the Gaudin-matrix can be expressed in terms of its principal minors and sequences of its matrix elements \cite{PST14}
 by the formula: 
\begin{equation} \label{tetel2}
\begin{split}
\text{det} \, \hat{\Phi}\! =\! \ell \, {\cal G}_c(H_i) \, \text{det} \, \hat{\Phi}(\{i\})\!+\!\!\!\!\!
\sum\limits_{n=1}^{m_H-1} \!\!\! \sum\limits_{\{\alpha \}} \,
\!\! (-1)^{n}  \hat{\Phi}_{i \alpha_1}  \hat{\Phi}_{\alpha_1 \alpha_2} \dots \!\! \hat{\Phi}_{\alpha_{n-1} \alpha_n} \,
 \ell \, {\cal G}_c(H_{\alpha_n})\, \text{det} \, \hat{\Phi}(\{i,\alpha_1,..., \! \alpha_n \}).
\end{split}
\end{equation}
}
Theorems 1. and 2. allow one to rewrite the double and single sums respectively in (\ref{o+}) and (\ref{o-})
into a more convenient form.
Using (\ref{tetel2}), the single sums of (\ref{o+}) and (\ref{o-}) can be represented as:
\begin{equation} \label{singlesum}
\begin{split}
&\sum\limits_{i=1}^{m_H} \frac{1}{\ell} {\cal G}_A(H_i)=\sum\limits_{i=1}^{m_H} \, {\cal G}_A(H_i) \,
{\cal G}_c(H_i) \,
\frac{\text{det} \, \hat{\Phi}(\{i \})}{\text{det} \, \hat{\Phi}}+ \\
&+\!\!\! \sum\limits_{i,j=1 \atop i \neq j}^{m_H} \!\! {\cal G}_A(H_i) 
{\cal G}_c(H_j) \!\!\! \sum\limits_{n=0}^{m_H-2} \!\!\! \sum\limits_{\{\alpha\}} (-1)^{n+1} 
{\cal G}_{\alpha_1}(H_i) {\cal G}_{\alpha_2}(H_{\alpha_1})...\!{\cal G}_{j}(H_{\alpha_n}) 
\frac{\mbox{det} \, \hat{\Phi}(\{i,j,\alpha_1,...,\alpha_n \}) }{\mbox{det} \, \hat{\Phi}},
\end{split}
\end{equation}
with $A\in\{s,c\}.$ 
The double sums of (\ref{o+}) and (\ref{o-}) can be represented in a very similar form:
\begin{equation} \label{doublesum}
\begin{split}
&\sum\limits_{i,j=1}^{m_H}  {\cal G}_A(H_i) \, \hat{\Phi}^{-1}_{ij}(\vec{H}) \, {\cal G}_B(H_j)
=\sum\limits_{i=1}^{m_H} \, {\cal G}_A(H_i) \,
{\cal G}_B(H_i) \,
\frac{\text{det} \, \hat{\Phi}(\{i \})}{\text{det} \, \hat{\Phi}}+ \\
&+\!\!\! \sum\limits_{i,j=1 \atop i \neq j}^{m_H} \!\! {\cal G}_A(H_i) 
{\cal G}_B(H_j) \!\!\! \sum\limits_{n=0}^{m_H-2} \!\!\! \sum\limits_{\{\alpha\}} (-1)^{n+1} 
{\cal G}_{\alpha_1}(H_i) {\cal G}_{\alpha_2}(H_{\alpha_1})...\!{\cal G}_{j}(H_{\alpha_n}) 
\frac{\mbox{det} \, \hat{\Phi}(\{i,j,\alpha_1,...,\alpha_n \}) }{\mbox{det} \, \hat{\Phi}},
\end{split}
\end{equation}
with $A,B \in \{s,c\}.$
It is convenient to determine first the large volume series expansion of the first terms in the right hand sides of 
(\ref{o+}) and (\ref{o-}). They are called the vacuum contributions \cite{Pozsg13,PST14} 
since they correspond to the $\{H_+\}=\emptyset$ case.
These terms can be rephrased as an infinite series similar to that of LeClair and 
Mussardo \cite{LM99,saleur,Pozsg11}. To get this series representation, first one has to 
construct the all order large volume solution of (\ref{lliegen}) for $A=s,c$. 
This can be obtained by a simple iterative solution of the equations. Then one has to insert 
these large volume series into (\ref{o+}),(\ref{o-}). 
  At the end of this process one gets a bulky sum of terms, such that a lot of terms 
are identical under certain permutations of the integrating variables. 
Taking into account these permutational symmetries by appropriate symmetry factors, 
one ends up with the formula for the vacuum contributions as follows:
\begin{equation} \label{Ovac}
\begin{split}
{\cal O}^\pm \big|_{vac}=\sum\limits_{n_+=0}^\infty \sum\limits_{n_-=0}^\infty \,
\frac{1}{n_+! \, n_-!} \, \int\limits_{-\infty}^{\infty} \prod\limits_{i=1}^{n_++n_-} \frac{d \theta_i}{2 \pi} \,
\prod\limits_{i=1}^{n_+} {\cal F}_+(\theta_i+i \, \eta) \, \prod\limits_{i=n_++1}^{n_++n_-} {\cal F}_-(\theta_i-i \, \eta)\\ \times
F^{O^\pm}_{n_++n_-,c}(\theta_1\!+\!i \, \eta,...,\theta_{n_+}\!+\!i \, \eta,\theta_{n_++1}\!-\!i \, \eta,...,\theta_{n_++n_-}\!-\!i \, \eta),
\end{split}
\end{equation}
where the "connected" diagonal form factors of $O^\pm$ are given by the definitions:
\begin{equation} \label{FO+c}
\begin{split}
F^{O^+}_{n,c}(\theta_1,\theta_2,...,\theta_n)\!\!=\!\!\! \sum\limits_{\sigma\in S_n} \!\!\!
\left( \cosh(\theta_{\sigma(1)})\,\cosh(\theta_{\sigma(n)})\!\!-\!\!
\sinh(\theta_{\sigma(1)})\,\sinh(\theta_{\sigma(n)})
\right) \!\! \prod\limits_{j=1}^{n-1} G(\theta_{\sigma(j)}-\theta_{\sigma(j+1)}), 
\end{split}
\end{equation}
\begin{equation} \label{FO-c}
\begin{split}
F^{O^-}_{n,c}(\theta_1,\theta_2,...,\theta_n)\!\!=\!\!\! \sum\limits_{\sigma\in S_n} \!\!\!
\left( \cosh(\theta_{\sigma(1)})\,\sinh(\theta_{\sigma(n)})\!\!-\!\!
\sinh(\theta_{\sigma(1)})\,\cosh(\theta_{\sigma(n)})
\right) \!\! \prod\limits_{j=1}^{n-1} G(\theta_{\sigma(j)}-\theta_{\sigma(j+1)}), 
\end{split}
\end{equation}
where $\sigma$ denotes the elements of the symmetric group $S_n.$ 
From the permutation symmetry of the summand in (\ref{FO-c}), it follows that
$F^{O^-}_{n,c}(\theta_1,\theta_2,...,\theta_n)=0,$ and consequently:
${\cal O}^-\!\big|_{vac}\!\!\!\!=0.$ 
The next step is to determine{\footnote{The necessary computations are almost literally the same as those presented in section 3. of ref. \cite{PST14}.}} the large volume series representations for the 
2nd and 3rd terms in (\ref{o+}) and (\ref{o-}). 
First one has to rewrite them 
with the help of the right hand sides of (\ref{singlesum}) and (\ref{doublesum}) taken at appropriate values 
of the indexes $A$  and $B$. 
Then the all order 
large volume series representation of the solution of (\ref{lliegen}) must be inserted into the result. 
Finally, the careful bookkeeping of terms being identical under certain permutations of the 
variables leads to the final formula:
\begin{equation} \label{opmfinal}
\begin{split}
{\cal O}^\pm =\frac{1}{\rho(H_1,..,H_{m_H})} 
 \sum\limits_{\{H_+\}\cup \{H_-\}} \!~\! {\cal D}^{{O^\pm}}(\{H_+\})
 \,\, \rho(\{H_-\}|\{H_+\}),
\end{split}
\end{equation}
where the so-called dressed form factors take the form:
\begin{equation} \label{Opmdress}
\begin{split}
{\cal D}^{O^\pm}(\{H_1,...,H_n\})=\sum\limits_{n_+=0}^\infty \sum\limits_{n_-=0}^\infty 
\frac{1}{n_+! \, n_-!} \! \int\limits_{-\infty}^{\infty} \! \prod\limits_{i=1}^{n_++n_-} \! \! \frac{d \theta_i}{2 \pi} \!
\prod\limits_{i=1}^{n_+} \! {\cal F}_+(\theta_i+i \, \eta) \! \! \! \prod\limits_{i=n_++1}^{n_++n_-} \! \!
{\cal F}_-(\theta_i-i \, \eta)\\
\times 
F^{O^\pm}_{n+n_++n_-,c}(H_1,H_2,...,H_n,\theta_1\!+\!i \, \eta,...,\theta_{n_+}\!+\!i \, \eta,\theta_{n_++1}\!-\!i \, \eta,...,\theta_{n_++n_-}\!-\!i \, \eta).
\end{split}
\end{equation}
Now we can discuss the results. First we discuss the case of ${\cal O}^-.$
(\ref{Opmdress}) and (\ref{FO-c}) implies that 
${\cal O}^-=0$ exactly. Taking into account the connection between ${\cal O}^-$ and 
 the fermionic expectation values 
of the pseudo-scalar fermion bilinear:
$$\langle \bar{\Psi} \gamma^5 \Psi \rangle =\tfrac{1}{a}  \langle O^-_{2n} \rangle_\lambda \sim 
a^{\tfrac{p-1}{p+1}} \, {\cal O^-}\!\! +\!\!...,$$
one{\footnote{Here dots stand for terms tending to zero when $a \to 0$ in the attractive
$0<p<1$ regime of the model.}} can conclude that the fermionic  expectation values of 
$ \bar{\Psi}\gamma^5 \Psi$ are zero.

Next we discuss the result obtained for ${\cal O}^+.$
According to (\ref{calO+teta}), ${\cal O}^+$ is proportional to 
the fermionic (solitonic) expectation values of $\Theta_T.$ This operator belongs to a
conserved current, this is why its connected diagonal form-factors between pure fermion 
(soliton) states can be determined by using the arguments of references 
\cite{LM99} and \cite{saleur}. The actual computations lead to the 
following simple result:
\begin{equation} \label{Tetaconn}
\begin{split}
F^{\Theta_T}_{n,c}(\theta_1,\theta_2,..,\theta_n)=2 \pi {\cal M}^2 
\, F^{O^+}_{n,c}(\theta_1,\theta_2,..,\theta_n).
\end{split}
\end{equation}
Then, (\ref{Tetaconn}) together with (\ref{calO+teta}), (\ref{opmfinal}) and (\ref{Opmdress}) 
imply that the conjecture (\ref{Ps1}) of ref. \cite{En} is valid for the 
diagonal fermionic (solitonic) matrix elements of the  trace of the
stress-energy tensor, too.

\section{Summary, outlook and discussion}

In this paper, using the light-cone lattice regularization, we computed the finite 
volume expectation values of the
composite operators $\bar{\Psi}\Psi$ and $\bar{\Psi}\gamma^5\Psi$
between multi-fermion (soliton) states in the Massive Thirring (sine-Gordon) model. 
In the light-cone regularized picture, these expectation values are related to such 
2-point functions of the lattice spin operators, in which the operators are located at
 neighboring sites of the lattice. The operator $\bar{\Psi}\Psi$ is a particularly
 interesting operator, since it is proportional to the trace of the stress-energy tensor.
Thus, its  continuum finite volume expectation values can be computed \cite{zamiz} also
from the set of non-linear integral equations (NLIE) governing the finite volume spectrum of 
the theory. The final result, which was obtained after a lengthy computation of the 
spin-spin 2-point functions of neighboring operators, reproduced the pure NLIE result.

In general the finite volume matrix elements of local operators are computed 
 via their large volume series.
Thinking in this framework, previously in \cite{En}, an all order large 
volume series representation similar to that of \cite{LM99,Pozsg13,PST14}
 was conjectured for 
the finite volume diagonal matrix elements of local operators between 
multi-fermion (soliton) states. To check the conjecture of \cite{En} we rephrased the 
diagonal multi-fermion (soliton) matrix elements of the trace of the stress-energy tensor 
as a large volume series. The form of the series was conform to the 
conjecture of \cite{En}. 

Nevertheless, one has to note that, so far the large volume 
series conjecture of \cite{En} for the finite volume diagonal matrix 
elements of local operators between 
pure fermion (soliton) states, has been checked in cases when the local operator 
belongs to a conserved quantity of the theory. This is why it would be interesting 
to test the conjecture for such operators, which do not belong to 
the conserved quantities of the model. The results of \cite{FT,FPT11} indicate that
the truncated conformal space approach could be an appropriate method for such 
investigations. 

Beyond the results of \cite{En} and the present paper, several questions are still open.
The light-cone lattice approach gives access to all eigenstates of the Hamiltonian. 
Thus, diagonal matrix elements between non-pure fermion or soliton states can also be 
computed in this framework. It would be interesting to see, how the large volume 
series representation of \cite{En} should be modified in that case.
 Finally, a much more difficult but still open problem is the determination of
 non-diagonal finite volume form factors.

Beyond our approach, there is another approach to the form-factors of the 
SG model in cylindrical geometry \cite{BS1,BS2,BS3,BS4,BS5,BS6,BS7}. 
We close the paper with some discussion concerning the comparison of our method 
with that of the series of papers \cite{BS1,BS2,BS3,BS4,BS5,BS6,BS7}. In \cite{BS1,BS2,BS3,BS4,BS5,BS6,BS7} 
the hidden Grassmannian structure of the XXZ model 
was exploited to determine the finite temperature 1-point functions \cite{BS6} and ratios of 
infinite volume form-factors \cite{BS7} of the local operators of the SG theory. 
In this approach the compactified direction is time and the compactification length 
corresponds to the inverse temperature. The 1-point functions and the form-factors 
are computed as the continuum limit of appropriate partition functions of the 6-vertex model.

Our approach is more conventional. We consider the inhomogeneous 6-vertex model as a lattice 
regularization of the MT (SG) model and the operators we consider is the set of 
composite operators of Fermi fields and their 
derivatives{\footnote{Or equivalently their bosonized counterparts in the SG model.}}. 
In our case the compactified direction is space which makes 
easy to consider matrix elements of operators between excited states of the model. The Fermi 
fields are expressed in terms of local spin operators and the form-factors are given by such 
correlation functions of the spins, in which the spins are located at neighboring 
positions on the lattice. The correlators are evaluated by usual Algebraic Bethe Ansatz methods 
\cite{KMT99}-\cite{MT00}.

\vspace{1cm}
{\tt Acknowledgments}

\noindent 
The author would like to thank Zolt\'an Bajnok and J\'anos Balog for useful discussions.
This work was supported by OTKA grant under K109312.
 The author also would like to thank the support of an MTA-Lend\"ulet Grant.

\appendix

\section{Integral representation of some typical sums} \label{appSUM}

In this appendix we summarize the integral representations of the typical
 sums (\ref{SUM1}), (\ref{SUM2}) arising in the computation of the expectation values
 (\ref{sigma_mp0}) and (\ref{sigma_pm0}).
For the sake of completeness we recall them in this appendix, too:
\begin{eqnarray} \label{SUM1appZZ}
\Sigma_\lambda^{(1)}[f](\xi)&=&\sum\limits_{a=1}^m \, f(\lambda_a) \, X(\lambda_a|\xi), \\
\label{SUM2app}
\Sigma_\lambda^{(2)}[f](\xi,\xi')&=&\sum\limits_{a,b=1}^m \, f(\lambda_a,\lambda_b) \, X(\lambda_a|\xi)
\, X(\lambda_b|\xi').
\end{eqnarray}
In order to be able to transform these sums into integral expressions, we require that
 in each of its arguments $f$ should not have worse than constant asymptotics at infinity.
In (\ref{SUM2app}) we also require for $f$ to be an antisymmetric function of its
 arguments.

First, let us consider the single sum (\ref{SUM1appZZ}). The straightforward application of (\ref{sumrule})
 to (\ref{SUM1appZZ}) gives the integral representation as follows:
\begin{eqnarray} \label{SUM1app}
\Sigma_\lambda^{(1)}[f](\xi)&=&-\sum\limits_{j=1}^{m_H} \, f(h_j) \, X(h_j|\xi)+
\int\limits_{-\infty}^{\infty} \frac{d \lambda}{2 \pi} \, f(\lambda) \, {\cal G}(\lambda|\xi) \\
&-&\sum\limits_{\alpha=\pm} \int\limits_{-\infty}^{\infty} \frac{d \lambda}{2 \pi} \, f(\lambda+i \, \alpha \, \eta) \, {\cal G}(\lambda+i \, \alpha \, \eta|\xi) \, {\cal F}^{(\lambda)}_\alpha(\lambda+i \, \alpha \, \eta). \nonumber
\end{eqnarray}
However the form (\ref{SUM1app}) is still not appropriate for our purposes, since it has no well defined continuum 
limit in the cases of our interest, when $\xi=\xi_\pm$.
If one tries to take the continuum limit in (\ref{SUM1app}), it becomes immediately obvious that the first integral 
term in (\ref{SUM1app}) will diverge in the continuum limit. The reason is as follows.  
On the one hand, as a consequence of (\ref{GXlead}) and (\ref{GPMllie}), it follows that 
in the continuum limit ${\cal G}(\lambda|\xi_\pm) \sim e^{\pm \tfrac{\pi}{\gamma} \lambda}$ at large $\lambda.$ 
On the other hand, as it was mentioned in section 3, the concrete functions{\footnote{We just recall, that in our actual computations $f$ can be $f_1,$ $f_2,$ or $f_3$ given by the formulas (\ref{fv1})-(\ref{fv3}).}}
 $f(\lambda)$ for which we should apply (\ref{SUM1app}) have constant asymptotics at infinity.
Thus, the integrand in $\int\limits_{-\infty}^{\infty} \frac{d \lambda}{2 \pi} \, f(\lambda) \, {\cal G}(\lambda|\xi)$ 
in the rhs. of (\ref{SUM1app}) blows up exponentially at infinity in the continuum limit,
 which implies that the integral itself diverges.

 However on the lattice each term of (\ref{SUM1app}) is well defined and convergent, 
because as a consequence of (\ref{llie}) and (\ref{ess0})
 on the lattice ${\cal G}(\lambda|\xi)$ decays exponentially at large $\lambda.$ This means, that in order to be able 
to define the continuum limit, further transformations of (\ref{SUM1app}) are required. By exploiting (\ref{llie}) 
one can make the following replacement into the first integral term of (\ref{SUM1app}):
\begin{equation} \label{llierepl}
\begin{split}
 {\cal G}(\lambda|\xi) \to&- \sum\limits_{\alpha=\pm} \int\limits_{-\infty}^{\infty} d\lambda' \,  G_{\lambda}(\lambda-\lambda'-i \, \alpha \, \eta) \, {\cal G}(\lambda'+i \, \alpha \, \eta|\xi) \, {\cal F}^{(\lambda)}_{\alpha}(\lambda'+i \, \alpha \, \eta)+\\
&+S_0(\lambda|\xi)+\sum\limits_{j=1}^{m_H} \,
2 \pi \, G_\lambda(\lambda-h_j) \, X_j(\xi),  
\end{split}
\end{equation}
Making the replacement (\ref{llierepl}) into the first integral term of (\ref{SUM1app}), 
one ends up with the formula as follows for $\Sigma_\lambda^{(1)}[f](\xi):$
\begin{equation} \label{SUM1form}
\begin{split}
\Sigma_\lambda^{(1)}[f](\xi)&={\cal J}_0[f](\xi)+{\cal S}_{X}[f](\xi)+{\cal S}_{\cal G}[f](\xi),
\end{split}
\end{equation}
where  ${\cal J}_0[f](\xi),$ ${\cal S}_{X}[f](\xi)$ and ${\cal S}_{\cal G}[f](\xi)$  are functionals of $f$ and are of the form:
\begin{eqnarray} \label{SX}
{\cal S}_{X}[f](\xi)&=&
\sum\limits_{j=1}^{m_H} \, X(h_j|\xi) \, \, {\cal J}_G[f](h_j) \\
{\cal S}_{\cal G}[f](\xi)&=&\sum\limits_{\alpha=\pm}\int\limits_{-\infty}^{\infty}  
\frac{d \lambda}{2 \pi} \,{\cal F}_{\alpha}^{(\lambda)}(\lambda+i\,  \alpha \,  \eta) \,
{\cal G}(\lambda+i \, \alpha \,  \eta|\xi)\, \, {\cal J}_G[f](\lambda+i\,  \alpha \, \eta),
\label{ScalG}
\end{eqnarray}
\begin{equation} \label{calJG}
\begin{split}
{\cal J}_0[f](\xi)=\int\limits_{-\infty}^{\infty} \, 
\frac{d \lambda}{2 \pi} \, f(\lambda) \, S_0(\lambda|\xi), \qquad \qquad 
{\cal J}_G[f](\lambda)=(f \star G_\lambda)(\lambda)-f(\lambda).
\end{split}
\end{equation}
Here $\star$ denotes convolution with conventions given by (\ref{Ffg}). 
In this final representation each term has a well defined continuum 
limit{\footnote{In the terms containing $S_0(\lambda|\xi)$ the continuum limit can be taken 
after evaluating the integral, because the naive expansion (\ref{S0kifejt}) of $S_0(\lambda|\xi)$ under the 
integral leads to incorrect results. The careful computations can be found in appendix \ref{appArg}.}}.

Similar, but more tedious computations lead to the following formula 
for $\Sigma_\lambda^{(2)}[f](\xi,\xi').$ It is composed of six terms:
\begin{equation} \label{SUM2form}
\begin{split}
\Sigma_\lambda^{(2)}[f](\xi,\xi')&=\Sigma_0[f](\xi,\xi')+\Sigma_X[f](\xi,\xi')
+\Sigma_{\cal G}[f](\xi,\xi')+\Sigma_{XX}[f](\xi,\xi') \\
&+\Sigma_{{\cal G}{\cal G}}[f](\xi,\xi')+\Sigma_{X{\cal G}}[f](\xi,\xi').
\end{split}
\end{equation}
 The lower index of each term on the right hand side of (\ref{SUM2form}) 
refers to the internal structure of the expression as it becomes clear from
 their explicit form: 
\begin{equation} \label{SUM0}
\begin{split}
\Sigma_0[f](\xi,\xi')=
\int\limits_{-\infty}^{\infty} \!\! \frac{d \lambda}{2 \pi} \int\limits_{-\infty}^{\infty} \frac{d \lambda'}{2 \pi} S_0(\lambda'|\xi) \, f(\lambda',\lambda) \, S_0(\lambda|\xi'),
\end{split}
\end{equation}
\begin{equation} \label{SUMX}
\begin{split}
\Sigma_X[f](\xi,\xi')=\sum\limits_{j=1}^{m_H} X(h_j|\xi) \, F_X[f](h_j|\xi')-
\sum\limits_{j=1}^{m_H} X(h_j|\xi') \, F_X[f](h_j|\xi),
\end{split}
\end{equation}
\begin{equation} \label{SUMF}
\begin{split}
\Sigma_{\cal G}[f](\xi,\xi')=\sum\limits_{\alpha=\pm}\int\limits_{-\infty}^{\infty}  
\frac{d \lambda}{2 \pi} \,{\cal F}_{\alpha}^{(\lambda)}(\lambda+i\,  \alpha \,  \eta) \,
&\left\{
{\cal G}(\lambda+i \, \alpha \,  \eta|\xi)\,F_X[f](\lambda+i\,  \alpha \,  \eta|\xi') \right. \\
-&\left. {\cal G}(\lambda+i \, \alpha \,  \eta|\xi')\,F_X[f](\lambda+i\,  \alpha \,  \eta|\xi) 
 \right\},
\end{split}
\end{equation}
\begin{equation} \label{SUMXF}
\begin{split}
\Sigma_{X{\cal G}}[f](\xi,\xi')=& \sum\limits_{\alpha=\pm}\int\limits_{-\infty}^{\infty}  
\frac{d \lambda}{2 \pi} \,{\cal F}_{\alpha}^{(\lambda)}(\lambda+i\,  \alpha \,  \eta) \,
\sum\limits_{j=1}^{m_H} F_{XX}[f](h_j,\lambda+i\,  \alpha \,  \eta) \times \\
&\times \left\{
X(h_j|\xi')\, {\cal G}(\lambda+i \, \alpha \,  \eta|\xi)-
X(h_j|\xi)\, {\cal G}(\lambda+i \, \alpha \,  \eta|\xi')
\right\},
\end{split}
\end{equation}
\begin{equation} \label{SUMXX}
\begin{split}
\Sigma_{XX}[f](\xi,\xi')=\sum\limits_{j=1}^{m_H}\sum\limits_{k=1}^{m_H} \, X(h_j|\xi') \, F_{XX}[f](h_j,h_k) \,
X(h_k|\xi), 
\end{split}
\end{equation}
\begin{equation} \label{SUMFF}
\begin{split}
\Sigma_{{\cal G}{\cal G}}[f](\xi,\xi')=\!\!\!
\sum\limits_{\alpha,\beta=\pm} 
\int\limits_{-\infty}^{\infty}\frac{d \lambda}{2 \pi} \!\!\! \int\limits_{-\infty}^{\infty}\frac{d \lambda'}{2 \pi}
 \! {\cal F}_{\alpha}^{(\lambda)}(\lambda+i\,  \alpha \,  \eta) \,{\cal F}_{\beta}^{(\lambda)}(\lambda'+i\,  \beta \,  \eta) \times \\
\times {\cal G}(\lambda+i \, \alpha \,  \eta|\xi')\, {\cal G}(\lambda'+i \, \beta \,  \eta|\xi) \,
F_{XX}[f](\lambda+i \, \alpha \,  \eta,\lambda'+i \, \beta \,  \eta),
\end{split}
\end{equation}
where the two functions $F_X$ and $F_{XX}$, which are functionals of $f,$ take the form: 
\begin{eqnarray}
F_X[f](\lambda|\xi)&=&J_S[f](\lambda|\xi)-J_{SG}[f](\lambda|\xi), \label{FX} \\
F_{XX}[f](\lambda,\lambda')&=&f(\lambda',\lambda)+J_G[f](\lambda,\lambda')-J_G[f](\lambda',\lambda)-
J_{GG}[f](\lambda,\lambda'),
\label{FXX}
\end{eqnarray}
with the "elementary" functionals:
\begin{eqnarray} \label{J_S}
J_S[f](\lambda|\xi)&=&\int\limits_{-\infty}^{\infty} \!\! \frac{d \lambda'}{2 \pi}
 \, S_0(\lambda'|\xi) \, f(\lambda',\lambda), 
\\ \label{J_SG}
J_{SG}[f](\lambda|\xi)&=&\int\limits_{-\infty}^{\infty} \!\! \frac{d \lambda'}{2 \pi}
\int\limits_{-\infty}^{\infty} \!\! d \lambda'' \, S_0(\lambda'|\xi) \, f(\lambda',\lambda'') \,
G_\lambda(\lambda''-\lambda),
\\ \label{J_G}
J_G[f](\lambda,\lambda')&=& \int\limits_{-\infty}^{\infty} \!\! d\lambda'' G_\lambda(\lambda-\lambda'') \, f(\lambda'',\lambda'),
\\ \label{J_GG}
J_{GG}[f](\lambda,\lambda')&=& \int\limits_{-\infty}^{\infty} \!\! d\lambda'' \int\limits_{-\infty}^{\infty} \!\! d\lambda''' G_\lambda(\lambda-\lambda'') \, f(\lambda'',\lambda''') \, G_\lambda(\lambda'''-\lambda').
\end{eqnarray}

Once again, we note, that at the derivation of (\ref{SUM1form}) and (\ref{SUM2form}), it was important
to bring the sums into a sum of such integral expressions  which contain ${\cal G}(\lambda)$ only in the combination 
${\cal G(\lambda)} \, {\cal F}^{(\lambda)}_{\alpha}(\lambda)$.
 The reason for the preference of such a form is, that in the continuum limit, 
this combination is integrable along the lines $\lambda\pm i \, \eta$ with $\lambda \in \mathbb{R}$ and with $\eta$ 
being a small positive contour deformation parameter.
This convenient form could be derived by eliminating the single ${\cal G}(\lambda)$ terms 
with the help of (\ref{llie}).

\section{Large $\rho_0$ expansions} \label{appArg}

In this appendix we summarize, how one can obtain the coefficient functions ${\cal K}_\pm(\lambda|\rho_0)$ 
(\ref{calKpm}) and  the bulk term (\ref{Opmbulk}) of (\ref{Opmlarge}) 
via the computation of the large $\rho_0$ (\ref{rho}) limit of the functionals 
(\ref{Cbulk}), (\ref{Cpmp}) and (\ref{Cpmm}).
The key point of the computations is that one should 
work in Fourier space. This is why as a first step we fix our conventions for the Fourier-transformations. 
The Fourier-transform of a function $f$ is given by:
\begin{equation} \label{Ff}
\tilde{f}(\omega)=\int\limits_{-\infty}^{\infty} \!\! dx\, e^{i \omega  x} \, f(x).
\end{equation}
The inverse transformation reads as:
\begin{equation} \label{iFf}
f(x)=\int\limits_{-\infty}^{\infty} \!\! \frac{d\omega}{2 \pi} e^{-i \omega  x} \, \tilde{f}(\omega).
\end{equation}
The Fourier-transform of the convolution of two functions $f$ and $g$ 
is given by the product of individual Fourier-transforms{\footnote{Provided they exist.}}:
 \begin{equation} \label{Ffg}
\begin{split}
(f \star g)(x)=\int\limits_{-\infty}^{\infty} \!\! dy \, f(x-y)\, g(y), \qquad \widetilde{(f \star g)}(\omega)=\tilde{f}(\omega) \, \tilde{g}(\omega).
\end{split}
\end{equation}

Formulas (\ref{Cbulk}), (\ref{Cpmp}) and (\ref{Cpmm}) imply that in order to 
derive the formulas (\ref{CcalK}), (\ref{calKpm}) and (\ref{Opmbulk}), 
the following functionals should be computed in the large 
$\rho_0$ limit:
\begin{itemize}
\item ${\cal J}_G[f](\lambda)$ of (\ref{calJG}) taken at the functions $f_1,f_2, \text{and} \, f_3$ given by (\ref{fv1}), (\ref{fv2}) and (\ref{fv3}).
\item ${\cal J}_0[f_1](\xi_-), \, {\cal J}_0[f_2](\xi_+), \,\text{and} \, {\cal J}_0[f_3](\xi_-)$ defined by (\ref{calJG}) with $\xi_\pm=\pm\rho_0-i \tfrac{\gamma}{2}.$
\item $F_X[f_\pm](\lambda|\xi_\pm)$ of (\ref{FX}) with $f_\pm$ given by (\ref{f_+}) and (\ref{f_-}).
\item $\Sigma_0[f_\pm](\xi_-,\xi_+)$ of (\ref{SUM0}).
\end{itemize}

The strategy of the large $\rho_0$ evaluation of the above listed functionals is as follows.
One can write them in a very special form, namely as a linear combination of convolutions, 
such that $\rho_0$ appears in the argument of the convolutions. This means, that the large $\rho_0$ 
expansion of the functionals of our interest is equivalent to determine the large argument
series expansion of the convolutions appearing in them. 
It is worth to represent these convolutions in Fourier-space, since by using the property (\ref{Ffg}) 
they can be written as a single Fourier-integral. The large argument series expansion of these 
Fourier-integral expressions can be computed by using the residue theorem. Thus the positions of
the poles of the integrand will determine the large argument decay of the convolutions of our interest.

To complete the concrete large $\rho_0$ computations, first one has to define some functions, which 
constitute the elementary building blocks of the calculations:
\begin{eqnarray}
F_c(\lambda)&=&\frac{1}{ 2 \gamma \, \cosh(\tfrac{\pi}{\gamma}\lambda)}, \qquad \qquad 
\qquad \qquad \qquad
\tilde{F}_c(\omega)=\frac{1}{2 \cosh(\tfrac{\gamma}{2}\omega)}, \label{Fcdef} \\
F^\pm_c(\lambda)&=&e^{\pm 2 \lambda} \, F_c(\lambda), \qquad \qquad
\text{in case} \,\, p>1: \, \, 
\tilde{F}^\pm_c(\omega)=\tilde{F}_c(\omega \mp 2 i), \label{Fcpmdef} \\
g(\lambda)&=&\frac{1}{\cosh(2 \lambda)-\cos(2 \gamma)}, \qquad \qquad \qquad \qquad
 \tilde{g}(\omega)=\frac{\pi}{\sin(2 \gamma)} \frac{\sinh(\tfrac{\omega}{2}(\pi-2\gamma))}{\sinh(\tfrac{\pi}{2}\omega)}, \label{gdef} \\
\tilde{g^2}(\omega)&=&\frac{\pi}{2 \sin^2(2 \gamma)}
\left\{
 \frac{2 \cot(2 \gamma) \sinh(\tfrac{\omega}{2}(\pi-2\gamma))}{\sinh(\tfrac{\pi}{2}\omega)}+
\frac{\omega \, \cosh(\tfrac{\omega}{2}(\pi-2\gamma))}{\sinh(\tfrac{\pi}{2}\omega)}
\right\}, \label{g2def} \\
g_\alpha(\lambda)&=&\frac{1}{\sinh^2(\lambda-i \, \alpha)}, \qquad \qquad \qquad \quad
\tilde{g}_\alpha(\omega)=-\pi\, \omega \frac{e^{(\tfrac{\pi}{2}-\alpha)\,\omega}}{\sinh(\tfrac{\pi}{2}\omega)}, \quad \alpha\in (0,\pi), \label{galfadef} \\
\psi(\lambda)&=&\frac{\sinh(2\lambda)}{\cosh(2 \lambda)-\cos(2 \gamma)}, \qquad \qquad \quad
 \tilde{\psi}'(\omega)=\frac{\pi \omega}{\sinh(\tfrac{\pi}{2}\omega)} \cosh(\tfrac{\omega}{2}(\pi-2\gamma)), \label{psidef} \\
G_\lambda(\lambda)&=&\int\limits_{-\infty}^{\infty} \frac{dx}{2 \pi} e^{-i \omega  x} \, \tilde{G}_\lambda(\omega), \qquad \qquad \quad \tilde{G}_\lambda(\omega)=\frac{1}{2} 
\frac{\sinh(\tfrac{\pi \omega}{2}(1-\tfrac{2 \gamma}{\pi}))}
{\cosh(\tfrac{\gamma \omega}{2}) \sinh(\tfrac{\pi \omega}{2}(1-\tfrac{ \gamma}{\pi}))}, \label{Gldef} \\
G^\pm_\lambda(\lambda)&=&e^{\pm 2 \lambda} \, G_\lambda(\lambda), \qquad \qquad \qquad
\quad \text{in case} \, \, p>1: \quad
\tilde{G}^\pm_\lambda(\omega)=\tilde{G}_\lambda(\omega \mp 2 i), \label{Glpmdef} \\
\chi_F(\lambda)&=&\int\limits_0^\lambda d\lambda' F_c(\lambda')
=\tfrac{1}{\pi} \arctan \left[ \tanh(\tfrac{\pi \lambda}{2 \gamma})\right], \qquad \qquad
\tilde{\chi}_F(\omega)=i \, \tilde{F}_c(\omega) \, {\frak r}(\omega), \label{chiFdef}\\ 
\chi(\lambda)&=&\int\limits_0^\lambda d\lambda' G_\lambda(\lambda'),
 \quad 
\tilde{\chi}(\omega)=i \, \tilde{G}_\lambda(\omega) \, {\frak r}(\omega),
\quad \text{with} \quad {\frak r}(\omega)=\tfrac{1}{2}\left( \tfrac{1}{\omega+i \, 0}+
\tfrac{1}{\omega-i \, 0} \right)\!,
 \label{chiLdef} 
\end{eqnarray}
where the $\pm i \, 0$ prescription in ${\frak r}(\omega)$ ensures the correct treatment of
 the $1/\omega$ singularity of the Fourier integral representations of
 (\ref{chiFdef}) and (\ref{chiLdef}).

Let us start with the computation of the bulk (\ref{Cbulk}) or in other words the 
global constant in rapidity term. 
The representation of the building blocks of (\ref{Cbulk}) 
as linear combinations of convolutions read as:  
\begin{eqnarray}
{\cal J}_0[f_1](\xi_-)&=&-\tfrac{1}{2}, \label{calJ1m} \\
{\cal J}_0[f_2](\xi_+)&=& -\sinh(2 \rho_0)\, (g_{\gamma/2} \star \chi_F)(2 \rho_0)-
\tfrac{1}{2} \, \cosh(2 \rho_0), \label{calJ2p} \\
{\cal J}_0[f_3](\xi_-)&=& \sinh(2 \rho_0+i \, \gamma)\,
 (g_{\gamma/2} \star \chi_F)(-2 \rho_0)-\tfrac{1}{2} \,
 \cosh(2 \rho_0+i \, \gamma), \label{calJ2p}
\end{eqnarray}
\begin{equation} \label{S0ppm}
\begin{split}
\Sigma_0[f_+](\xi_+,\xi_-)=\frac{\cos \gamma}{2 \gamma^2} \left[ \cosh(2 \rho_0) \,{\cal T}_\gamma(\rho_0)-I(\rho_0) \right],
\end{split}
\end{equation}
\begin{equation} \label{S0mpm}
\begin{split}
\Sigma_0[f_-](\xi_+,\xi_-)=i \, \frac{\cos \gamma}{2 \gamma^2} \cosh(2 \rho_0) \, \sin \gamma \, {\cal T}_0(\rho_0),
\end{split}
\end{equation}
where ${\cal T}_\alpha(\rho_0)$ and $I(\rho_0)$ are given by:
\begin{equation} \label{Irho0}
\begin{split}
I(\rho_0)&=4 \, \gamma^2 \, (F_c \star \psi' \star \chi_F)(2 \rho_0),
\end{split}
\end{equation}
\begin{equation} \label{Talfa0}
\begin{split}
{\cal T}_\alpha (\rho_0)=\cos \alpha \, {\cal T}_0(\rho_0)=4 \gamma^2 \, \cos \alpha \left[ e^{2 \rho_0} \, (F_c^- \star g \star F_c)(2 \rho_0)-e^{- 2 \rho_0} (F_c^+ \star g \star F_c)(2 \rho_0) 
\right].
\end{split}
\end{equation}
The large argument expansions of appendix \ref{largeARG} 
lead to the following leading order large 
$\rho_0$ expressions for these building blocks in the attractive regime:
\begin{equation} \label{calJ2kifejt}
\begin{split}
{\cal J}_0[f_2](\xi_+)=-\frac{e^{-i \tfrac{p \pi}{2}}}{2} \,
\frac{1+p}{\cos(\tfrac{p \pi}{2})} \, e^{-2 \, p \, \rho_0}+O(e^{-2 \rho_0}),
\end{split}
\end{equation}
\begin{equation} \label{calJ3kifejt}
\begin{split}
{\cal J}_0[f_3](\xi_-)=-\frac{e^{i\, \gamma +i \, \tfrac{p \pi}{2}}}{2} \,
\frac{1+p}{\cos(\tfrac{p \pi}{2})} \, e^{-2 \, p \, \rho_0}+O(e^{-2 \rho_0}),
\end{split}
\end{equation}
\begin{equation} \label{Sigma0+kifejt}
\begin{split}
\Sigma_0[f_+](\xi_+,\xi_-)=\frac{\pi}{2 \gamma} \cos \gamma 
\left[\cot \gamma \, \cot \tfrac{\pi^2}{2 \gamma}-1  
\right]\, 
e^{2 (1-p) \, \rho_0}+O(e^{-2 \rho_0}),
\end{split}
\end{equation}
\begin{equation} \label{Sigma0-kifejt}
\begin{split}
\Sigma_0[f_-](\xi_+,\xi_-)=\tfrac{i}{2}\sin \gamma+i \sin(2 \gamma) \frac{\pi}{4 \gamma} \,
\left[ \tfrac{1}{\sin \gamma} \, \cot \tfrac{\pi^2}{2 \gamma}-\tfrac{1}{\cos \gamma}  
\right]\, 
e^{2 (1-p) \, \rho_0}+O(e^{-2 \rho_0}).
\end{split}
\end{equation}
Finally, inserting (\ref{calJ2kifejt}), (\ref{calJ3kifejt}), (\ref{Sigma0+kifejt}) and (\ref{Sigma0-kifejt}) 
into (\ref{Cbulk}) one ends up with the final result given by (\ref{Opmbulk}).

Now, we can continue with the computation of the coefficient functions $C^\pm_\pm(\lambda)$ given in (\ref{Cpmp}) and (\ref{Cpmm}). These functions can 
 also be represented as appropriate linear combinations of some convolutions. 
 Such convolution type representations of the elementary building blocks of $C^\pm_\pm(\lambda)$ are given by the 
 following formulas:
\begin{eqnarray} \label{JGf1}
{\cal J}_G[f_1](\lambda)&=&-\tfrac{p+1}{ 2\, p}, \\
{\cal J}_G[f_2](\lambda)&=& 2 \cosh(2 \rho_0) \, \chi(\infty)+\sinh(2 \rho_0) \, (g_{\gamma/2} \star \chi)(\lambda+\rho_0)-f_2(\lambda|\rho_0), \label{JGf2} \\
{\cal J}_G[f_3](\lambda)&=& 2 \cosh(2 \rho_0+i \, \gamma) \, \chi(\infty)\!-\!\sinh(2 \rho_0+i \gamma) \, (g_{\gamma/2} \star \chi)(\lambda-\rho_0)\!-\!f_3(\lambda|\rho_0), \label{JGf3}
\end{eqnarray}
\begin{equation} \label{Jsfppm}
\begin{split}
{J}_{S}[f_+](\lambda|\xi_\pm)=\cos \gamma \, \cosh(2 \rho_0) \left\{
e^{i\gamma} \left[ e^{2 \lambda} ( g \star F_c)(\lambda \mp \rho_0)-
e^{\pm 2 \rho_0} ( g \star F_c^+)(\lambda\mp \rho_0)
\right] \right. \\
-\left. e^{-i \gamma} \left[ e^{-2 \lambda}( g \star F_c)(\lambda\mp \rho_0)
-e^{\mp 2 \rho_0}( g \star F_c^-)(\lambda\mp \rho_0)\right] \right\}-2 \cos \gamma
\, ( \psi' \star \chi_F)(\lambda \mp \rho_0),
\end{split}
\end{equation}
\begin{equation} \label{Jsgfppm}
\begin{split}
{J}_{SG}[f_+](\lambda|\xi_\pm)\! =\! \cos \gamma  \cosh(2 \rho_0) \left\{
e^{i\gamma} \left[ e^{2 \lambda} (G^- \!\! \star g \star F_c)(\lambda \mp \rho_0)-
e^{\pm 2 \rho_0} (G\star g \star F_c^+)(\lambda\mp \rho_0)
\right] \right. \\
-\!\! \left. e^{-i \gamma} \left[ e^{-2 \lambda}(G^+ \!\! \star  g \star F_c)(\lambda\mp \rho_0)
\!-e^{\mp 2 \rho_0}(G \star g \star F_c^-)(\lambda\mp \rho_0)\right] \right\}\!\!- \!\!2 \cos \gamma
 (G\star \psi' \star \chi_F)(\lambda \mp \rho_0).
\end{split}
\end{equation}
\begin{equation} \label{Jsfmpm}
\begin{split}
{J}_S[f_-](\lambda|\xi_\pm)=\cos \gamma \, \sinh(2 \rho_0) \left\{ 
e^{i\, \gamma} \left[e^{2 \lambda} (g \star F_c)(\lambda\mp\rho_0)-e^{\pm 2 \rho_0} (g \star F_c^+)(\lambda\mp \rho_0)
 \right]-\right. \\
\left. 
+e^{-i\, \gamma} \left[ e^{-2 \lambda } (g\star F_c)(\lambda \mp \rho_0)-
e^{\mp 2 \rho_0} (g \star F_c^-)(\lambda\mp \rho_0)\right]
\right\},
\end{split}
\end{equation}
\begin{equation} \label{Jsgfmpm}
\begin{split}
{J}_{SG}[f_-](\lambda|\xi_\pm)\!=\! \cos \gamma  \sinh(2 \rho_0) \left\{ 
e^{i\, \gamma} \! \left[e^{2 \lambda} (G^- \!\! \star g \star F_c)(\lambda\mp\rho_0)\!-\!e^{\pm 2 \rho_0} (G \star g \star F_c^+)(\lambda\mp \rho_0)
 \right]\right. \\
\left. 
+e^{-i\, \gamma} \left[ e^{-2 \lambda } (G^+ \! \! \star g\star F_c)(\lambda \mp \rho_0)-
e^{\mp 2 \rho_0} (G \star g \star F_c^-)(\lambda\mp \rho_0)\right]
\right\}.
\end{split}
\end{equation}
The large argument expansions presented in appendix \ref{largeARG} lead to the following 
leading order large $\rho_0$ forms for the functionals (\ref{JGf1})-(\ref{Jsgfmpm}):
\begin{eqnarray} \label{JGf2kifejt}
{\cal J}_G[f_2](\lambda)=-e^{(1-p) \rho_0} e^{-(1+p) \lambda} \, e^{-i \tfrac{p \pi}{2}} \, \frac{1+p}{2 \sin(\tfrac{p \pi}{2})}+O(e^{-2 \rho_0}), \\
{\cal J}_G[f_3](\lambda)=-e^{(1-p) \rho_0} e^{(1+p) \lambda} \, e^{i \, \gamma+i \tfrac{p \pi}{2}} \, \frac{1+p}{2 \sin(\tfrac{p \pi}{2})}+O(e^{-2 \rho_0}), \label{JGf3kifejt}
\end{eqnarray}
\begin{equation} \label{Jsf+kifejt}
\begin{split}
\frac{{J}_S[f_+](\lambda|\xi_\pm)}{\sinh(2 \rho_0)}=\mp \cos \gamma \, e^{\pm (2 \lambda+i\, \gamma)}+O(e^{-2 \rho_0}),
\end{split}
\end{equation}
\begin{equation} \label{Jsgf+kifejt}
\begin{split}
\frac{{J}_{SG}[f_+](\lambda|\xi_\pm)}{\sinh(2 \rho_0)}=\pm e^{\pm i \, \gamma} \,
\frac{\pi}{2 \gamma} \left[ \tan \tfrac{\pi^2}{2 \gamma}-\cot \gamma
\right]\, e^{\pm (1+p) \lambda} \, e^{(1-p) \rho_0}
\mp \cos \gamma \, e^{\pm (2 \lambda+i\, \gamma)}+O(e^{-2 \rho_0}),
\end{split}
\end{equation}
\begin{equation} \label{Jsf+kifejt}
\begin{split}
\frac{{J}_S[f_-](\lambda|\xi_\pm)}{\sinh(2 \, \rho_0)}=- \cos \gamma \, e^{\pm (2 \lambda+i\, \gamma)}+O(e^{-2 \rho_0}),
\end{split}
\end{equation}
\begin{equation} \label{Jsgf-kifejt}
\begin{split}
\frac{{J}_{SG}[f_-](\lambda|\xi_\pm)}{\sinh(2 \rho_0)}=e^{\pm i \, \gamma} \,
\frac{\pi}{2 \gamma} \left[ \tan \tfrac{\pi^2}{2 \gamma}-\cot \gamma
\right]\, e^{\pm (1+p) \lambda} \, e^{(1-p) \rho_0}
- \cos \gamma \, e^{\pm (2 \lambda+i\, \gamma)}+O(e^{-2 \rho_0}).
\end{split}
\end{equation}
Finally inserting these leading order expressions into (\ref{Cpmp}) and (\ref{Cpmm}), 
one ends up with (\ref{CcalK}) with ${\cal K}_\pm(\lambda|\rho_0)$ given by (\ref{calKpm}).

\subsection{The large $N$ magnitude of the terms ${\cal O}^{\pm}_{XX},$ ${\cal O}^{\pm}_{X{\cal G}},$ 
and ${\cal O}^{\pm}_{{\cal G}{\cal G}}$} \label{multilinQ}

At the derivation of (\ref{Opmlarge}) we omitted the contributions of the terms ${\cal O}^\pm_{XX},$ 
${\cal O}^{\pm}_{X{\cal G}}$ and ${\cal O}^{\pm}_{{\cal G}{\cal G}}$ from (\ref{Opm0XG}).
The reason for this was that, according to the anticipated result (\ref{roughexact}),
 these terms are next to leading order ones with respect to 
${\cal O}^{\pm}_{0},$ ${\cal O}^{\pm}_{X},$ and ${\cal O}^{\pm}_{\cal G}$
in the continuum limit. In this subsection we present the proof of the second line of 
(\ref{roughexact}). Namely, we show , that the multilinear or quadratic 
in ${\cal G}$ and $X$ terms are indeed of order $\tfrac{1}{N^2}$ in the 
large $N$ limit and so they are really negligible with respect to the constant and purely linear terms.

Formulas (\ref{OXX}), (\ref{OXG}) and (\ref{OGG}) together with (\ref{SUMXF})-(\ref{SUMFF}) imply the following 
representations for ${\cal O}^{\pm}_{XX},$ ${\cal O}^{\pm}_{X{\cal G}},$ and ${\cal O}^{\pm}_{{\cal G}{\cal G}}:$

\begin{equation} \label{OXGS1}
\begin{split}
{\cal O}^\pm_{X{\cal G}}=& \tfrac{1}{\sinh(2 \rho_0)} \sum\limits_{\alpha=\pm}\int\limits_{-\infty}^{\infty}  
\frac{d \lambda}{2 \pi} \,{\cal F}_{\alpha}^{(\lambda)}(\lambda+i\,  \alpha \,  \eta) \,
\sum\limits_{j=1}^{m_H} F_{XX}[f_\pm](h_j,\lambda+i\,  \alpha \,  \eta) \times \\
&\times \left\{
X(h_j|\xi_-)\, {\cal G}(\lambda+i \, \alpha \,  \eta|\xi_+)-
X(h_j|\xi_+)\, {\cal G}(\lambda+i \, \alpha \,  \eta|\xi_-)
\right\},
\end{split}
\end{equation}
\begin{equation} \label{OXXS1}
\begin{split}
{\cal O}^\pm_{X{X}}=\tfrac{1}{\sinh(2 \rho_0)} \sum\limits_{j=1}^{m_H}\sum\limits_{k=1}^{m_H} \, X(h_j|\xi_-) \, F_{XX}[f_\pm](h_j,h_k) \,
X(h_k|\xi_+), 
\end{split}
\end{equation}
\begin{equation} \label{OGGS1}
\begin{split}
{\cal O}^\pm_{{\cal G}{\cal G}}=\tfrac{1}{\sinh(2 \rho_0)} \!\!\!
\sum\limits_{\alpha,\beta=\pm} 
\int\limits_{-\infty}^{\infty}\frac{d \lambda}{2 \pi} \!\!\! \int\limits_{-\infty}^{\infty}\frac{d \lambda'}{2 \pi}
 \! {\cal F}_{\alpha}^{(\lambda)}(\lambda+i\,  \alpha \,  \eta) \,{\cal F}_{\beta}^{(\lambda)}(\lambda'+i\,  \beta \,  \eta) \times \\
\times {\cal G}(\lambda+i \, \alpha \,  \eta|\xi_-)\, {\cal G}(\lambda'+i \, \beta \,  \eta|\xi_+) \,
F_{XX}[f_\pm](\lambda+i \, \alpha \,  \eta,\lambda'+i \, \beta \,  \eta),
\end{split}
\end{equation}
where according to (\ref{FXX}), (\ref{J_G}) and (\ref{J_GG}):
\begin{equation}
\begin{split} \label{FXXpm}
F_{XX}[f_\pm](\lambda,\lambda')&=f_\pm(\lambda',\lambda)+\int\limits_{-\infty}^{\infty} \!\! d\lambda'' G_\lambda(\lambda-\lambda'') \, f(\lambda'',\lambda')\\
-\int\limits_{-\infty}^{\infty} \!\! d\lambda'' G_\lambda(\lambda'-\lambda'') \, f(\lambda'',\lambda)&-
\int\limits_{-\infty}^{\infty} \!\! d\lambda'' \int\limits_{-\infty}^{\infty} \!\! d\lambda''' G_\lambda(\lambda-\lambda'') \, f(\lambda'',\lambda''') \, G_\lambda(\lambda'''-\lambda').
\end{split}
\end{equation}
The function $G_\lambda(\lambda)$ is given by (\ref{Gldef}) and 
$f_\pm(\lambda,\lambda')$ is defined  in (\ref{f_+}) and (\ref{f_-}).
To make apparent the $\rho_0$ dependence of $f_\pm,$ we rephrase them as follows:
\begin{eqnarray}
f_+(\lambda,\lambda')&=&\cosh(2 \rho_0) \, f_+^{(1)}(\lambda,\lambda')+f_+^{(2)}(\lambda,\lambda'), \label{fprho} \\
f_-(\lambda,\lambda')&=&\sinh(2 \rho_0) \, f_-^{(1)}(\lambda,\lambda'), \label{fmrho}
\end{eqnarray}
with
\begin{eqnarray}
f^{(1)}_+(\lambda,\lambda')&=& 2 \cos(\gamma) \,
 \frac{ \left[\sinh(2\lambda+i \, \gamma)-\sinh(2\lambda'+i \, \gamma)  \right]}{\cosh(2(\lambda-\lambda'))-\cos(2 \gamma)},
\label{f_+1} \\
\qquad f^{(2)}_+(\lambda,\lambda')&=& -2 \cos(\gamma) \,\frac{ \sinh(2(\lambda-\lambda'))}{\cosh(2(\lambda-\lambda'))-\cos(2 \gamma)},
\label{f_+2} \\
f^{(1)}_-(\lambda,\lambda')&=& 2 \cos(\gamma) \,
 \frac{\cosh(2\lambda+i \, \gamma)-\cosh(2\lambda'+i \, \gamma)}{\cosh(2(\lambda-\lambda'))-\cos(2 \gamma)}. 
\label{f_-1}
\end{eqnarray}
The point in the representations (\ref{fprho}) and (\ref{fmrho}) is that the $\rho_0$ dependence is lifted as a prefactor, and the coefficient functions $f^{(1)}_\pm$ and $f^{(2)}_+$ are $\rho_0$ independent.

To determine the magnitude of  
${\cal O}^{\pm}_{XX},$ ${\cal O}^{\pm}_{X{\cal G}},$ and ${\cal O}^{\pm}_{{\cal G}{\cal G}},$ first one has to 
compute the large $N$ magnitudes of each building blocks of the formulas (\ref{OXGS1}), (\ref{OXXS1}) and (\ref{OGGS1}).

In all the three quantities the functional $F_{XX}[f_\pm](\lambda,\lambda')$ 
given by (\ref{FXXpm}) arises. Now we show , that it is of order $e^{2 \rho_0}$ at large $\rho_0.$ 
Formulas (\ref{f_+1}), (\ref{f_+2}) and ((\ref{f_-1})) for the functions 
$f^{(1)}_\pm,$ $f^{(2)}_+,$ imply that these functions have constant asymptotics 
 at infinity in each of their variables. On the other hand (\ref{Gldef}) implies that $G_\lambda(\lambda)$ 
is $\rho_0$ independent and has the large $\lambda$ asymptotics:
\begin{equation}
\begin{split}
G_\lambda(\lambda)\sim e^{-\alpha_G \, |\lambda|}, \qquad 
\text{with:} \quad \alpha_G=1+\text{Min}\left(p,1+\tfrac{2}{p}\right), \qquad 0<p,
\end{split}
\end{equation}
with $p$ being the coupling constant defined by (\ref{csatrel1}).
These large $\lambda$ and $\lambda'$ asymptotics ensure, that all integrals will converge in (\ref{FXXpm}).
Furthermore, if one uses the representations (\ref{fprho}) and (\ref{fmrho}) in (\ref{FXXpm}), then 
it becomes obvious that the $\rho_0$ dependence is only given by the trivial factors $\cosh(2 \rho_0)$ 
or $\sinh(2 \rho_0)$ of (\ref{fprho}) and (\ref{fmrho}), respectively, such that these factors 
can be lifted in front of the convergent $\rho_0$ independent integrals. This implies, that 
$F_{XX}[f_\pm]\sim e^{2 \rho_0}$ in the large $\rho_0$ limit.

The next common building block in (\ref{OXGS1}), (\ref{OXXS1}) and (\ref{OGGS1}) is a trivial $\rho_0$ dependent prefactor: $\frac{1}{\sinh(2 \rho_0)}$ which is of order $e^{-2 \rho_0},$ 
when $\rho_0\to \infty.$

Formulas (\ref{OXGS1}), (\ref{OXXS1}) and (\ref{OGGS1}) are multilinear in ${\cal G}(\lambda|\xi_\pm)$ and $X(\lambda|\xi_\pm).$ Their large $N$ magnitudes can be read off from (\ref{GXNa}) 
and they both turn to be of order $\tfrac{1}{N}$ in the large $N$ limit.

The amount of information provided so far is enough to give the large $N$ estimate for ${\cal O}^\pm_{XX}$ 
given by (\ref{OXXS1}). Multiplying the magnitudes of the building blocks immediately leads to the 
large $N$ estimate: ${\cal O}^\pm_{XX} \sim \tfrac{1}{N^2}.$

To prove that  ${\cal O}^\pm_{X{\cal G}}$ and  ${\cal O}^\pm_{{\cal G}{\cal G}}$ are also of order 
$\tfrac{1}{N^2},$ one should deal with the large $N$ limit of ${\cal F}^{(\lambda)}_\pm(\lambda\pm i \, \eta),$ too.
This function is defined in (\ref{calFl}) and due to (\ref{Zlcont}) 
it becomes of order one in the large $N$ limit. This is why, at leading order its continuum counterpart 
${\cal F}^{(\lambda)}_{\pm,c}(\lambda\pm i \, \eta)$ can be substituted into the formulas (\ref{OXGS1}) 
and (\ref{OGGS1}). 
Equation (\ref{DDVcont}) implies, that at large $\lambda$ this function decays as: 
${\cal F}^{(\lambda)}_{\pm,c}(\lambda\pm i \, \eta)\sim e^{-\ell \, \sinh \tfrac{\pi}{\gamma}(\lambda \pm i \eta)}.$
This extremely rapid decay at infinity ensures the convergence of all integrals entering the expressions 
occurring in (\ref{OXGS1}) and (\ref{OGGS1}). This implies that, the large $N$ magnitudes of 
${\cal O}^{\pm}_{X{\cal G}}$ and ${\cal O}^{\pm}_{{\cal G}{\cal G}}$ are given by the product of the 
magnitudes of the basic building blocks determined above.

For completeness we summarize below the magnitudes of the important building blocks of 
${\cal O}^{\pm}_{XX},$ ${\cal O}^{\pm}_{X{\cal G}},$ 
and ${\cal O}^{\pm}_{{\cal G}{\cal G}}$ together with the relation between $\rho_0$ and $N$ 
given by (\ref{rho}):
\begin{equation} \label{magnitudelist}
\begin{split}
\tfrac{1}{N}\sim e^{-(1+p) \rho_0}, \qquad \tfrac{1}{\sinh(2 \rho_0)} \sim e^{-2 \rho_0}\qquad F_{XX}[f_\pm](\lambda,\lambda')\sim f_{\pm}(\lambda,\lambda') \sim e^{2 \rho_0}, \\
\qquad {\cal G}(\lambda|\xi_\pm)\sim e^{-(1+p)\rho_0}, \qquad X(\lambda|\xi_\pm)\sim e^{-(1+p)\rho_0}.
\end{split}
\end{equation} 
We emphasize, that these $\rho_0$ dependences are not entangled with the $\lambda$ dependence of the quantities
 they belong to, thus they can be lifted in front of $\rho_0$ independent convergent sums and integrals 
entering the formulas (\ref{OXGS1}), (\ref{OXXS1}) and (\ref{OGGS1}) after replacing (\ref{FXXpm}) with the 
representations (\ref{fprho}) and (\ref{fmrho}) into them. 
This implies that the magnitude of the individual building blocks listed in (\ref{magnitudelist}) 
determine that large $N$ magnitudes of ${\cal O}^{\pm}_{XX},$ ${\cal O}^{\pm}_{X{\cal G}}$ 
and ${\cal O}^{\pm}_{{\cal G}{\cal G}}:$
\begin{eqnarray}
{\cal O}^{\pm}_{XX} &\sim&  \,\, \tfrac{1}{\sinh(2 \rho_0)} \,\, X \,\, X \,\, F_{XX}[f_\pm] \sim e^{-2 \,(1+p) \rho_0} \sim \tfrac{1}{N^2}, \nonumber \\
{\cal O}^{\pm}_{X{\cal G}} &\sim&  \,\, \tfrac{1}{\sinh(2 \rho_0)} \,\, X \,\, {\cal G} \,\, F_{XX}[f_\pm] \,
\sim e^{-2 \,(1+p) \rho_0} \sim \tfrac{1}{N^2}, \label{OXXXGGG} \\
{\cal O}^{\pm}_{{\cal G}{\cal G}} &\sim& \,\, \tfrac{1}{\sinh(2 \rho_0)} \, \, {\cal G}\, \, {\cal G} \,\, F_{XX}[f_\pm] \sim e^{-2 \,(1+p) \rho_0} \sim \tfrac{1}{N^2}. \nonumber 
\end{eqnarray}
One can see from (\ref{OXXXGGG}) that each expression is of order $\tfrac{1}{N^2}\sim a^2$ in the large $N$ limit,  
consequently they are negligible in the continuum limit 


\section{Large argument series representations}\label{largeARG}
In this appendix we list the large argument expansions of the convolutions being necessary for the 
explicit computations presented in appendix \ref{appArg}. 
As it was mentioned in appendix \ref{appArg}, the 
series representations listed below, can be obtained by evaluating the Fourier-representations of the convolutions 
with the help of the residue theorem. The constituent functions of the relevant convolutions together with their 
Fourier-transforms are listed in (\ref{Fcdef})-(\ref{chiLdef}). 
The definitions of the necessary convolutions together with their large argument expansion read as follows:
\begin{equation} \label{GgFcP}
\begin{split}
(G_\lambda\star g \star F_c^+)(\lambda)\stackrel{\lambda \to \pm \infty}{=}\sum\limits_{k=0}^\infty \left\{ 
U_{1,k}^{(\pm)} \, e^{\mp 2 (1+k)  \lambda}\right. &+
U_{2,k}^{(\pm)} \, e^{\mp (1+2 k) \tfrac{\pi}{\gamma} \lambda}+
U_{3,k}^{(\pm)} \, e^{2 \lambda \mp (1+2 k) \tfrac{\pi}{\gamma} \lambda}  \\
 &+ \left. U_{4,k}^{(\pm)} \, e^{\mp 2 (1+k) \tfrac{\pi}{\pi-\gamma} \lambda} \right\}, \qquad \text{with:}
\end{split}
\end{equation}
\begin{eqnarray}
U_{1,k}^{(+)}&=&\csc (2 \gamma ) \sin (2 \gamma  (k+1)) \sec (\gamma  (k+2)),   
\label{U1kp} \nonumber \\
U_{2,k}^{(+)}&=&   \frac{\pi  (-1)^{-k} \csc ^2(\gamma ) \sec (\gamma ) \tan \left(\frac{\pi ^2 (2 k+1)}{2 \gamma
   }\right)}{4 \gamma },
\label{U2kp} \nonumber \\
U_{3,k}^{(+)}&=& \frac{\pi  \csc ^2(\gamma ) \sec (\gamma ) \sin ^2\left(\frac{(\pi -2 \gamma ) (-2 \gamma +2 \pi
    k+\pi )}{2 \gamma }\right) \csc \left(\frac{(\pi -\gamma ) (-2 \gamma +2 \pi  k+\pi )}{2
   \gamma }\right) \sec \left(\frac{\pi  (\gamma +2 \pi  k+\pi )}{2 \gamma }\right)}{4 \gamma },  
\label{U3kp} \nonumber \\
U_{4,k}^{(+)}&=&\frac{\pi  (-1)^{-k} \csc (2 \gamma ) \sin ^2\left(\frac{\pi  (\pi -2 \gamma ) (k+1)}{\pi
   -\gamma }\right) \csc \left(\frac{\pi ^2 (k+1)}{\pi -\gamma }\right) \sec \left(\frac{\pi 
   \gamma  (k+1)}{\pi -\gamma }\right) \sec \left(\gamma +\frac{\pi  \gamma  (k+1)}{\pi -\gamma
   }\right)}{2 (\pi -\gamma )},  
\label{U4kp} \nonumber 
\end{eqnarray}
\begin{eqnarray}
U_{1,k}^{(-)}&=&\csc (2 \gamma ) \sin (2 \gamma  (k+1)) \sec (\gamma  k),   
\label{U1km} \\
U_{2,k}^{(-)}&=& -\frac{\pi  (-1)^k \csc ^2(\gamma ) \sec (\gamma ) \tan \left(\frac{\pi ^2 (2 k+1)}{2 \gamma
   }\right)}{4 \gamma }, 
\label{U2km} \nonumber \\
U_{3,k}^{(-)}&=& \frac{\pi  \csc ^2(\gamma ) \sec (\gamma ) \sin ^2\left(\frac{(\pi -2 \gamma ) (2 \gamma +2 \pi 
   k+\pi )}{2 \gamma }\right) \csc \left(\frac{(\pi -\gamma ) (2 \gamma +2 \pi  k+\pi )}{2
   \gamma }\right) \sec \left(\frac{\pi  (-\gamma +2 \pi  k+\pi )}{2 \gamma }\right)}{4 \gamma },
\label{U3km} \nonumber \\
U_{4,k}^{(-)}&=&\frac{\pi  (-1)^k \csc (\gamma ) \sec (\gamma ) \sin ^2\left(\frac{\pi  (\pi -2 \gamma )
   (k+1)}{\pi -\gamma }\right) \csc \left(\frac{\pi ^2 (k+1)}{\pi -\gamma }\right) \sec
   \left(\frac{\pi  \gamma  (k+1)}{\pi -\gamma }\right) \sec \left(\frac{\gamma  (\gamma +\pi 
   k)}{\pi -\gamma }\right)}{4 \pi -4 \gamma }. \nonumber
\label{U4km}
\end{eqnarray}
\begin{equation} \label{GgFcM}
\begin{split}
(G_\lambda\star g \star F_c^-)(\lambda)\stackrel{\lambda \to \pm \infty}{=}\sum\limits_{k=0}^\infty \left\{ 
Z_{1,k}^{(\pm)} \, e^{\mp 2 (1+k)  \lambda}\right. &+
Z_{2,k}^{(\pm)} \, e^{\mp (1+2 k) \tfrac{\pi}{\gamma} \lambda}+
Z_{3,k}^{(\pm)} \, e^{-2 \lambda \mp (1+2 k) \tfrac{\pi}{\gamma} \lambda}  \\
 &+ \left. Z_{4,k}^{(\pm)} \, e^{\mp 2 (1+k) \tfrac{\pi}{\pi-\gamma} \lambda} \right\}, \qquad \text{with:}
\end{split}
\end{equation}
\begin{eqnarray}
Z_{1,k}^{(+)}&=&  \csc (2 \gamma ) \sin (2 \gamma  (k+1)) \sec (\gamma  k), 
\label{Z1kp} \nonumber \\
Z_{2,k}^{(+)}&=&   \frac{\pi  (-1)^{-k} \csc ^2(\gamma ) \sec (\gamma ) \cot \left(\frac{\pi  (\gamma +2 \pi  k+\pi
   )}{2 \gamma }\right)}{4 \gamma },
\label{Z2kp} \nonumber \\
Z_{3,k}^{(+)}&=&  -\frac{\pi  (-1)^{-k} \csc ^2(\gamma ) \sec (\gamma ) \sin ^2\left(\frac{\pi ^2 (2 k+1)}{2
   \gamma }-2 \gamma \right) \sec \left(\frac{\pi ^2 (2 k+1)}{2 \gamma }-\gamma \right) \sec
   \left(\frac{\pi  (\gamma +2 \pi  k+\pi )}{2 \gamma }\right)}{4 \gamma },
\label{Z3kp} \nonumber \\
Z_{4,k}^{(+)}&=&  \frac{\pi  (-1)^{-k} \csc (\gamma ) \sec (\gamma ) \sin ^2\left(\frac{\pi  (\pi -2 \gamma )
   (k+1)}{\pi -\gamma }\right) \csc \left(\frac{\pi ^2 (k+1)}{\pi -\gamma }\right) \sec
   \left(\frac{\pi  \gamma  (k+1)}{\pi -\gamma }\right) \sec \left(\frac{\gamma  (\gamma +\pi 
   k)}{\pi -\gamma }\right)}{4 \pi -4 \gamma },
\label{Z4kp} \nonumber 
\end{eqnarray}
\begin{eqnarray} 
Z_{1,k}^{(-)}&=&   \csc (2 \gamma ) \sin (2 \gamma  (k+1)) \sec (\gamma  (k+2)),
\label{Z1km} \nonumber \\
Z_{2,k}^{(-)}&=&  \frac{\pi  (-1)^k \csc ^2(\gamma ) \sec (\gamma ) \tan \left(\frac{\pi ^2 (2 k+1)}{2 \gamma
   }\right)}{4 \gamma }
\label{Z2km} \nonumber \\
Z_{3,k}^{(-)}&=&  -\frac{\pi  (-1)^k \csc ^2(\gamma ) \sec (\gamma ) \sin ^2\left(2 \gamma +\frac{\pi ^2 (2
   k+1)}{2 \gamma }\right) \sec \left(\frac{\pi  (-\gamma +2 \pi  k+\pi )}{2 \gamma }\right)
   \sec \left(\gamma +\frac{\pi ^2 (2 k+1)}{2 \gamma }\right)}{4 \gamma }, 
\label{Z3km} \nonumber \\
Z_{4,k}^{(-)}&=& \frac{\pi  (-1)^k \csc (2 \gamma ) \sin ^2\left(\frac{\pi  (\pi -2 \gamma ) (k+1)}{\pi -\gamma
   }\right) \csc \left(\frac{\pi ^2 (k+1)}{\pi -\gamma }\right) \sec \left(\frac{\pi  \gamma 
   (k+1)}{\pi -\gamma }\right) \sec \left(\gamma +\frac{\pi  \gamma  (k+1)}{\pi -\gamma
   }\right)}{2 (\pi -\gamma )},
\label{Z4km} \nonumber
\end{eqnarray}
\begin{equation} \label{GPgFc}
\begin{split}
(G^+_\lambda\star g \star F_c)(\lambda)\stackrel{\lambda \to \pm \infty}{=}\sum\limits_{k=0}^\infty \left\{ 
V_{1,k}^{(\pm)} \, e^{\mp 2 (1+k)  \lambda}\right. &+
V_{2,k}^{(\pm)} \, e^{\mp (1+2 k) \tfrac{\pi}{\gamma} \lambda}+
V_{3,k}^{(\pm)} \, e^{2 \lambda \mp (1+2 k) \tfrac{\pi}{\gamma} \lambda}  \\
 &+ \left. V_{4,k}^{(\pm)} \, e^{2 \lambda \mp 2 (1+k) \tfrac{\pi}{\pi-\gamma} \lambda} \right\}, \qquad \text{with:}
\end{split}
\end{equation}
\begin{eqnarray}
V_{1,k}^{(+)}&=&  \csc (\gamma ) \sec (\gamma ) \sin (\gamma  (k+1)), 
\label{V1kp} \nonumber \\
V_{2,k}^{(+)}&=&   \frac{\pi  (-1)^{-k} \csc ^2(\gamma ) \sec (\gamma ) \sin \left(\frac{\pi ^2 (2 k+1)}{2 \gamma
   }-2 \gamma \right) \sec \left(\frac{\pi ^2 (2 k+1)}{2 \gamma }-\gamma \right)}{4 \gamma },
\label{V2kp} \nonumber \\
V_{3,k}^{(+)}&=&  -\frac{\pi  (-1)^{-k} \csc ^2(\gamma ) \sec (\gamma ) \sin \left(2 \gamma +\frac{\pi ^2 (2
   k+1)}{2 \gamma }\right) \csc \left(\frac{\pi  (\gamma +2 \pi  k+\pi )}{2 \gamma }\right)}{4
   \gamma },
\label{V3kp} \nonumber \\
V_{4,k}^{(+)}&=&  -\frac{\pi  (-1)^{-k} \csc (\gamma ) \sec (\gamma ) \sin \left(\frac{\pi  (\pi -2 \gamma )
   (k+1)}{\pi -\gamma }\right) \sin \left(\frac{(\pi -2 \gamma ) (\gamma +\pi  k)}{\pi -\gamma
   }\right) }{4 (\pi -\gamma )\, \sin\left(\frac{\pi ^2 (k+1)}{\pi -\gamma }\right) \,
\cos \left(\frac{\pi  \gamma 
   (k+1)}{\pi -\gamma }\right) \, \cos \left(\frac{\gamma  (\gamma +\pi  k)}{\pi -\gamma
   }\right)
},
\label{V4kp} \nonumber 
\end{eqnarray}
\begin{eqnarray}
V_{1,k}^{(-)}&=&   \csc (\gamma ) \sec (\gamma ) \sin (\gamma  (k+1))- \delta_{k,0} \,\frac{\pi  \sec (\gamma )}{2 \pi -2 \gamma },
\label{V1km} \nonumber \\
V_{2,k}^{(-)}&=& -\frac{\pi  (-1)^k \csc ^2(\gamma ) \sec (\gamma ) \sin \left(2 \gamma +\frac{\pi ^2 (2 k+1)}{2
   \gamma }\right) \sec \left(\gamma +\frac{\pi ^2 (2 k+1)}{2 \gamma }\right)}{4 \gamma },
\label{V2km} \nonumber \\
V_{3,k}^{(-)}&=&   \frac{\pi  (-1)^k \csc ^2(\gamma ) \sec (\gamma ) \sin \left(\frac{\pi ^2 (2 k+1)}{2 \gamma }-2
   \gamma \right) \sec \left(\frac{\pi ^2 (2 k+1)}{2 \gamma }\right)}{4 \gamma },
\label{V3km} \nonumber \\
V_{4,k}^{(-)}&=& -\frac{\pi  (-1)^k \csc (\gamma ) \sec (\gamma ) \sin \left(\frac{\pi  (\pi -2 \gamma )
   (k+1)}{\pi -\gamma }\right) \sin \left(\frac{(\pi -2 \gamma ) (\pi  (k+2)-\gamma )}{\pi
   -\gamma }\right) }{4 (\pi -\gamma ) \, \sin \left(\frac{\pi ^2 (k+1)}{\pi -\gamma }\right) \cos \left(\frac{\pi 
   \gamma  (k+1)}{\pi -\gamma }\right) \cos \left(\frac{\gamma  (\pi  (k+2)-\gamma )}{\pi
   -\gamma }\right) }.
\label{V4km} \nonumber
\end{eqnarray}
\begin{equation} \label{GMgFc}
\begin{split}
(G^-_\lambda\star g \star F_c)(\lambda)\stackrel{\lambda \to \pm \infty}{=}\sum\limits_{k=0}^\infty \left\{ 
J_{1,k}^{(\pm)} \, e^{\mp 2 (1+k)  \lambda}\right. &+
J_{2,k}^{(\pm)} \, e^{\mp (1+2 k) \tfrac{\pi}{\gamma} \lambda}+
J_{3,k}^{(\pm)} \, e^{-2 \lambda \mp (1+2 k) \tfrac{\pi}{\gamma} \lambda}  \\
 &+ \left. J_{4,k}^{(\pm)} \, e^{-2 \lambda \mp 2 (1+k) \tfrac{\pi}{\pi-\gamma} \lambda} \right\}, \qquad \text{with:}
\end{split}
\end{equation}
\begin{eqnarray}
J_{1,k}^{(+)}&=&  \csc (\gamma ) \sec (\gamma ) \sin (\gamma  (k+1))- \delta_{k,0} \, \frac{\pi  \sec (\gamma )}{2 \pi -2 \gamma },
\label{J1kp} \nonumber \\
J_{2,k}^{(+)}&=&  -\frac{\pi  (-1)^{-k} \csc ^2(\gamma ) \sec (\gamma ) \sin \left(2 \gamma +\frac{\pi ^2 (2
   k+1)}{2 \gamma }\right) \sec \left(\gamma +\frac{\pi ^2 (2 k+1)}{2 \gamma }\right)}{4 \gamma
   }, 
\label{J2kp} \nonumber \\
J_{3,k}^{(+)}&=&  \frac{\pi  (-1)^{-k} \csc ^2(\gamma ) \sec (\gamma ) \sin \left(\frac{\pi ^2 (2 k+1)}{2 \gamma
   }-2 \gamma \right) \sec \left(\frac{\pi ^2 (2 k+1)}{2 \gamma }\right)}{4 \gamma }
\label{J3kp} \nonumber \\
J_{4,k}^{(+)}&=& \frac{\pi  (-1)^{-k} \csc (2 \gamma ) \sin \left(\frac{\pi  (\pi -2 \gamma ) (k+1)}{\pi -\gamma
   }-2 \gamma \right) \sin \left(\frac{\pi  (\pi -2 \gamma ) (k+1)}{\pi -\gamma }\right)}{2 (\pi -\gamma )\, \sin
   \left(\frac{\pi  (\gamma +\pi  (-k-2))}{\pi -\gamma }\right) \cos \left(\frac{\pi  \gamma 
   (k+1)}{\pi -\gamma }\right) \cos \left(\frac{\gamma  (\pi  (k+2)-\gamma )}{\pi -\gamma
   }\right)},
\label{J4kp} \nonumber 
\end{eqnarray}
\begin{eqnarray}
J_{1,k}^{(-)}&=&  \csc (\gamma ) \sec (\gamma ) \sin (\gamma  k),
\label{J1km} \nonumber \\
J_{2,k}^{(-)}&=&  \frac{\pi  (-1)^k \csc ^2(\gamma ) \sec (\gamma ) \sin \left(\frac{\pi ^2 (2 k+1)}{2 \gamma }-2
   \gamma \right) \sec \left(\frac{\pi ^2 (2 k+1)}{2 \gamma }-\gamma \right)}{4 \gamma },
\label{J2km} \nonumber \\
J_{3,k}^{(-)}&=& -\frac{\pi  (-1)^k \csc ^2(\gamma ) \sec (\gamma ) \sin \left(2 \gamma +\frac{\pi ^2 (2 k+1)}{2
   \gamma }\right) \sec \left(\frac{\pi ^2 (2 k+1)}{2 \gamma }\right)}{4 \gamma }, 
\label{J3km} \nonumber \\
J_{4,k}^{(-)}&=& \frac{\pi  (-1)^k \csc (\gamma ) \sec (\gamma ) \sin \left(\frac{(\pi -2 \gamma ) (\gamma +\pi 
   k)}{\pi -\gamma }\right) \sin \left(\frac{\pi  (\gamma +2 \gamma  k-\pi  k)}{\pi -\gamma
   }\right) }{4 (\pi - \gamma) \,\sin \left(\frac{\pi  (\gamma +\pi  k)}{\pi -\gamma }\right) \cos \left(\frac{\pi 
   \gamma  (k+1)}{\pi -\gamma }\right) \cos \left(\frac{\gamma  (\gamma +\pi  k)}{\pi -\gamma
   }\right)  }.
\label{J4km} \nonumber
\end{eqnarray}
\begin{equation} \label{GPsiChiF}
\begin{split}
(G_\lambda\star\psi' \star \chi_F)(\lambda)\stackrel{\lambda \to \pm \infty}{=}\sum\limits_{k=1}^\infty \left\{ 
\left(2 \lambda \, \hat{I}^{(\pm)}_{1,k}+I^{(\pm)}_{1,k} \right) \, e^{\mp (2 k-1) \tfrac{\pi}{\gamma} \lambda}+
I_{2,k}^{(\pm)} \, e^{\mp 2 k  \lambda} \right.\\
 \left.+I_{3,k}^{(\pm)} \, e^{\mp 2 k \tfrac{\pi}{\pi-\gamma} \lambda} \right\}, 
\end{split}
\end{equation}
with:
\begin{eqnarray}
\hat{I}_{1,k}^{(+)}&=&  \frac{(-1)^{k+1} (p+1)^2}{\pi }, 
\label{hI1kp} \nonumber \\
I_{1,k}^{(+)}&=& \frac{1}{2} (-1)^k (p+1) (3 \cos (\pi  (1-2 k) p)-1) \csc (\pi  (1-2 k) p),  
\label{I1kp} \nonumber \\
I_{2,k}^{(+)}&=& \frac{1}{4} (-1)^k \sin \left(\frac{2 \pi  k (p-1)}{p+1}\right) \csc \left(\frac{\pi  k
   p}{p+1}\right) \sec ^2\left(\frac{\pi  k}{p+1}\right), 
\label{I2kp} \nonumber \\
I_{3,k}^{(+)}&=& \frac{(p+1) \sec (\pi  k) \sin \left(\frac{2 \pi  k (p-1)}{p}\right) \csc \left(\frac{\pi  k
   p+\pi  k}{p}\right) \sec ^2\left(\frac{\pi  k}{p}\right)}{4 p},
\label{I4kp} \nonumber 
\end{eqnarray}
\begin{eqnarray}
\hat{I}_{1,k}^{(-)}&=&   \frac{(-1)^{k+1} (p+1)^2}{\pi },
\label{hI1km} \nonumber \\
I_{1,k}^{(-)}&=& -\frac{1}{2} (-1)^{k+1} (p+1) (3 \cos (\pi  (2 (k+1)+1) p)-1) \csc (\pi  (2 (k+1)+1) p), 
\label{I1km} \nonumber \\
I_{2,k}^{(-)}&=&   -\frac{1}{4} (-1)^k \sin \left(\frac{2 \pi  k (p-1)}{p+1}\right) \csc \left(\frac{\pi  k
   p}{p+1}\right) \sec ^2\left(\frac{\pi  k}{p+1}\right),
\label{I2km} \nonumber \\
I_{3,k}^{(-)}&=& \frac{(p+1) \sec \left(\frac{\pi  k}{p}\right)}{2 p}.
\label{I3km} \nonumber
\end{eqnarray}
\begin{equation} \label{gFc}
\begin{split}
(g \star F_c)(\lambda)\stackrel{\lambda \to \pm \infty}{=}\sum\limits_{k=0}^\infty \left\{ 
H_{1,k}^{(\pm)} \, e^{\mp 2 (k+1)  \lambda}+H^{(\pm)}_{2,k} \, e^{\mp (2 k+1) \tfrac{\pi}{\gamma} \lambda}
  \right\}, \qquad \text{with:} 
\end{split}
\end{equation}
\begin{eqnarray}
H_{1,k}^{(+)}&=&   \csc (\gamma ) \sec (\gamma ) \sin (\gamma  k),
\label{H1kp} \nonumber \\
H_{2,k}^{(+)}&=&  -\frac{\pi  (-1)^k \csc (2 \gamma )}{\gamma }, 
\label{H2kp} \nonumber \\
H_{1,k}^{(-)}&=& \csc (\gamma ) \sec (\gamma ) \sin (\gamma  k),
\label{H1km} \nonumber \\
H_{2,k}^{(-)}&=& -\frac{\pi  (-1)^k \csc (2 \gamma )}{\gamma }, 
\label{H2km} \nonumber
\end{eqnarray}

\begin{equation} \label{gFcP}
\begin{split}
(g \star F_c^+)(\lambda)\stackrel{\lambda \to \pm \infty}{=}\sum\limits_{k=0}^\infty \left\{ 
K_{1,k}^{(\pm)} \, e^{\mp 2 (k+1)  \lambda}+K^{(\pm)}_{2,k} \, e^{2 \lambda \mp (2 k+1) \tfrac{\pi}{\gamma} \lambda}
  \right\}, \qquad \text{with:} 
\end{split}
\end{equation}
\begin{eqnarray}
K_{1,k}^{(+)}&=&  \csc (2 \gamma ) \sin (2 \gamma  k) \sec (\gamma +\gamma  k),
\label{K1kp} \nonumber \\
K_{2,k}^{(+)}&=& -\frac{\pi  (-1)^{-k} \csc (2 \gamma ) \sin \left(\frac{(\pi -2 \gamma ) (2 \gamma -2 \pi 
   (k+1)+\pi )}{2 \gamma }\right) \sec \left(\frac{\pi  (\gamma +2 \pi  k+\pi )}{2 \gamma
   }\right)}{\gamma },
\label{K2kp} \nonumber \\
K_{1,k}^{(-)}&=& \csc (2 \gamma ) \sin (2 \gamma  k) \sec (\gamma -\gamma  k),
\label{K1km} \nonumber \\
K_{2,k}^{(-)}&=& -\frac{\pi  (-1)^k \csc (2 \gamma ) \sin \left(\frac{(\pi -2 \gamma ) (2 \gamma +2 \pi  k+\pi
   )}{2 \gamma }\right) \sec \left(\frac{\pi  (-\gamma +2 \pi  k+\pi )}{2 \gamma
   }\right)}{\gamma }.
\label{K2km} \nonumber
\end{eqnarray}

\begin{equation} \label{gFcM}
\begin{split}
(g \star F_c^-)(\lambda)\stackrel{\lambda \to \pm \infty}{=}\sum\limits_{k=0}^\infty \left\{ 
L_{1,k}^{(\pm)} \, e^{\mp 2 (k+1)  \lambda}+L^{(\pm)}_{2,k} \, e^{-2 \lambda \mp (2 k+1) \tfrac{\pi}{\gamma} \lambda}
  \right\}, \qquad \text{with:} 
\end{split}
\end{equation}
\begin{eqnarray}
L_{1,k}^{(+)}&=&  \csc (2 \gamma ) \sin (2 \gamma  k) \sec (\gamma -\gamma  k),
\label{L1kp} \nonumber \\
L_{2,k}^{(+)}&=& -\frac{\pi  (-1)^k \csc (2 \gamma ) \sin \left(\frac{(\pi -2 \gamma ) (2 \gamma +2 \pi  k+\pi
   )}{2 \gamma }\right) \sec \left(\frac{\pi  (-\gamma +2 \pi  k+\pi )}{2 \gamma
   }\right)}{\gamma },
\label{L2kp} \nonumber \\
L_{1,k}^{(-)}&=& \csc (2 \gamma ) \sin (2 \gamma  k) \sec (\gamma +\gamma  k),
\label{L1km} \nonumber \\
L_{2,k}^{(-)}&=& -\frac{\pi  (-1)^k \csc (2 \gamma ) \sin \left(2 \gamma +\frac{\pi ^2 (2 k+1)}{2 \gamma }\right)
   \sec \left(\frac{\pi  (-\gamma +2 \pi  k+\pi )}{2 \gamma }\right)}{\gamma }.
\label{L2km} \nonumber
\end{eqnarray}

\begin{equation} \label{gachi}
\begin{split}
(g_{\gamma/2} \star \chi)(\lambda) \! \stackrel{\lambda \to \pm \infty}{=} \mp 2 \chi(\infty)+
\sum\limits_{k=1}^{\infty} \left \{ e^{(\pm)}_{1,k} \, e^{\mp 2 k \lambda} +e^{(\pm)}_{2,k} \, e^{(1\mp 2 k) \tfrac{\pi}{\gamma}\lambda}
+e^{(\pm)}_{3,k} \, e^{\mp  \tfrac{2 k \, \pi}{\pi-\gamma}\lambda}
\right \},
\end{split}
\end{equation}
\begin{eqnarray}
e_{1,k}^{(+)}&=& -2 e^{i \gamma  k}, \qquad   \qquad \qquad \qquad   \qquad \qquad \qquad e_{1,k}^{(-)}=-e_{1,-k}^{(+)},
\label{e1kp} \nonumber \\
e_{2,k}^{(+)}&=& -\frac{i \pi  (-1)^k e^{-\frac{i \pi ^2 (2 k-1)}{2 \gamma }} \sec \left(\frac{\pi ^2 (2 k-1)}{2
   \gamma }\right)}{\gamma }, \quad \quad e_{2,k}^{(-)}=-e_{2,-k}^{(+)},
\label{e2kp} \nonumber \\
e_{3,k}^{(+)}&=& \frac{\pi  \sec \left(\frac{\pi  \gamma  k}{\pi -\gamma }\right)}{\pi -\gamma }, \qquad \qquad \qquad  \qquad \qquad \qquad e_{3,k}^{(-)}=-e_{3,-k}^{(+)}. 
\label{e3kp} \nonumber 
\end{eqnarray}
Then 
\begin{equation} \label{f2G}
\begin{split}
(f_2 \star G)(\lambda)&=2 \cosh(2 \rho_0) \, \chi(\infty)+\sinh(2 \rho_0) \, (g_{\gamma/2} \star \chi)(\lambda+\rho_0),\\
f_2(\lambda|\rho_0)&=e^{-2 \rho_0}+\sinh(2 \rho_0) \,\sum\limits_{k=1}^{\infty}  e_{1,k}^{(+)} \, e^{-2 k (\lambda+\rho_0)},
\end{split}
\end{equation}
\begin{equation} \label{f3G}
\begin{split}
(f_3 \star G)(\lambda)&=2 \cosh(2 \rho_0+i \, \gamma) \, \chi(\infty)-\sinh(2 \rho_0+i \gamma) \, (g_{\gamma/2} \star \chi)(\lambda-\rho_0),\\
f_3(\lambda|\rho_0)&=e^{-2 \rho_0-i \, \gamma}-\sinh(2 \rho_0+i \, \gamma) \, \sum\limits_{k=1}^{\infty} e_{1,k}^{(-)} \, e^{2 k (\lambda-\rho_0)}.
\end{split}
\end{equation}
For $\alpha \in (0,\pi):$
\begin{equation} \label{Hapm}
\begin{split}
(g_\alpha \star \chi_F)(\pm 2 \rho_0)=\mp \tfrac{1}{2}+\sum\limits_{k=0}^{\infty} \hat{a}_k^+  e^{\mp \tfrac{2 \pi}{\gamma}(1+2k) \rho_0}
+\sum\limits_{k=1}^{\infty} \hat{b}_k^+  e^{\mp 4 k \rho_0},
\end{split}
\end{equation}
\begin{eqnarray}
\hat{a}_{k}^{(+)}&=&\-\frac{\pi  (-1)^k e^{-\frac{i \pi  (\pi -2 \alpha ) (2 k-3)}{2 \gamma }} \sec \left(\frac{\pi 
   (\gamma +\pi  (2 k-3))}{2 \gamma }\right)}{\gamma },    \qquad \hat{a}_k^-=-(\hat{a}_{k}^{(+)})^*,
\label{ahkp} \nonumber \\
\hat{b}_{k}^{(+)}&=&-e^{2 \, i \, \alpha \,  k} \sec (\gamma  k),\quad \qquad \qquad \qquad \qquad \qquad \qquad \, \, \hat{b}_k^-=-(\hat{b}_{k}^{(+)})^*,  
\label{bhkp} \nonumber 
\end{eqnarray}
where here $*$ denotes complex conjugation. Finally, we close this appendix with the large $\rho_0$ series 
representation of $I(\rho_0)$ and ${\cal T}_0 (\rho_0)$ given by (\ref{Irho0}) and (\ref{Talfa0}), respectively.
\begin{equation} \label{Irho}
\begin{split}
I(\rho_0)=\gamma^2+\sum\limits_{k=0}^\infty \hat{I}_{1,k} \, e^{-4 \, k \, \rho_0}+
\sum\limits_{k=0}^\infty \left\{ (\hat{I}_{2,k} +\hat{I}_{3,k} \, \rho_0) \, e^{-\tfrac{2 \pi}{\gamma}\,(1+2 k) \rho_0}
\right\},
\end{split}
\end{equation}
\begin{eqnarray}
\hat{I}_{1,k}&=&  -2 \gamma ^2 \left(\tan ^2(\gamma  k)-1\right), 
\label{Ih1k} \nonumber \\
\hat{I}_{2,k}&=&  2 \pi  \left(\pi  \csc ^2\left(\frac{\pi ^2 (2 k+1)}{2 \gamma }\right)-2 \gamma \right), 
\label{Ih1k} \nonumber \\
\hat{I}_{3,k}&=&   8 \pi  \cot \left(\frac{\pi ^2 (2 k+1)}{2 \gamma }\right).
\label{Ih3k} \nonumber 
\end{eqnarray}

\begin{equation} \label{calT0}
\begin{split}
{\cal T}_0 (\rho_0)=e^{-2 \rho_0} \, t_0+\sum\limits_{k=0}^\infty \left\{
t_{1,k} \, e^{2 \rho_0-\tfrac{2 \pi}{\gamma}(1+2k) \, \rho_0}+
t_{2,k} \, e^{-2 \rho_0-\tfrac{2 \pi}{\gamma}(1+2k) \, \rho_0}+
t_{3,k} \, e^{ -2\, (k+2)\, \rho_0}   \right \},
\end{split}
\end{equation}
with:
\begin{eqnarray}
t_{0}&=& 2 \gamma ^2 \sec (\gamma ),  
\label{t0} \nonumber \\
t_{1,k}&=&   2 \pi  \gamma  \left(\csc (\gamma ) \cot \left(\frac{\pi ^2 (2 k+1)}{2 \gamma }\right)-\sec
   (\gamma )\right),
\label{t1k} \nonumber \\
t_{2,k}&=&   -2 \pi  \gamma  \left(\sec (\gamma )+\csc (\gamma ) \cot \left(\frac{\pi ^2 (2 k+1)}{2 \gamma
   }\right)\right),
\label{t2k} \nonumber \\
t_{3,k}&=&   2 \gamma ^2 \left[2 \sec (\gamma )-\sec (\gamma  (k+1)) \sec (\gamma  (k+2))\right].
\label{t3k} \nonumber 
\end{eqnarray}


\end{document}